\documentclass[prd,preprintnumbers,twocolumn,amsmath,nofootinbib,amssymb]{revtex4}

\usepackage[T1]{fontenc}
\usepackage{graphicx,color,dcolumn,booktabs,bm}
\usepackage{longtable,lscape}
\usepackage{makecell}
\usepackage{txfonts}
\usepackage{overpic}
\usepackage{amssymb}
\usepackage{amsmath}
\usepackage{epstopdf}
\usepackage{indentfirst}
\usepackage{feynmf}   %{feynmp}
\usepackage{slashed}  %for Feynman symbols
\usepackage{cases}
\usepackage{color}
\usepackage{float}
\usepackage{multirow}
\usepackage{ulem}
\usepackage{enumerate}
\usepackage{siunitx}
\usepackage{braket}
\usepackage{graphicx,color,dcolumn,booktabs,bm}
\usepackage{epsfig,dsfont,amssymb,amsmath,amsfonts,amsbsy,mathrsfs}
\graphicspath{{Figures/}} %

\usepackage{hyperref}
\usepackage{bbding}
\newcommand{\wignerthreej}[6]{%
	\left( \begin{array}{ccc}
		#1 & #2 & #3 \\
		#4 & #5 & #6
	\end{array} \right)
}

\hypersetup{colorlinks,citecolor=blue,anchorcolor=red,menucolor=red, linkcolor=red,filecolor=red,runcolor=red,urlcolor=blue,frenchlinks=true}

\makeatletter
\@addtoreset{equation}{section}
\makeatother

\allowdisplaybreaks

\begin{document}
%DT18082	
\title{Mass spectra and electromagnetic characteristics of the $K^{(*)}\bar D^{(*)}$ and $K^{(*)}{D}^{(*)}$ molecular tetraquarks \\from the coupled-channel dynamics}
	
\author{Jia-Ning Cui$^{1}$}
\author{Fu-Lai Wang$^{1,2,3,4}$}
\email{wangfulai@lzu.edu.cn}
\affiliation{$^1$School of Physical Science and Technology, Lanzhou University, Lanzhou 730000, China\\
$^2$Lanzhou Center for Theoretical Physics, Key Laboratory of Theoretical Physics of Gansu Province, Key Laboratory of Quantum Theory and Applications of MoE, Gansu Provincial Research Center for Basic Disciplines of Quantum Physics, Lanzhou University, Lanzhou 730000, China\\
$^3$MoE Frontiers Science Center for Rare Isotopes, Lanzhou University, Lanzhou 730000, China\\
$^4$Research Center for Hadron and CSR Physics, Lanzhou University and Institute of Modern Physics of CAS, Lanzhou 730000, China}

\begin{abstract}
Motivated by the observation of numerous charmed-strange hadrons lying close to the $K^{(*)}\bar D^{(*)}$ and $K^{(*)}D^{(*)}$ thresholds, we systematically investigate their mass spectra and electromagnetic characteristics, explicitly incorporating the $S$-$D$ wave mixing and coupled-channel effects. For the mass spectra, we employ the one-boson-exchange model and obtain several promising $K^{(*)}\bar D^{(*)}$ and $K^{(*)}D^{(*)}$ molecular candidates awaiting future experimental confirmation. The coupled-channel dynamics is found to play an essential role: it not only enhances binding in existing channels but also generates new loosely bound states that are absent in the single-channel approximation. Regarding the electromagnetic characteristics, we discuss the M1 radiative decay widths and magnetic moments of these charmed-strange molecular tetraquarks based on our obtained mass spectra and spatial wave functions within the constituent quark model. Our results demonstrate that the electromagnetic observables serve as valuable discriminants for clarifying the nature of these hadronic molecules. We encourage experimental collaborations to focus on such predicted charmed-strange molecular tetraquark candidates, especially the genuinely exotic $K^{(*)}\bar D^{(*)}$ states comprising four different flavors with valence quark content $\bar c \bar s u d$.
\end{abstract}
\maketitle
	
\section{Introduction}\label{sec1}

Since the discovery of the $X(3872)$ by the Belle Collaboration in 2003 \cite{Belle:2003nnu}, the study of the exotic hadrons has become a central theme in hadron physics \cite{Liu:2013waa,Hosaka:2016pey,Chen:2016qju,Richard:2016eis,Lebed:2016hpi,Liu:2019zoy,Brambilla:2019esw,Chen:2022asf,Olsen:2017bmm,Guo:2017jvc,Meng:2022ozq,Liu:2024uxn,Wang:2025sic,Wang:2025dur,Bai:2026atm}. Unlike the conventional mesons and baryons, these states resist interpretation within the simple quark-antiquark or three-quark pictures of the conventional quark model, suggesting more complex configurations such as hadronic molecules, compact multiquarks, glueballs, hybrids, and others. They offer a unique window into the hadron structures and the non-perturbative dynamics of quantum chromodynamics (QCD). Among the various proposals, the hadronic molecule picture, where two or more conventional hadrons bind via the residual strong interactions, has gained considerable traction, largely due to the observation of numerous near-threshold hadronic states \cite{Brambilla:2019esw}.

\begin{figure}[htbp]
\centering
\includegraphics[width=8.6cm]{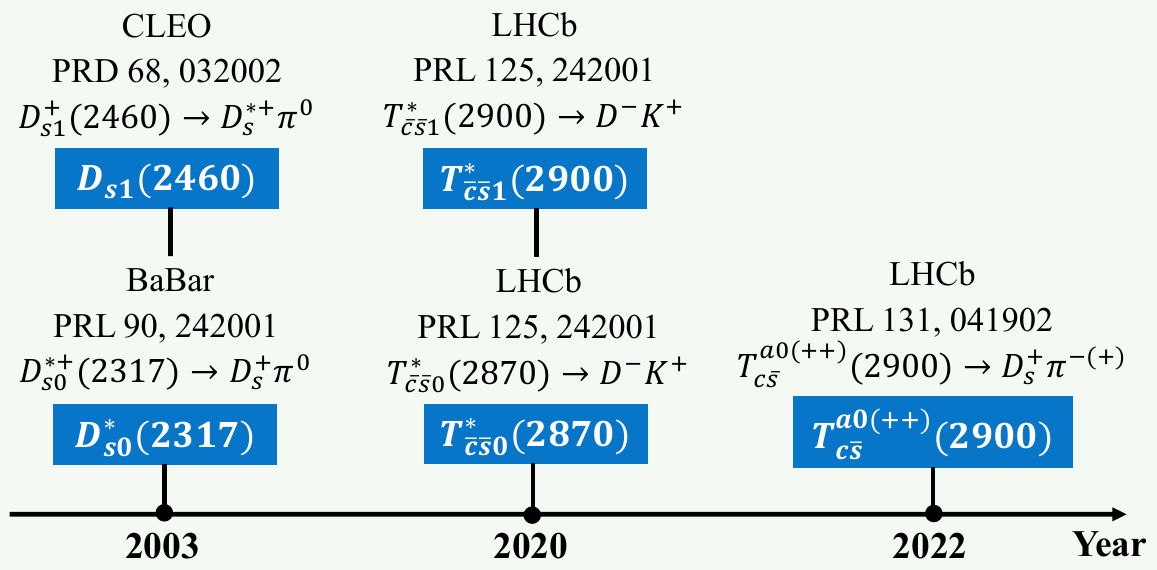}
\caption{Experimental discoveries of the charmed-strange hadrons.}\label{timeline}
\end{figure}

Within this landscape, the charm-strange molecular tetraquarks have attracted widespread attention as an important platform for searching for the tetraquark hadrons. The simplest configurations in this sector are the $S$-wave loosely bound states formed by a ground-state kaon ($K^{(*)}$) and a ground-state anti-charmed or charmed meson ($\bar D^{(*)}$ or $D^{(*)}$). Notably, the valence quark content of the $K^{(*)}\bar D^{(*)}$ systems ($\bar c\bar s u d$) is manifestly exotic, comprising four different flavors, which unambiguously signals new hadronic degrees of freedom beyond the conventional hadrons. Experimentally, a series of charmed-strange hadrons have been observed near the corresponding thresholds. The investigation traces back to 2003, when the BaBar Collaboration discovered the $D_{s0}^*(2317)$ in the $D_s^+\pi^0$ invariant mass distribution \cite{BaBar:2003oey}, followed shortly by the observation of the $D_{s1}(2460)$ by the CLEO Collaboration \cite{CLEO:2003ggt}. Both states exhibit masses anomalously lower than theoretical predictions for the conventional $c\bar s$ mesons \cite{Godfrey:1985xj,Godfrey:1986wj,Godfrey:2015dva,Ebert:2009ua,Song:2015nia}, and their proximity to the $KD$ and $KD^*$ thresholds leads to their interpretation as the $KD$ and $KD^*$ molecular candidates \cite{Barnes:2003dj, Kolomeitsev:2003ac, Hofmann:2003je, Chen:2004dy, Sassen:2005ej, Zhang:2006ix, Guo:2006fu, Guo:2006rp, Shen:2007zza, Flynn:2007ki, Gamermann:2007fi, Faessler:2007us, Gamermann:2006nm, Xie:2010zza, Wang:2012bu, Feng:2012zze, Feng:2012zzf, Liu:2012zya, Agadjanov:2014ana, MartinezTorres:2014kpc, Navarra:2015iea, Guo:2015dha, Ortega:2016mms, Albaladejo:2016hae, Albaladejo:2016jhb, Du:2017ttu, Bali:2017pdv, Albaladejo:2018mhb, Guo:2018tjx, Matuschek:2020gqe, Cheung:2020mql, Kong:2021ohg, Xiao:2024kfq, Huang:2021fdt, Albaladejo:2022sux, Guo:2022izw, Chen:2022svh, Liu:2023uly, Ikeno:2023ojl, Gil-Dominguez:2023puj, Kim:2023htt, Montesinos:2024uhq, Yeo:2024chk, Li:2024rlw, Shen:2025qpj, Zhou:2025yjb}. A major breakthrough occurred in 2020 when the LHCb Collaboration reported two new structures, $T_{\bar c \bar s 0}^{*}(2870)$ and $T_{\bar c \bar s 1}^{*}(2900)$, in the $D^-K^+$ invariant mass spectrum of the $B^+\to D^+D^-K^+$ decay \cite{LHCb:2020bls,LHCb:2020pxc}. The $T_{\bar c \bar s 0}^{*}(2870)$, with valence quark content $\bar c\bar s u d$, lies remarkably close to the $K^*\bar D^*$ threshold, supporting the $K^*\bar D^*$ molecular assignment \cite{Kong:2021ohg,Xiao:2024kfq,Molina:2020hde, Liu:2020nil, Chen:2020aos, Hu:2020mxp, Agaev:2020nrc, He:2020btl, Wang:2021lwy, Dai:2022htx, Chen:2023syh, Ding:2024dif, Ding:2025uhh, Ke:2022ocs, Molina:2010tx}. In 2022, the isovector partners $T_{c\bar s}^{a0(++)}(2900)$  were observed by LHCb in the $D_s^+\pi^{-(+)}$ channels \cite{LHCb:2022sfr,LHCb:2022lzp}. Sharing $I(J^P)=1(0^+)$ quantum numbers and lying close to the $K^*D^*$ threshold, the $T_{c\bar s}^{a0(++)}(2900)$ have been discussed as candidates for isovector $K^*D^*$ molecules \cite{Chen:2022svh,Agaev:2022eyk, Duan:2023lcj, Wang:2023hpp, Molina:2023ghu, Duan:2023qsg, Yeo:2026dgo}. These experimental discoveries is summarized in Fig. \ref{timeline}. These discoveries clearly indicate that the $K^{(*)}\bar D^{(*)}$ and $K^{(*)}D^{(*)}$ systems are ideal platforms for investigating the charmed-strange molecular tetraquarks.

Extensive theoretical studies of the $K^{(*)}\bar D^{(*)}$ and $K^{(*)}D^{(*)}$ molecular candidates have been performed using a variety of methods, such as the one-boson-exchange (OBE) model \cite{Chen:2016qju,Guo:2017jvc,Liu:2019zoy,Chen:2022asf,Meng:2022ozq}. However, most of these works are limited to the single-channel approximations. Recent advances in the hadronic molecular states have shown that the coupled-channel dynamics can be crucial, and sometimes even essential, for generating the loosely bound states. For example, the $P_c(4312)$ \cite{LHCb:2019kea} and $P_{cs}(4459)$ \cite{LHCb:2020jpq} cannot be properly described without incorporating couplings to nearby channels \cite{Chen:2019asm,Chen:2020kco,Wang:2022mxy}. Therefore, a comprehensive investigation that fully accounts for the coupled-channel effects, including the $S$-$D$ wave mixing and couplings among all relevant $K^{(*)}\bar D^{(*)}$ or $K^{(*)}D^{(*)}$ channels, is needed to provide reliable predictions for the mass spectra of these molecular candidates.

Furthermore, the internal structures of the hadronic states typically cannot be deciphered from their mass spectra alone. Therefore, it is essential to investigate other properties, which serve as a crucial complement to the mass spectra. Among them, the electromagnetic  characteristics have long been recognized as powerful probes of the hadron structures. In particular, the M1 radiative decay widths and magnetic moments of the hadrons are highly sensitive to the spatial distribution, spin composition, and flavor content of their wave functions, and they differ markedly among molecular states, compact multiquarks, and conventional mesons. In recent years, such observables have been widely used to discriminate between different configurations of the hadronic states \cite{Ozdem:2018qeh, Xu:2020flp, Ozdem:2021btf, Ozdem:2021ugy, Ozdem:2021yvo, Ozdem:2021hka, Ozdem:2021vry, Ozdem:2022ylm, Ozdem:2022yhi, Ozdem:2023eyz, Mutuk:2024jxf, Mutuk:2024elj, Guo:2023fih, Li:2024wxr, Lei:2024geu, Lei:2023ttd, Ozdem:2024ydl, Ozdem:2024jty, Ozdem:2024yel, Ozdem:2025olj, Deng:2021gnb, Wu:2022gie, Schlumpf:1993rm, Schlumpf:1992vq, Cheng:1997kr, Ha:1998gf, Liu:2003ab, Huang:2004tn, Zhu:2004xa, Wang:2016dzu, Li:2021ryu, Wang:2022tib, Zhou:2022gra, Wang:2022nqs, Wang:2023aob, An:2022qpt, Wang:2023bek, Wang:2023ael, Gao:2021hmv, Zhang:2025ame, Lai:2024jfe, Wang:2024sbw, Zhu:2025abk, Wang:2024kke, Sheng:2024hkf, Li:2024jlq, Mutuk:2024ltc, Yue:2023qgx, Fu:2025lfo, Li:2023wgq, Zhang:2024usz, Li:2025ddx, Ozdem:2026wmf}. In particular, lattice QCD has recently begun to focus on the electromagnetic  characteristics of multiquark states. For example, Ref. \cite{Vujmilovic:2025czt} discusses the internal structure of the $T_{bb}$ tetraquark through its electromagnetic  characteristics. Nevertheless, a systematic discussion of these electromagnetic observables for the $K^{(*)}\bar D^{(*)}$ and $K^{(*)}D^{(*)}$ molecular candidates is still lacking at present.

Motivated by the above considerations, in this work we perform a systematic study of both the mass spectra and the electromagnetic  characteristics of the $K^{(*)}\bar D^{(*)}$ and $K^{(*)}D^{(*)}$ molecular candidates, explicitly incorporating the $S$-$D$ wave mixing and coupled-channel effects. Our goals are twofold: (i) to derive the effective interactions for the $K^{(*)}\bar D^{(*)}$ and $K^{(*)}D^{(*)}$ systems using the OBE model and to solve the coupled-channel Schr\"odinger equation, thereby systematically predicting such promising $K^{(*)}\bar D^{(*)}$ and $K^{(*)}D^{(*)}$ charmed-strange molecular tetraquark candidates, and (ii) to evaluate the M1 radiative decay widths and magnetic moments of these candidates within the constituent quark model, thereby providing valuable discriminants for clarifying the nature of these charmed-strange hadronic molecules.

The remainder of this paper is organized as follows. In Sec. \ref{sec2}, we present the OBE effective potentials for the $K^{(*)}\bar D^{(*)}$ and $K^{(*)}D^{(*)}$ systems and discuss the bound-state properties obtained from the coupled-channel Schr\"odinger equation. In Sec. \ref{sec3}, we evaluate the M1 radiative decay widths and magnetic moments of the identified molecular candidates. Finally, a summary and outlook are given in Sec. \ref{sec4}.

\section{Mass spectra of the $K^{(*)}\bar{D}^{(*)}$ and $K^{(*)} D^{(*)}$ molecular candidates}\label{sec2}
	
In this section, we systematically investigate the mass spectra of the $K^{(*)}\bar{D}^{(*)}$ and $K^{(*)} D^{(*)}$ molecular candidates. Starting from the OBE model, we derive the effective potentials for the $K^{(*)}\bar{D}^{(*)}$ systems by exchanging light scalar ($\sigma$), pseudoscalar ($\pi$, $\eta$), and vector ($\rho$, $\omega$) mesons \cite{Chen:2016qju}. The effective potentials for the $K^{(*)}D^{(*)}$ systems are then obtained from those of the $K^{(*)}\bar{D}^{(*)}$ systems via the $G$-parity rule \cite{Klempt:2002ap}. These effective potentials are used to solve the coupled-channel Schr\"odinger equation, where both the $S$-$D$ wave mixing and coupled-channel effects are considered. The obtained bound-state properties yield the corresponding mass spectra that not only provide guidance for future experimental searches for such hadronic molecules, but also supply numerically determined their spatial wave functions. The latter serve as the essential inputs for subsequent studies of their electromagnetic characteristics.
	
\subsection{OBE effective potentials for the $K^{(*)}\bar{D}^{(*)}$ and $K^{(*)}D^{(*)}$ systems}
	
To derive the OBE effective potentials for the $K^{(*)}\bar{D}^{(*)}$ systems, we start from the effective Lagrangians describing couplings of the $S$-wave anti-charmed mesons to the light mesons. Heavy-quark symmetry, chiral symmetry, and hidden local symmetry lead to the following form \cite{Ding:2008gr}:
\begin{eqnarray}\label{effectiveLagrangians1}
\mathcal{L}_{\bar D^{(*)} \bar D^{(*)}\mathcal{E}}&=&		-2g_{\sigma}\bar{D}^{\dag}_{a}\bar{D}_{a}\sigma+2g_{\sigma}\bar{D}^{*\dag}_{a\mu}\bar{D}^{*\mu}_{a}\sigma\nonumber\\
&&+i\frac{2g}{f_{\pi}}v^{\alpha}\varepsilon_{\alpha\mu\nu\lambda}\bar{D}^{*\mu\dag}_{a}\bar{D}^{*\lambda}_{b}\partial^{\nu}\mathbb{P}_{ab}\nonumber\\
&&+\frac{2g}{f_{\pi}}(\bar{D}^{*\mu\dag}_{a}\bar{D}_{b}+\bar{D}^{\dag}_{a}\bar{D}^{*\mu}_{b})\partial_{\mu}\mathbb{P}_{ab}\nonumber\\
&&+\sqrt{2}\beta g_{V}\bar{D}_{a}\bar{D}^{\dag}_{b}v\cdot\mathbb{V}_{ab}-\sqrt{2}\beta g_{V}\bar{D}^{*}_{a\mu}\bar{D}^{*\mu\dag}_{b}v\cdot\mathbb{V}_{ab}\nonumber\\
&&-2\sqrt{2}i\lambda g_{V}\bar{D}^{*\mu\dag}_{a}\bar{D}^{*\nu}_{b}(\partial_{\mu}\mathbb{V}_{\nu}-\partial_{\nu}\mathbb{V}_{\mu})_{ab}\nonumber\\
&&-2\sqrt{2}\lambda g_{V}v^{\lambda}\varepsilon_{\lambda\mu\alpha\beta}(\bar{D}^{*\mu\dag}_{a}\bar{D}_{b}+\bar{D}^{\dag}_{a}\bar{D}^{*\mu}_{b})\partial^{\alpha}\mathbb{V}^{\beta}_{ab},
\end{eqnarray}
where $\mathcal{E}$ denotes the light mesons. The $S$-wave anti-charmed mesons are normalized as \cite{Ding:2008gr}
\begin{eqnarray}
\bra{0} \bar{D} \ket{\bar c{q}(0^{-})}&=&\sqrt{m_{{D}}},\\
\bra{0} \bar{D}^{*\mu} \ket{\bar c{q}(1^{-})}&=&
\sqrt{m_{{D}^{*}}}\epsilon^{\mu}.
\end{eqnarray}
In the static limit, the polarization vector $\epsilon^{\mu}$ for $\bar{D}^{*}$ is taken as $\epsilon^{\mu}_0 = (0,0,0,-1)$ and $\epsilon^{\mu}_{\pm1} = (0,\pm1/\sqrt{2},i/\sqrt{2},0)$. Under the SU(2) flavor symmetry, the pseudoscalar and vector matrices $\mathbb{P}$ and $\mathbb{V}$ are given by \cite{Ding:2008gr}
\renewcommand{\arraystretch}{1.50}
\begin{eqnarray*}
\mathbb{P}=
\begin{pmatrix}
\frac{\pi^{0}}{\sqrt{2}}+\frac{\eta}{\sqrt{6}}&\pi^{+}\\
\pi^{-}&-\frac{\pi^{0}}{\sqrt{2}}+\frac{\eta}{\sqrt{6}}
\end{pmatrix},\\[0.8ex]
\mathbb{V}=
\begin{pmatrix}
\frac{\rho^{0}}{\sqrt{2}}+\frac{\omega}{\sqrt{2}}&\rho^{+}\\
\rho^{-}&-\frac{\rho^{0}}{\sqrt{2}}+\frac{\omega}{\sqrt{2}}
\end{pmatrix}.
\end{eqnarray*}
	
In accordance with Refs. \cite{Liu:2020nil,Wang:2024ukc,Wang:2024kke,Wang:2025wpc}, the effective Lagrangians describing the interactions of the $S$-wave kaons with light mesons are given by
\begin{eqnarray}\label{effectiveLagrangians2}
\mathcal{L}_{K^{(*)}K^{(*)}\mathcal{E}}&=&-2g_{\sigma}^{\prime}K^{\dag}_{a}K_{a}\sigma+2g_{\sigma}^{\prime}K^{*\dag}_{a\mu}K^{*\mu}_{a}\sigma\nonumber\\
&&+i\frac{2g^{\prime}}{f_{\pi}}v^{\alpha}\varepsilon_{\alpha\mu\nu\lambda}K^{*\mu\dag}_{a}K^{*\lambda}_{b}\partial^{\nu}\mathbb{P}_{ab}\nonumber\\
&&+\frac{2g^{\prime}}{f_{\pi}}(K^{*\mu\dag}_{a}K_{b}+K^{\dag}_{a}K^{*\mu}_{b})\partial_{\mu}\mathbb{P}_{ab}\nonumber\\
&&+\sqrt{2}\beta^{\prime} g_{V}^{\prime}K_{a}K^{\dag}_{b}v\cdot\mathbb{V}_{ab}-\sqrt{2}\beta^{\prime} g_{V}^{\prime}K^{*}_{a\mu}K^{*\mu\dag}_{b}v\cdot\mathbb{V}_{ab}\nonumber\\
&&-2\sqrt{2}i\lambda^{\prime} g_{V}^{\prime}K^{*\mu\dag}_{a}K^{*\nu}_{b}(\partial_{\mu}\mathbb{V}_{\nu}-\partial_{\nu}\mathbb{V}_{\mu})_{ab}\nonumber\\
&&-2\sqrt{2}\lambda^{\prime} g_{V}^{\prime}v^{\lambda}\varepsilon_{\lambda\mu\alpha\beta}(K^{*\mu\dag}_{a}K_{b}+K^{\dag}_{a}K^{*\mu}_{b})\partial^{\alpha}\mathbb{V}^{\beta}_{ab},
\end{eqnarray}
where $K^{(*)}$ represents the doublet $(K^{(*)+},\,K^{(*)0})$, with normalizations \cite{Wang:2025wpc,Wang:2024kke}
\begin{eqnarray}
\bra{0} K \ket{q\bar{s}(0^{-})}&=&\sqrt{m_{K}},\\
\bra{0} K^{*\mu} \ket{q\bar{s}(1^{-})}&=&
\sqrt{m_{K^{*}}}\epsilon^{\mu}.
\end{eqnarray}
	
Coupling constants in Eqs. (\ref{effectiveLagrangians1}) and (\ref{effectiveLagrangians2}) are fixed from either the experimental data or the theoretical estimates. Guided by the quark model \cite{Riska:2000gd}, we take $g_{\sigma}=g_{\sigma}^{\prime}=-2.82$ \cite{Wang:2025wpc,Sheng:2024hkf} and set $\lambda^{\prime}=\SI{0.56}{GeV^{-1}}$ \cite{Wang:2024ukc}. Vector meson dominance gives $\beta=0.90$ \cite{Isola:2003fh,Cleven:2016qbn}, while $\lambda=\SI{0.56}{GeV^{-1}}$ is extracted by comparing lattice QCD and light-cone sum rule form factors \cite{Isola:2003fh}. The decays $D^{*}\to D\pi$ and $K^{*}\to K\pi$ \cite{ParticleDataGroup:2022pth} yield $g=0.59$ and $g^{\prime}=1.12$ \cite{Wang:2024ukc}, respectively. Hidden-gauge symmetry of vector mesons implies $\beta^{\prime}=0.835$ \cite{Molina:2010tx}. In addition, we adopt $f_{\pi}=\SI{132}{MeV}$ and $g_{V}=g_{V}^{\prime}=5.83$ \cite{Isola:2003fh,Bando:1987br}.
	
The scattering amplitude $\mathcal{M}^{K^{(*)}\bar{D}^{(*)}\to K^{(*)}\bar{D}^{(*)}}(\boldsymbol{q})$ is evaluated from the above effective Lagrangians. Using the Breit approximation \cite{Berestetskii:1982qgu}, the momentum-space potential $\mathcal{V}^{K^{(*)}\bar{D}^{(*)}\to K^{(*)}\bar{D}^{(*)}}(\boldsymbol{q})$ reads
\begin{eqnarray}
\mathcal{V}^{K^{(*)}\bar{D}^{(*)}\to K^{(*)}\bar{D}^{(*)}}(\boldsymbol{q}) =
 -\frac{\mathcal{M}^{K^{(*)}\bar{D}^{(*)}\to K^{(*)}\bar{D}^{(*)}}(\boldsymbol{q})}{\sqrt{\prod_{i}2m_{i}\prod_{f}2m_{f}}},
\end{eqnarray}
where $m_{i,\,f}$ denote the masses of the initial and final particles. A Fourier transform then gives the coordinate-space potential $\mathcal{V}^{K^{(*)}\bar{D}^{(*)}\to K^{(*)}\bar{D}^{(*)}}(\boldsymbol{r})$. To account for the hadron finite size, we introduce a monopole form factor $\mathcal{F}_{M}(q,m_{\mathcal{E}}) = (\Lambda^{2}-m_{\mathcal{E}}^{2})/(\Lambda^{2}-q^{2})$ at each vertex \cite{Tornqvist:1993ng,Tornqvist:1993vu}, with cutoff $\Lambda$. Following the deuteron description \cite{Tornqvist:1993ng,Tornqvist:1993vu} and successful molecular candidates such as $P_c$ and $T_{cc}$ \cite{Chen:2016qju,Liu:2019zoy,Chen:2022asf}, a typical scale is $\Lambda\sim 1$ GeV, i.e., a loosely bound state obtained with this scale is considered a viable hadronic molecule \cite{Chen:2016qju}. In this work, $\Lambda$ is scanned from $0.8$ to $2.5$ GeV. Including the form factor, the coordinate-space potential reads
\begin{eqnarray}\label{Fourier transformation}
&&\mathcal{V}^{K^{(*)}\bar{D}^{(*)}\to K^{(*)}\bar{D}^{(*)}}(\boldsymbol{r})\nonumber\\
&=&\int\frac{d^{3}\boldsymbol{q}e^{i\boldsymbol{q}\cdot\boldsymbol{r}}}{(2\pi)^{3}} \mathcal{V}^{K^{(*)}\bar{D}^{(*)}\to K^{(*)}\bar{D}^{(*)}}(\boldsymbol{q}) \,\mathcal{F}^{2}_{M}(q,m_{\mathcal{E}}).
\end{eqnarray}
	
The flavor wave functions for the $K^{(*)}\bar{D}^{(*)}$ systems, with total isospin $I=1$ or $0$, are
\begin{eqnarray}
\ket{1,1}&=&\left|K^{(*)+}\bar{D}^{(*)0}\right\rangle,\\
\ket{1,0}&=&\frac{1}{\sqrt{2}}\left(\left|K^{(*)+}D^{(*)-}\right\rangle + \left|K^{(*)0}\bar{D}^{(*)0}\right\rangle \right),\\ \label{flavor10}
\ket{1,-1}&=&\left|K^{(*)0}D^{(*)-}\right\rangle,\\
\ket{0,0}&=&\frac{1}{\sqrt{2}}\left(\left|K^{(*)+}D^{(*)-}\right\rangle - \left|K^{(*)0}\bar{D}^{(*)0}\right\rangle \right). \label{flavor00}
\end{eqnarray}
We also incorporate the $S$-$D$ wave mixing and coupled-channel effects, and the spin-orbital wave functions $\ket{{}^{2S+1}L_J}$ are
\begin{eqnarray}
K\bar{D}&:&\left|Y_{L,m_{L}}\right\rangle,\label{spin0}\\   K^{*}\bar{D}&:&\sum_{m_{S},m_{L}}\mathbb{C}_{1m_{S},Lm_{L}}^{J,m_{J}}\epsilon^{\mu}_{m_{S}}\left|Y_{L,m_{L}}\right\rangle,\label{spin1}\\ K\bar{D}^{*}&:&\sum_{m_{S},m_{L}}\mathbb{C}_{1m_{S},Lm_{L}}^{J,m_{J}}\epsilon^{\mu}_{m_{S}}\left|Y_{L,m_{L}}\right\rangle,\label{spin1}\\K^{*}\bar{D}^{*}&:&\sum_{m,m^{\prime},m_{S},m_{L}}\mathbb{C}_{1m,1m^{\prime}}^{S,m_{S}}\mathbb{C}_{Sm_{S},Lm_{L}}^{J,m_{J}}\epsilon^{\mu}_{m}\epsilon^{\nu}_{m^{\prime}}\left|Y_{L,m_{L}}\right\rangle\label{spin2}
\end{eqnarray}
with $\mathbb{C}_{ab,cd}^{e,f}$ the Clebsch-Gordan coefficients and $\ket{Y_{L,m_L}}$ the spherical harmonics. Including the $S$-$D$ wave mixing effect, the relevant configurations are
\begin{eqnarray*}
K\bar{D}&:&J^{P}=0^{+}~\left|^{1}S_{0}\right\rangle,\\    K^{*}\bar{D}&:&J^{P}=1^{+}~\left|^{3}S_{1}\right\rangle,~\left|^{3}D_{1}\right\rangle,\\
K\bar{D}^{*}&:&J^{P}=1^{+}~\left|^{3}S_{1}\right\rangle,~\left|^{3}D_{1}\right\rangle,\\K^{*}\bar{D}^{*}&:&J^{P}=0^{+}~\left|^{1}S_{0}\right\rangle,~\left|^{5}D_{0}\right\rangle,\\   &&J^{P}=1^{+}~\left|^{3}S_{1}\right\rangle,~\left|^{3}D_{1}\right\rangle,~\left|^{5}D_{1}\right\rangle,\\  &&J^{P}=2^{+}~\left|^{5}S_{2}\right\rangle,~\left|^{1}D_{2}\right\rangle,~\left|^{3}D_{2}\right\rangle,~\left|^{5}D_{2}\right\rangle.
\end{eqnarray*}
	
With the flavor and spin-orbital wave functions established, the full OBE potentials for the $K^{(*)}\bar{D}^{(*)}$ and $K^{(*)}D^{(*)}$ systems are given in Appendix \ref{The effective potentials}. Solving the coupled-channel Schr\"odinger equation yields possible bound states. The resulting mass spectra are presented in the following subsection.
	
\subsection{Mass spectra of the $K^{(*)}\bar{D}^{(*)}$ molecular candidates}

In the following, we first analyze the bound state properties of the $K^{(*)}\bar{D}^{(*)}$ systems in the single-channel approach with the $S$-$D$ wave mixing effect explicitly included. Table \ref{KstbarD} summarizes the binding energies $E$, the root-mean-square radii $r_{\text{RMS}}$, and the probability weights $P$ for the bound states obtained in this framework. Among them, the single-channel potentials supply insufficient attraction to produce the loosely bound states for the $S$-wave $K\bar{D}$ and $K\bar{D}^*$ systems.
	
\begin{table}[!htbp]
\renewcommand\tabcolsep{0.08cm}
\renewcommand{\arraystretch}{1.50}
\caption{Single-channel results for the bound state properties in the $K^{(*)}\bar{D}^{(*)}$ systems. Here, $P$ denotes the probability weight of each component (in \%).}\label{KstbarD}
\begin{tabular*}{86mm}{@{\extracolsep{\fill}}ccccc}
\toprule[1pt]
\toprule[1pt]
\multirow{4}{*}{$K^*\bar{D}~[0(1^{+})]$}&$\Lambda~(\rm{GeV})$ &$E~(\rm {MeV})$ &$r_{\rm RMS}~(\rm {fm})$&$P(^{3}{S}_{1}/^{3}{D}_{1})$\\
\Xcline{2-5}{0.75pt}
&$2.14$&$-0.47$&$4.98$&\textbf{100.00}/0\\
&$2.32$&$-1.89$&$3.12$&\textbf{100.00}/0\\
&$2.50$&$-3.92$&$2.26$&\textbf{100.00}/0\\
\midrule[1.0pt]
\multirow{4}{*}{$K^*\bar{D}^*~[0(0^{+})]$}&$\Lambda~(\rm{GeV})$ &$E~(\rm {MeV})$ &$r_{\rm RMS}~(\rm {fm})$&$P(^{1}{S}_{0}/^{5}{D}_{0})$\\
\Xcline{2-5}{0.75pt}
&$0.86$&$-0.49$&$4.79$&\textbf{99.56}/0.44\\
&$0.94$&$-8.33$&$1.58$&\textbf{99.24}/0.76\\
&$1.01$&$-24.21$&$1.01$&\textbf{99.23}/0.77\\
\midrule[1.0pt]
\multirow{4}{*}{$K^*\bar{D}^*~[0(1^{+})]$}&$\Lambda~(\rm{GeV})$ &$E~(\rm {MeV})$ &$r_{\rm RMS}~(\rm {fm})$&$P(^{3}{S}_{1}/^{3}{D}_{1}/^{5}{D}_{1})$\\
\Xcline{2-5}{0.75pt}
&$1.08$&$-0.60$&$4.63$&\textbf{99.27}/0.73/$\mathcal{O}(0)$\\
&$1.18$&$-8.09$&$1.62$&\textbf{98.59}/1.41/$\mathcal{O}(0)$\\
&$1.28$&$-24.75$&$1.01$&\textbf{98.41}/1.59/$\mathcal{O}(0)$\\
\midrule[1.0pt]
\multirow{4}{*}{$K^*\bar{D}^*~[0(2^{+})]$}&$\Lambda~(\rm{GeV})$ &$E~(\rm {MeV})$ &$r_{\rm RMS}~(\rm {fm})$&$P(^{5}{S}_{2}/^{1}{D}_{2}/^{3}{D}_{2}/^{5}{D}_{2})$\\
\Xcline{2-5}{0.75pt}
&$2.04$&$-0.54$&$4.99$&\textbf{98.03}/0.13/$\mathcal{O}(0)$/1.84\\
&$2.27$&$-2.81$&$2.78$&\textbf{96.22}/0.24/$\mathcal{O}(0)$/3.54\\
&$2.50$&$-6.86$&$1.91$&\textbf{94.67}/0.31/$\mathcal{O}(0)$/5.02\\
\midrule[1.0pt]
\multirow{4}{*}{$K^*\bar{D}^*~[1(2^{+})]$}&$\Lambda~(\rm{GeV})$ &$E~(\rm {MeV})$ &$r_{\rm RMS}~(\rm {fm})$&$P(^{5}{S}_{2}/^{1}{D}_{2}/^{3}{D}_{2}/^{5}{D}_{2})$\\
\Xcline{2-5}{0.75pt}
&$2.14$&$-0.45$&$4.81$&\textbf{99.71}/0.04/$\mathcal{O}(0)$/0.25\\
&$2.22$&$-6.26$&$1.63$&\textbf{99.43}/0.08/$\mathcal{O}(0)$/0.49\\
&$2.29$&$-17.35$&$1.00$&\textbf{99.38}/0.09/$\mathcal{O}(0)$/0.53\\
\bottomrule[1pt]\bottomrule[1pt]
\end{tabular*}
\end{table}

The $K^*\bar{D}$ state with $I(J^P)=0(1^{+})$ does not experience the tensor force, and consequently no $D$-wave admixture appears in its wave function. A loosely bound state first appears at $\Lambda = 2.14$ GeV with the binding energy $E = -0.47$ MeV and a large root-mean-square radius $r_{\text{RMS}} = 4.98$ fm. Increasing $\Lambda$ to $2.32$ GeV and $2.50$ GeV deepens the binding to $-1.89$ MeV and $-3.92$ MeV, respectively, while the corresponding radius decreases to $3.12$ fm and $2.26$ fm. This behaviour reflects the enhanced effective attraction at larger cutoffs, which reduces the spatial extension of the bound state. The $^3S_1$ component remains $100\%$, confirming the absence of the $D$-wave mixing.

For the $K^*\bar{D}^*$ state with the quantum numbers $I(J^P)=0(0^{+})$, a very shallow bound state emerges at $\Lambda = 0.86$ GeV, with the binding energy $E = -0.49$ MeV and a root-mean-square radius $r_{\text{RMS}} = 4.79$ fm. The wave function is almost entirely $^1S_0$ ($99.42\%$), accompanied by a tiny $^5D_0$ admixture of only $0.44\%$. As $\Lambda$ increases to $0.94$ GeV and $1.01$  GeV, the attraction becomes significantly stronger, deepening the binding to $-8.33$ MeV and $-24.21$ MeV, while the radius shrinks to $1.58$ fm and $1.01$ fm, respectively. The $^5D_0$ component remains below $1\%$ throughout, indicating that the $S$-wave channel dominates this loosely bound state.

For the $K^*\bar{D}^*$ state with $I(J^P)=0(1^{+})$, a loosely bound state first appears at $\Lambda = 1.08$ GeV with $E = -0.60$ MeV and $r_{\text{RMS}} = 4.63$ fm. The wave function is dominated by the $^3S_1$ partial wave ($99.27\%$), while the $^3D_1$ component contributes only $0.73\%$ and the $^5D_1$ weight is negligible. This indicates that the tensor force, although present, induces only a tiny $D$-wave admixture. As $\Lambda$ increases to $1.18$ GeV and $1.28$ GeV, the enhanced attraction deepens the binding energy to $-8.09$ MeV and $-24.75$ MeV, respectively, while the radius shrinks to $1.62$ fm and $1.01$ fm. Despite the stronger binding, the $^3S_1$ probability remains above $98\%$, confirming that this loosely bound state retains a predominantly $S$-wave molecular nature.

For the $K^*\bar{D}^*$ state with $I(J^P)=0(2^{+})$, a loosely bound state emerges at $\Lambda = 2.04$ GeV with $E = -0.54$ MeV and $r_{\text{RMS}} = 4.99$ fm. The wave function is dominated by the $^5S_2$ partial wave ($98.03\%$), accompanied by a small $^5D_2$ component ($1.84\%$) and a negligible $^1D_2$ admixture ($0.13\%$). Raising $\Lambda$ to $2.27$ GeV and $2.50$ GeV gradually increases the binding energy to $-2.81$ MeV and $-6.86$ MeV, respectively, while the corresponding radius reduces to $2.78$ fm and $1.91$ fm. The $^5S_2$ weight decreases to $94.67\%$ and the $^5D_2$ share rises to $5.02\%$ as the cutoff increases. This trend reflects the growing influence of the tensor force, which admixes more $D$-wave into the predominantly $S$-wave state at stronger couplings. Nevertheless, the $S$-wave remains dominant.

For the $K^*\bar{D}^*$ state with $I(J^P)=1(2^{+})$, a shallow bound state first appears at $\Lambda = 2.14$ GeV with $E = -0.45$ MeV and $r_{\text{RMS}} = 4.81$ fm. The wave function is overwhelmingly dominated by the $^5S_2$ component ($99.71\%$), with a tiny $^5D_2$ admixture ($0.25\%$) and a negligible $^1D_2$ part ($0.04\%$). As $\Lambda$ increases to $2.22$ GeV and $2.29$ GeV, the binding energy deepens significantly to $-6.26$ MeV and $-17.35$ MeV, while the radius shrinks to $1.63$ fm and $1.00$ fm, respectively. The $^5S_2$ probability remains nearly $99.4\%$ and the $^5D_2$ fraction stays below $0.6\%$ throughout. The $\Lambda$ values required to produce a loosely bound state are somewhat larger than those for the isoscalar partners, consistent with weaker attraction in the isovector sector \cite{Klempt:2002ap}.

We now incorporate the coupled-channel effect to discuss how they modify the bound-state properties of the $K^{(*)}\bar{D}^{(*)}$ systems. Table \ref{KstbarDstCP} presents the numerical results for five distinct channel combinations. A comparison with the single-channel analysis reveals that the coupled-channel dynamics plays a crucial role in the formation of several loosely bound states: it not only enhances the attraction in already bound states but also generates new loosely bound states that do not appear in the single-channel treatment.

\renewcommand\tabcolsep{0.20cm}
\renewcommand{\arraystretch}{1.50}
\begin{table}[!htbp]
\caption{Coupled-channel results for the bound state properties in the $K^{(*)}\bar{D}^{(*)}$ systems. Here, $P$ denotes the probability weight of each component (in \%).}\label{KstbarDstCP}
\begin{tabular*}{86mm}{@{\extracolsep{\fill}}cccc}
\toprule[1pt]
\toprule[1pt]
\multicolumn{4}{c}{$K\bar{D}/K^*\bar{D}^*$ coupled system with $I(J^P)=0(0^+)$}\\
\midrule[1.0pt]
$\Lambda~(\rm{GeV})$ &$E~(\rm {MeV})$ &$r_{\rm RMS}~(\rm {fm})$&$P(K\bar{D}/K^*\bar{D}^*)$\\
$1.19$&$-0.91$&$4.58$&\textbf{95.58}/4.42\\
$1.22$&$-12.09$&$1.47$&\textbf{82.44}/17.56\\
$1.24$&$-27.56$&$0.97$&\textbf{72.96}/27.04\\
\midrule[1.0pt]
\multicolumn{4}{c}{$K\bar{D}^*/K^*\bar{D}/K^*\bar{D}^*$ coupled system with $I(J^P)=0(1^+)$}\\
\midrule[1.0pt]
$\Lambda~(\rm{GeV})$ &$E~(\rm {MeV})$ &$r_{\rm RMS}~(\rm {fm})$&$P(K\bar{D}^*/K^*\bar{D}/K^*\bar{D}^*)$\\
$1.26$&$-1.28$&$4.19$&\textbf{95.60}/0.22/4.18\\
$1.29$&$-8.06$&$1.82$&\textbf{84.30}/0.89/14.81\\
$1.32$&$-25.21$&$0.98$&\textbf{64.41}/3.54/32.05\\
\midrule[1.0pt]
\multicolumn{4}{c}{$K\bar{D}^*/K^*\bar{D}/K^*\bar{D}^*$ coupled system with $I(J^P)=1(1^+)$}\\
\midrule[1.0pt]
$\Lambda~(\rm{GeV})$ &$E~(\rm {MeV})$ &$r_{\rm RMS}~(\rm {fm})$&$P(K\bar{D}^*/K^*\bar{D}/K^*\bar{D}^*)$\\
$1.77$&$-1.71$&$3.50$&\textbf{89.06}/6.89/4.05\\
$1.79$&$-8.38$&$1.60$&\textbf{77.96}/13.88/8.16\\
$1.81$&$-19.09$&$1.03$&\textbf{69.60}/19.12/11.28\\
\midrule[1.0pt]
\multicolumn{4}{c}{$K^*\bar{D}/K^*\bar{D}^*$ coupled system with $I(J^P)=0(1^+)$}\\
\midrule[1.0pt]
$\Lambda~(\rm{GeV})$ &$E~(\rm {MeV})$ &$r_{\rm RMS}~(\rm {fm})$&$P(K^*\bar{D}/K^*\bar{D}^*)$\\
$1.04$&$-0.70$&$4.43$&\textbf{95.25}/4.75\\
$1.08$&$-6.74$&$1.67$&\textbf{84.42}/15.58\\
$1.12$&$-19.61$&$1.01$&\textbf{73.46}/26.54\\
\midrule[1.0pt]
\multicolumn{4}{c}{$K^*\bar{D}/K^*\bar{D}^*$ coupled system with $I(J^P)=1(1^+)$}\\
\midrule[1.0pt]
$\Lambda~(\rm{GeV})$ &$E~(\rm {MeV})$ &$r_{\rm RMS}~(\rm {fm})$&$P(K^*\bar{D}/K^*\bar{D}^*)$\\
$2.30$&$-0.36$&$5.07$&\textbf{97.52}/2.48\\
$2.35$&$-4.91$&$1.81$&\textbf{91.94}/8.06\\
$2.40$&$-14.67$&$1.04$&\textbf{86.84}/13.16\\
\bottomrule[1pt]\bottomrule[1pt]
\end{tabular*}
\end{table}
	
For the $K\bar{D}/K^*\bar{D}^*$ coupled system with $I(J^P)=0(0^+)$, a loosely bound state first appears at $\Lambda = 1.19$ GeV, with the binding energy $E = -0.91$ MeV and the radius $r_{\text{RMS}} = 4.58$ fm. The $K\bar{D}$ component dominates ($95.58\%$), while the $K^*\bar{D}^*$ admixture is only $4.42\%$. As $\Lambda$ increases to $1.22$ GeV and $1.24$ GeV, the binding deepens significantly to $-12.09$ MeV and $-27.56$ MeV, respectively, and the corresponding radius shrinks to $1.47$ fm and $0.97$ fm. Concurrently, the $K^*\bar{D}^*$ fraction rises to $17.56\%$ and $27.04\%$, indicating that the coupled-channel dynamics becomes increasingly important at larger cutoffs. The substantial growth of the $K^*\bar{D}^*$ component suggests a strong mixing between such two channels, which enhances the effective attraction and leads to a loosely bound state. This behavior contrasts with the single-channel $K\bar{D}$ state with $I(J^P)=0(0^+)$, where no bound state existed, highlighting the crucial role of the $K^*\bar{D}^*$ coupling in forming the loosely bound state for the $K\bar{D}/K^*\bar{D}^*$ coupled system with $I(J^P)=0(0^+)$.

For the $K\bar{D}^*/K^*\bar{D}/K^*\bar{D}^*$ coupled system with $I(J^P)=0(1^+)$, a loosely bound state first appears at $\Lambda = 1.26$ GeV, with $E = -1.28$ MeV and $r_{\text{RMS}} = 4.19$ fm. The wave function is dominated by the $K\bar{D}^*$ channel ($95.60\%$), accompanied by tiny contributions from $K^*\bar{D}$ ($0.22\%$) and $K^*\bar{D}^*$ ($4.18\%$). As $\Lambda$ increases to $1.29$ GeV and $1.32$ GeV, the binding energy deepens to $-8.06$ MeV and $-25.21$ MeV, respectively, while the radius shrinks to $1.82$ fm and $0.98$ fm. Notably, the $K^*\bar{D}^*$ component grows significantly to $14.81\%$ and $32.05\%$, whereas the $K\bar{D}^*$ fraction decreases to $84.30\%$ and $64.41\%$. The $K^*\bar{D}$ channel remains negligible throughout. This behavior indicates that the coupled-channel mixing between the $K\bar{D}^*$ and $K^*\bar{D}^*$ becomes increasingly important as the cutoff increases. The appearance of a shallow bound state at a moderate cutoff ($\Lambda\approx1.26$ GeV) suggests that the $K\bar{D}^*/K^*\bar{D}/K^*\bar{D}^*$ coupled system with $I(J^P)=0(1^+)$ is a promising hadronic molecule candidate, with the $K^*\bar{D}^*$ component playing a nontrivial role in the binding mechanism.

For the $K\bar{D}^*/K^*\bar{D}/K^*\bar{D}^*$ coupled system with $I(J^P)=1(1^+)$, a shallow bound state first appears at $\Lambda = 1.77$ GeV, with the binding energy $E = -1.71$ MeV and the radius $r_{\text{RMS}} = 3.50$ fm. The $K\bar{D}^*$ channel dominates ($89.06\%$), followed by $K^*\bar{D}$ ($6.89\%$) and $K^*\bar{D}^*$ ($4.05\%$). As $\Lambda$ increases to $1.79$ GeV and $1.81$ GeV, the binding deepens to $-8.38$ MeV and $-19.09$ MeV, the radius shrinks to $1.60$ fm and $1.03$ fm, and the weights of $K^*\bar{D}$ and $K^*\bar{D}^*$ rise significantly, reaching $13.88\%$/$19.12\%$ and $8.16\%$/$11.28\%$, respectively. This evolution indicates that the $K^*\bar{D}$ and $K^*\bar{D}^*$ channels, though initially subdominant, become increasingly important as the cutoff grows, thereby enhancing the overall attraction. Compared with the $K\bar{D}^*/K^*\bar{D}/K^*\bar{D}^*$ coupled system with $I(J^P)=0(1^+)$, the $K\bar{D}^*/K^*\bar{D}/K^*\bar{D}^*$ coupled system with $I(J^P)=1(1^+)$ requires a larger cutoff ($\Lambda \sim 1.77$ GeV) to become bound, reflecting weaker effective attraction in the $I=1$ sector \cite{Klempt:2002ap}. Nevertheless, the existence of a loosely bound state at a phenomenologically acceptable cutoff makes the $K\bar{D}^*/K^*\bar{D}/K^*\bar{D}^*$ coupled system with $I(J^P)=1(1^+)$ the isovector hadronic molecule candidate.

For the $K^*\bar{D}/K^*\bar{D}^*$ coupled system with $I(J^P)=0(1^+)$, a loosely bound state appears at $\Lambda = 1.04$ GeV, with the binding energy $E = -0.70$ MeV and the radius $r_{\text{RMS}} = 4.43$ fm. The $K^*\bar{D}$ channel dominates ($95.25\%$), while the $K^*\bar{D}^*$ admixture is $4.75\%$. Raising $\Lambda$ to $1.08$ GeV and $1.12$ GeV deepens the binding to $-6.74$ MeV and $-19.61$ MeV, reduces the radius to $1.67$ fm and $1.01$ fm, and increases the $K^*\bar{D}^*$ fraction to $15.58\%$ and $26.54\%$, respectively. This behavior illustrates that the $K^*\bar{D}^*$ channel, although initially small, grows rapidly with $\Lambda$. Remarkably, the required cutoff for binding ($\Lambda \approx 1.04$ GeV) is significantly lower than that in the single-channel $K^*\bar{D}$ state with $I(J^P)=0(1^+)$ ($\Lambda \approx 2.14$ GeV), indicating that the $K^*\bar{D}^*$ coupling greatly enhances the effective interaction and leads to the existence of the loosely bound state at a more natural scale.

For the $K^*\bar{D}/K^*\bar{D}^*$ coupled system with $I(J^P)=1(1^+)$, a loosely bound state first appears at $\Lambda = 2.30$ GeV, with the binding energy $E = -0.36$ MeV and the radius $r_{\text{RMS}} = 5.07$ fm. The $K^*\bar{D}$ channel dominates ($97.52\%$), while the $K^*\bar{D}^*$ admixture is $2.48\%$. As $\Lambda$ increases to $2.35$ GeV and $2.40$ GeV, the binding deepens to $-4.91$ MeV and $-14.67$ MeV, the radius shrinks to $1.81$ fm and $1.04$ fm, and the $K^*\bar{D}^*$ fraction grows to $8.06\%$ and $13.16\%$, respectively. This behavior shows that, similar to the isoscalar counterpart, the $K^*\bar{D}^*$ channel provides additional attraction that becomes more effective at larger cutoffs. However, the $K^*\bar{D}/K^*\bar{D}^*$ coupled system with $I(J^P)=1(1^+)$ requires a considerably higher cutoff ($\Lambda \sim 2.30$ GeV) to form a loosely bound state, in contrast to the $K^*\bar{D}/K^*\bar{D}^*$ coupled system with $I(J^P)=0(1^+)$ where binding occurs already at $\Lambda \approx 1.04$ GeV. This difference reflects the weaker attraction in the isovector sector, consistent with the expected suppression of the $\rho/\pi$-exchange contributions for the $I=1$ system \cite{Klempt:2002ap}. 	

In this subsection, we have systematically investigated the $K^{(*)}\bar{D}^{(*)}$ systems as the potential hadronic molecules within both the single-channel and coupled-channel frameworks, with the cutoff parameters in the range $\Lambda \sim 0.8$-$2.5$ GeV. The single-channel analysis yields the loosely bound states for the $K^*\bar{D}$ state with $I(J^P)=0(1^+)$ and for several $K^*\bar{D}^*$ states with $I(J^P)=0(0^+), 0(1^+), 0(2^+), 1(2^+)$. In contrast, the $S$-wave $K\bar{D}$ and $K\bar{D}^*$ single-channel potentials do not support any bound state. When the coupled-channel effect is switched on, new loosely bound states emerge that are absent in the single-channel treatment, namely: the $K\bar{D}/K^*\bar{D}^*$ coupled system with $I(J^P)=0(0^+)$, the $K\bar{D}^*/K^*\bar{D}/K^*\bar{D}^*$ coupled system with $I(J^P)=0(1^+)$, the $K\bar{D}^*/K^*\bar{D}/K^*\bar{D}^*$ coupled system with $I(J^P)=1(1^+)$, and the $K^*\bar{D}/K^*\bar{D}^*$ coupled system with $I(J^P)=1(1^+)$. Notably, the $K^*\bar{D}/K^*\bar{D}^*$ coupled system with $I(J^P)=0(1^+)$ becomes bound at a much lower cutoff ($1.06$ GeV) than its single-channel $K^*\bar{D}$ state with $I(J^P)=0(1^+)$ ($2.10$ GeV), indicating that the $K^*\bar{D}$-$K^*\bar{D}^*$ coupling significantly enhances the effective attraction. A comparison with the single-channel results reveals that the coupled-channel dynamics plays a crucial role in the formation of several loosely bound states. Overall, the present analysis demonstrates that a number of $K^{(*)}\bar{D}^{(*)}$ states can form the loosely bound states with cutoffs in the range $\Lambda \sim 0.8$-$2.5$ GeV, and the following appear as the promising hadronic molecule candidates worthy of further experimental scrutiny:
\begin{itemize}
\item[(i)] $I(J^P)=0(0^+)$ $K\bar{D}/K^*\bar{D}^*$ coupled system,
\item[(ii)] $I(J^P)=0,\,1(1^+)$ $K\bar{D}^*/K^*\bar{D}/K^*\bar{D}^*$ coupled systems,
\item[(iii)] $I(J^P)=0,\,1(1^+)$ $K^*\bar{D}/K^*\bar{D}^*$ coupled systems,
\item[(iv)] $I(J^P)=0(0^+,\,1^+,\,2^+),\,1(2^+)$ $K^*\bar{D}^*$ systems.
\end{itemize}
These findings provide a clear roadmap for future experimental searches for the $K^{(*)}\bar{D}^{(*)}$ molecular tetraquark candidates.

\subsection{Mass spectra of the $K^{(*)}D^{(*)}$ molecular candidates}
	
As a natural extension of our study of the $K^{(*)}\bar{D}^{(*)}$ systems, we now turn to the $K^{(*)}D^{(*)}$ systems. The effective potentials for the $K^{(*)}D^{(*)}$ interactions are obtained from those of the $K^{(*)}\bar{D}^{(*)}$ systems via the $G$-parity rule \cite{Klempt:2002ap}. Under this rule, exchange of mesons with negative $G$-parity ($\pi$, $\omega$) receives an extra minus sign, while those with positive $G$-parity ($\sigma$, $\eta$, $\rho$) remain unchanged. With these effective potentials, we solve the coupled-channel Schr\"odinger equation to determine the bound-state properties of the $K^{(*)}D^{(*)}$ systems.

We first perform a single-channel analysis that explicitly includes the $S$-$D$ wave mixing effect. Table \ref{KD} presents the binding energies $E$, the root-mean-square radii $r_{\text{RMS}}$, and the probability weights $P$ for the resulting bound states. The $G$-parity transformation flips the signs of the $\pi$ and $\omega$ exchange contributions, thereby modifying the overall attractive strength relative to the $K^{(*)}\bar{D}^{(*)}$ systems.
	
\renewcommand\tabcolsep{0.08cm}
\renewcommand{\arraystretch}{1.50}
\begin{table}[!htbp]
\caption{Single-channel bound-state properties of the $K^{(*)}D^{(*)}$ systems. $P$ denotes the probability weight (in \%) of each component.}\label{KD}
\begin{tabular*}{86mm}{@{\extracolsep{\fill}}ccccc}
\toprule[1pt]
\toprule[1pt]
\multirow{4}{*}{$KD~[0(0^{+})]$}&$\Lambda~(\rm{GeV})$ &$E~(\rm {MeV})$ &$r_{\rm RMS}~(\rm {fm})$&$P(^1{S}_0)$\\
\Xcline{2-5}{0.75pt}
&2.24&-0.78&4.90&\textbf{100.00}\\
&2.37&-2.29&3.45&\textbf{100.00}\\
&2.50&-4.34&2.61&\textbf{100.00}\\
\midrule[1.0pt]
\multirow{4}{*}{$KD^*~[0(1^{+})]$}&$\Lambda~(\rm{GeV})$ &$E~(\rm {MeV})$ &$r_{\rm RMS}~(\rm {fm})$&$P(^3{S}_1/^3{D}_1)$\\
\Xcline{2-5}{0.75pt}
&2.20&-0.78&4.88&\textbf{100.00}/0\\
&2.35&-2.65&3.22&\textbf{100.00}/0\\
&2.50&-5.28&2.37&\textbf{100.00}/0\\
\midrule[1.0pt]
\multirow{4}{*}{$K^*D~[0(1^{+})]$}&$\Lambda~(\rm{GeV})$ &$E~(\rm {MeV})$ &$r_{\rm RMS}~(\rm {fm})$&$P(^3{S}_1/^3{D}_1)$\\
\Xcline{2-5}{0.75pt}
&1.52&-0.43&5.10&\textbf{100.00}/0\\
&1.79&-9.95&1.52&\textbf{100.00}/0\\
&2.05&-26.65&1.00&\textbf{100.00}/0\\
\midrule[1.0pt]
\multirow{4}{*}{$K^*D^*~[0(0^{+})]$}&$\Lambda~(\rm{GeV})$ &$E~(\rm {MeV})$ &$r_{\rm RMS}~(\rm {fm})$&$P(^{1}{S}_{0}/^{5}{D}_{0})$\\
\Xcline{2-5}{0.75pt}
&1.44&-0.65&4.79&\textbf{95.91}/4.09\\
&1.55&-9.72&1.64&\textbf{85.68}/14.32\\
&1.66&-34.24&1.00&\textbf{76.89}/23.11\\
\midrule[1.0pt]
\multirow{4}{*}{$K^*D^*~[0(1^{+})]$}&$\Lambda~(\rm{GeV})$ &$E~(\rm {MeV})$ &$r_{\rm RMS}~(\rm {fm})$&$P(^{3}{S}_{1}/^{3}{D}_{1}/^{5}{D}_{1})$\\
\Xcline{2-5}{0.75pt}
&1.40&-0.63&4.78&\textbf{97.60}/2.40/$\mathcal{O}(0)$\\
&1.54&-10.20&1.59&\textbf{92.39}/7.61/$\mathcal{O}(0)$\\
&1.67&-32.54&1.00&\textbf{88.54}/11.46/$\mathcal{O}(0)$\\
\midrule[1.0pt]
\multirow{4}{*}{$K^*D^*~[0(2^{+})]$}&$\Lambda~(\rm{GeV})$ &$E~(\rm {MeV})$ &$r_{\rm RMS}~(\rm {fm})$&$P(^{5}{S}_{2}/^{1}{D}_{2}/^{3}{D}_{2}/^{5}{D}_{2})$\\
\Xcline{2-5}{0.75pt}
&1.00&-0.57&4.76&\textbf{98.50}/0.24/$\mathcal{O}(0)$/1.26\\
&1.12&-11.05&1.49&\textbf{96.10}/0.64/$\mathcal{O}(0)$/3.26\\
&1.23&-31.33&1.00&\textbf{94.69}/0.87/$\mathcal{O}(0)$/4.44\\
\midrule[1.0pt]
\multirow{4}{*}{$K^*D^*~[1(0^{+})]$}&$\Lambda~(\rm{GeV})$ &$E~(\rm {MeV})$ &$r_{\rm RMS}~(\rm {fm})$&$P(^{1}{S}_{0}/^{5}{D}_{0})$\\
\Xcline{2-5}{0.75pt}
&1.86&-0.41&5.02&\textbf{99.86}/0.14\\
&2.15&-6.84&1.67&\textbf{99.64}/0.36\\
&2.43&-21.31&1.00&\textbf{99.50}/0.50\\
\bottomrule[1pt]\bottomrule[1pt]
\end{tabular*}
\end{table}

For the $KD$ state with $I(J^P)=0(0^+)$, a loosely bound state emerges at $\Lambda = 2.24$ GeV, with the binding energy $E = -0.78$ MeV and the root-mean-square radius $r_{\text{RMS}} = 4.90$ fm. As the cutoff increases to $2.37$ GeV and $2.50$ GeV, the binding deepens to $-2.29$ MeV and $-4.34$ MeV, respectively, while the radius decreases to $3.45$ fm and $2.61$ fm. Throughout this range, the $^1S_0$ component remains $100\%$.

For the $KD^*$ state with $I(J^P)=0(1^+)$, a loosely bound state first appears at $\Lambda = 2.20$ GeV, characterized by $E = -0.78$ MeV and $r_{\text{RMS}} = 4.88$ fm. No $D$-wave admixture is present, as the tensor force does not contribute in this channel. As $\Lambda$ increases to $2.35$ GeV and $2.50$ GeV, the binding deepens to $-2.65$ MeV and $-5.28$ MeV, respectively, while the corresponding radius shrinks to $3.22$ fm and $2.37$ fm. The $^3S_1$ component remains $100\%$ across the considered cutoff range.

For the $K^*D$ state with $I(J^P)=0(1^+)$, the attraction is substantially stronger than that in the corresponding $K^*\bar{D}$ channel. A loosely bound state first appears at $\Lambda = 1.52$ GeV, with $E = -0.43$ MeV and $r_{\text{RMS}} = 5.10$ fm. Raising the cutoff to $1.79$ GeV and $2.05$ GeV deepens the binding energy to $-9.95$ MeV and $-26.65$ MeV, respectively, while the corresponding radius shrinks to $1.52$ fm and $1.00$ fm. The $^3S_1$ component remains $100\%$, reflecting the absence of the tensor-induced $D$-wave mixing.

For the $K^*D^*$ state with $I(J^P)=0(0^+)$, a loosely bound state first appears at $\Lambda = 1.44$ GeV, with $E = -0.65$ MeV and $r_{\text{RMS}} = 4.79$ fm. The wave function is dominated by the $^1S_0$ component ($95.91\%$), while the $^5D_0$ admixture amounts to $4.09\%$, indicating a non-negligible but still subleading $D$-wave contribution arising from the tensor force. As $\Lambda$ increases to $1.55$ GeV and $1.66$ GeV, the binding energy deepens to $-9.72$ MeV and $-34.24$ MeV, and the radius shrinks to $1.64$ fm and $1.00$ fm, respectively. Concurrently, the $^5D_0$ fraction grows to $14.32\%$ and $23.11\%$, reflecting the increasing influence of the tensor interaction at larger cutoffs.

For the $K^*D^*$ state with $I(J^P)=0(1^+)$, a loosely bound state first appears at $\Lambda = 1.40$ GeV, with $E = -0.63$ MeV and $r_{\text{RMS}} = 4.78$ fm. The wave function is dominated by the $^3S_1$ partial wave ($97.60\%$), accompanied by a small $^3D_1$ admixture ($2.40\%$), indicating that the tensor force, although present, induces only a modest $D$-wave component. As $\Lambda$ increases to $1.54$ GeV and $1.67$ GeV, the enhanced attraction deepens the binding energy to $-10.20$ MeV and $-32.54$ MeV, respectively, while the radius shrinks to $1.59$ fm and $1.00$ fm. The $^3D_1$ fraction grows to $7.61\%$ and $11.46\%$, reflecting the increased role of the tensor interaction at larger cutoffs. Conversely, the $^5D_1$ component remains negligible throughout. This behavior suggests that the $^3S_1$ channel provides the dominant binding mechanism, with the tensor force playing a supportive but non-negligible role, especially at stronger couplings.

For the $K^*D^*$ state with $I(J^P)=0(2^+)$, a loosely bound state emerges at $\Lambda = 1.00$ GeV, with $E = -0.57$ MeV and $r_{\text{RMS}} = 4.76$ fm. The $^5S_2$ partial wave overwhelmingly dominates ($98.50\%$), while the $^5D_2$ component contributes only $1.26\%$ and the $^1D_2$ admixture is negligible ($0.24\%$). As $\Lambda$ increases to $1.12$ GeV and $1.23$ GeV, the binding energy deepens to $-11.05$ MeV and $-31.33$ MeV, respectively, and the corresponding radius shrinks to $1.49$ fm and $1.00$ fm. Concurrently, the $^5D_2$ share grows to $3.26\%$ and $4.44\%$, indicating a modest but increasing role of the tensor force at larger cutoffs. Nevertheless, the $^5S_2$ wave remains dominant, with its weight staying above $94\%$ throughout.

The $K^*D^*$ state with $I(J^P)=1(0^+)$ first develops a loosely bound state at $\Lambda = 1.86$ GeV, characterized by $E = -0.41$ MeV and $r_{\text{RMS}} = 5.02$ fm. The wave function is overwhelmingly dominated by the $^1S_0$ component ($99.86\%$), with a tiny $^5D_0$ admixture ($0.14\%$). As $\Lambda$ increases to $2.15$ GeV and $2.43$ GeV, the binding energy deepens to $-6.84$ MeV and $-21.31$ MeV, the radius shrinks to $1.67$ fm and $1.00$ fm, and the $^5D_0$ fraction rises only modestly to $0.36\%$ and $0.50\%$, confirming the persistence of $S$-wave dominance.

We now turn to the coupled-channel analysis. Table \ref{KstDstCP} summarizes the numerical results for three distinct channel combinations in the $K^{(*)}D^{(*)}$ systems. The inclusion of the coupled-channel dynamics enhances the effective attraction in already bound single-channel configurations, reduces the cutoffs required to achieve binding. This pattern is particularly evident in the $KD/K^*D^*$ and $KD^*/K^*D/K^*D^*$ coupled systems.
	
\renewcommand\tabcolsep{0.20cm}
\renewcommand{\arraystretch}{1.50}
\begin{table}[!htbp]
\caption{Coupled-channel bound-state properties of the $K^{(*)}D^{(*)}$ systems. $P$ denotes the probability weight (in \%) of each component.}\label{KstDstCP}
\begin{tabular*}{86mm}{@{\extracolsep{\fill}}cccc}
\toprule[1pt]
\toprule[1pt]
\multicolumn{4}{c}{$KD/K^*D^*$ coupled system with $I(J^P)=0(0^+)$}\\
\midrule[1.0pt]
$\Lambda~(\rm{GeV})$ &$E~(\rm {MeV})$ &$r_{\rm RMS}~(\rm {fm})$&$P(KD/K^*D^*)$\\
1.26&-1.01&4.60&\textbf{98.61}/1.39\\
1.34&-12.94&1.62&\textbf{94.51}/5.49\\
1.42&-40.40&1.01&\textbf{89.38}/10.62\\
\midrule[1.0pt]
\multicolumn{4}{c}{$KD^*/K^*D/K^*D^*$ coupled system with $I(J^P)=0(1^+)$}\\
\midrule[1.0pt]
$\Lambda~(\rm{GeV})$ &$E~(\rm {MeV})$ &$r_{\rm RMS}~(\rm {fm})$&$P(KD^*/K^*D/K^*D^*)$\\
1.10&-0.68&5.03&\textbf{96.92}/1.96/1.12\\
1.17&-11.81&1.63&\textbf{86.86}/8.44/4.70\\
1.24&-36.84&0.99&\textbf{76.37}/15.23/8.40\\
\midrule[1.0pt]
\multicolumn{4}{c}{$K^*D/K^*D^*$ coupled system with $I(J^P)=0(1^+)$}\\
\midrule[1.0pt]
$\Lambda~(\rm{GeV})$ &$E~(\rm {MeV})$ &$r_{\rm RMS}~(\rm {fm})$&$P(K^*D/K^*D^*)$\\
1.10&-0.46&5.06&\textbf{98.70}/1.30\\
1.20&-9.24&1.60&\textbf{94.50}/5.50\\
1.30&-28.93&1.01&\textbf{90.52}/9.48\\
\bottomrule[1pt]\bottomrule[1pt]
\end{tabular*}
\end{table}

For the $KD/K^*D^*$ coupled system with $I(J^P)=0(0^+)$, a loosely bound state first appears at $\Lambda = 1.26$ GeV, with $E = -1.01$ MeV and $r_{\text{RMS}} = 4.60$ fm. The $KD$ component dominates ($98.61\%$), while the $K^*D^*$ admixture is only $1.39\%$. As $\Lambda$ increases to $1.34$ GeV and $1.42$ GeV, the binding deepens to $-12.94$ MeV and $-40.40$ MeV, the radius shrinks to $1.62$ fm and $1.01$ fm, and the $K^*D^*$ fraction rises to $5.49\%$ and $10.62\%$, respectively. This evolution demonstrates that the coupled-channel mixing significantly enhances the effective attraction, leading to reduce the cutoffs required to achieve binding.

For the $KD^*/K^*D/K^*D^*$ coupled system with $I(J^P)=0(1^+)$, a loosely bound state first emerges at $\Lambda = 1.10$ GeV, with $E = -0.68$ MeV and $r_{\text{RMS}} = 5.03$ fm. The $KD^*$ channel dominates the wave function ($96.92\%$), while the $K^*D$ and $K^*D^*$ components contribute only $1.96\%$ and $1.12\%$, respectively. As $\Lambda$ increases to $1.17$ GeV and $1.24$ GeV, the binding deepens to $-11.81$ MeV and $-36.84$ MeV, and the radius shrinks to $1.63$ fm and $0.99$ fm, respectively. Concomitantly, the $K^*D$ and $K^*D^*$ fractions rise significantly: the $K^*D$ component grows to $8.44\%$ and $15.23\%$, and the $K^*D^*$ component to $4.70\%$ and $8.40\%$. In contrast, the $KD^*$ fraction decreases to $86.86\%$ and $76.37\%$. This redistribution of probabilities indicates that the channel mixing becomes increasingly important at larger cutoffs, enhancing the overall attraction.

For the $K^*D/K^*D^*$ coupled system with $I(J^P)=0(1^+)$, a loosely bound state emerges at $\Lambda = 1.10$ GeV, with $E = -0.46$ MeV and $r_{\text{RMS}} = 5.06$ fm. The $K^*D$ component dominates ($98.70\%$), while the $K^*D^*$ admixture is only $1.30\%$. As $\Lambda$ increases to $1.20$ GeV and $1.30$ GeV, the binding energy deepens to $-9.24$ MeV and $-28.93$ MeV, the radius shrinks to $1.60$ fm and $1.01$ fm, and the $K^*D^*$ fraction grows to $5.50\%$ and $9.48\%$, respectively. Notably, this coupled system becomes bound at a significantly lower cutoff ($\Lambda = 1.10$ GeV) compared to its single-channel $K^*D$ state with $I(J^P)=0(1^+)$ ($\Lambda = 1.52$ GeV). This reduction in the required cutoff clearly demonstrates that the $K^*D^*$ channel provides additional attractive interaction through channel coupling.

Using both the single-channel (with the $S$-$D$ mixing effect) and coupled-channel approaches, we have carried out a comprehensive investigation of the $K^{(*)}D^{(*)}$ systems as the hadronic molecule candidates, varying the cutoff $\Lambda$ from $0.8$ to $2.5$ GeV. The single-channel treatment identifies the loosely bound states in the $KD$ state with $I(J^P)=0(0^+)$, the $KD^*$ state with $I(J^P)=0(1^+)$, the $K^*D$ state with $I(J^P)=0(1^+)$, and the $K^*D^*$ states with $I(J^P)=0(0^+),\,0(1^+),\,0(2^+),\,1(0^+)$.  The inclusion of the coupled-channel dynamics further strengthens the effective attraction, resulting in the loosely bound states at lower cutoffs and highlighting the pivotal role of channel mixing.  Taking together the single-channel and coupled-channel results, the following systems stand out as the promising hadronic molecule candidates, warranting dedicated experimental attention:
\begin{itemize}
\item[(i)] $I(J^P)=0(0^+)$ $KD/K^*D^*$ coupled system,
\item[(ii)] $I(J^P)=0(1^+)$ $K{D}^*/K^*{D}/K^*{D}^*$ coupled system,
\item[(iii)] $I(J^P)=0(1^+)$ $K^*{D}/K^*{D}^*$ coupled system,
\item[(iv)] $I(J^P)=0(0^+,\,1^+,\,2^+),\,1(0^+)$ $K^*{D}^*$ systems.
\end{itemize}
Collectively, these findings chart a clear course for future experimental searches for the $K^{(*)}{D}^{(*)}$ molecular tetraquark candidates.

\section{Electromagnetic characteristics of the $K^{(*)}\bar{D}^{(*)}$ and $K^{(*)}D^{(*)}$ molecular candidates}\label{sec3}

To elucidate the internal structures of the $K^{(*)}\bar{D}^{(*)}$ and $K^{(*)}D^{(*)}$ molecular candidates and to provide essential guidance for future experiments, we systematically investigate their radiative decay properties and magnetic moments. Electromagnetic characteristics of the hadrons have been extensively explored using various theoretical approaches \cite{Meng:2022ozq}. Among these, the constituent quark model has proved successful in describing the measured magnetic moments of the octet and decuplet baryons \cite{Schlumpf:1993rm, Kumar:2005ei, Ramalho:2009gk}. Building on this success, the present work adopts the constituent quark model to perform a systematic analysis of the electromagnetic characteristics of the $K^{(*)}\bar{D}^{(*)}$ and $K^{(*)}D^{(*)}$ molecular candidates. The formalism used in our calculations is detailed below.

\subsection{Theoretical frameworks}
	
For a transition from an initial state $H$ to a final state $H'$ accompanied by the photon emission, the M1 radiative decay width is given by \cite{Wang:2023bek,Lai:2024jfe,Zhang:2025ame,Zhu:2025abk,Wang:2024kke,Sheng:2024hkf,Wang:2023ael}
\begin{align}\label{M1 radiative decay width}
\Gamma_{H\rightarrow H^{\prime}\gamma}=
\frac{k^{3}}{m_{N}^{2}}
\frac{\alpha_{EM}}{2J_{H}+1}
\frac{\textstyle\sum\limits_{J_{H^{\prime}z},J_{Hz}}
\wignerthreej{J_{H^{\prime}} }{1}{J_{H} }{-J_{H^{\prime}z }}{0}{J_{Hz} }^{2} }{
\wignerthreej{J_{H^{\prime}} }{1}{J_{H} }{-J_{z }}{0}{J_{z} }^{2}}
\frac{|\mu_{H\rightarrow H^{\prime}\gamma} |^{2} }{\mu_{N}^{2}},
\end{align}
where $\alpha_{\text{EM}}\approx 1/137$ is the electromagnetic fine structure constant, and the photon momentum is $k = (m_H^2 - m_{H'}^2)/(2m_H)$. The Wigner $3j$ symbols enforce the angular momentum conservation, $J_H$ and $J_{H'}$ are the total angular momenta of the initial and final states, and $J_z = \min\{J_H, J_{H'}\}$. The nucleon mass $m_N = 0.938$ GeV \cite{ParticleDataGroup:2024cfk} together with the nuclear magneton $\mu_N = e/(2m_N)$ sets the energy scale. The transition magnetic moment $\mu_{H\rightarrow H^{\prime}\gamma}$ is the key dynamical quantity, defined as \cite{Li:2021ryu,Zhou:2022gra,Wang:2022tib,Wang:2023bek,Lai:2024jfe,Zhang:2025ame,Zhu:2025abk}
\begin{align}\label{transition magnetic moment}
\mu_{H\rightarrow H^{\prime}\gamma}=
\left\langle J_{H^{\prime}},J_{z}\left|
\sum_{j}\hat{\mu}^{S}_{zj}e^{-i\boldsymbol{k\cdot r}_{j}}+\hat{\mu}^{L}_{z}
\right|J_{H},J_{z}\right\rangle.
\end{align}
The spin and orbital magnetic moment operators are respectively \cite{Liu:2003ab,Huang:2004tn,Zhu:2004xa,Wang:2016dzu,Li:2021ryu,Zhou:2022gra,Wang:2022tib,Gao:2021hmv,Wang:2023bek,Wang:2022nqs,Mutuk:2024ltc,Mutuk:2024jxf,Mutuk:2024elj,Guo:2023fih,Li:2024wxr,Lei:2024geu,Lei:2023ttd,Lai:2024jfe,Zhang:2025ame,Zhu:2025abk}
\begin{eqnarray}
\hat{\mu}_{zj}^{S} &=& \frac{e_j}{2m_j}\hat{\sigma}_{zj},\\
\hat{\mu}^{L}_{z}&=&\frac{m_{\alpha}}{m_{\alpha}+m_{\beta}}\mu_{\beta}\hat{L}_{z}+\frac{m_{\beta}}{m_{\alpha}+m_{\beta}}\mu_{\alpha}\hat{L}_{z},
\end{eqnarray}
where $e_j$ and $m_j$ are the charge and the mass of the $j$-th constituent, $\hat{\sigma}_{zj}$ is the $z$-component of the Pauli spin operator, $m_{\alpha(\beta)}$ and $\mu_{\alpha(\beta)}$ are the masses and the magnetic moments of the two constituent hadrons, and $\hat{L}_z$ is the $z$-component of the relative orbital angular momentum. The factor $e^{-i\boldsymbol{k\cdot r}_{j}}$ in Eq. (\ref{transition magnetic moment}) originates from the spatial wave function of the emitted photon. To facilitate analytical integration over coordinates, it is expanded in the spherical Bessel functions $j_l(k r_j)$ and the spherical harmonics $Y_{lm}$ \cite{Quantum Theory of Angular Momentum}:
\begin{align}
e^{-i\boldsymbol{k\cdot r}_{j}}=
4\pi\sum_{l=0}^{\infty}\sum_{m=-l}^{l}
(-i)^{l}j_{l}(kr_{j})Y^{*}_{lm}(\Omega_{\boldsymbol{k} })Y_{lm}(\Omega_{\boldsymbol{r}_{j} }).
\end{align}
This expansion enables us to account for the effects of the finite spatial extension of the molecular state and its constituent hadrons.

The matrix element in Eq.~(\ref{transition magnetic moment}) is evaluated using the total wave function, which factorizes into spatial, flavor, color, and spin components. Among these, the flavor, color, and spin parts are explicitly constructed, whereas the spatial part involves overlap integrals that are essential for determining radiative transitions. The spatial wave function itself consists of two distinct factors: one describing the internal structure of each constituent hadron $H_1$ and $H_2$, and the other characterizing their relative motion. Explicitly,
\begin{equation}\label{spacewavefunction}
\ket{\psi^{\rm{space}}} = \ket{\phi_{H_1}} \otimes \ket{\phi_{H_2}} \otimes \ket{\phi_{\rm{mol}}},
\end{equation}
where $\ket{\phi_{H_1}}$ and $\ket{\phi_{H_2}}$ denote the internal spatial wave functions of the two constituent hadrons, and $\ket{\phi_{\rm{mol}}}$ is the spatial wave function for their relative motion. The computation of the factor $\langle \psi^{\mathrm{space}}(f)|e^{-i\boldsymbol{k\cdot r}_{j}}|\psi^{\mathrm{space}}(i)\rangle$ is detailed in Appendix B of Ref. \cite{Wang:2022nqs}.

To compute the spatial overlap integrals entering the transition amplitudes, we need the explicit spatial wave functions $\ket{\phi_{\text{mol}}}$ and $\ket{\phi_{H_i}}$ defined above. For the $K^{(*)}\bar{D}^{(*)}$ and $K^{(*)}D^{(*)}$ molecular candidates, the relative-motion wave function $\ket{\Psi_{\text{mol}}}$ is directly taken from the numerical spatial wave functions obtained from solving the coupled-channel Schr\"odinger equation in the above section, which incorporate the $S$-$D$ wave mixing and coupled-channel effects. The internal spatial wave functions $\ket{\phi_{H_i}}$ of the constituent hadrons ($K^{(*)}$, $\bar{D}^{(*)}$, $D^{(*)}$) are described by the simple harmonic oscillator wave functions, widely employed in quark-model studies of the electromagnetic  characteristics of the hadronic molecules \cite{Wang:2024kke,Wang:2022tib,Zhou:2022gra,Wang:2022nqs,Wang:2023aob,Wang:2023bek,Wang:2023ael,Lai:2024jfe,Sheng:2024hkf,Wang:2024sbw,Zhu:2025abk,Zhang:2025ame}. These functions take the form
\begin{equation}
\phi_{n,l,m}(\beta,\boldsymbol{r})=\sqrt{\frac{2n!}{\Gamma(n+l+\frac{3}{2})}} L_n^{l+\frac{1}{2}}(\beta^2 r^2) \beta^{l+\frac{3}{2}}e^{-\frac{\beta^2 r^2}{2}} r^l Y_{lm}(\Omega),
\end{equation}
where $\beta$ is the oscillator parameter, $n$, $l$, and $m$ are the radial, orbital, and magnetic quantum numbers. $L_{n}^{l+\frac{1}{2}}(x)$ denotes the associated Laguerre polynomial. For the specific hadrons involved, we adopt the oscillator parameters determined from the hadron radii and consistently used in Refs. \cite{Wang:2024kke,Godfrey:2015dva}: $\beta_D = 0.601$ GeV, $\beta_{D^*} = 0.516$ GeV, $\beta_K = 0.460$ GeV, and $\beta_{K^*} = 0.320$ GeV. With the molecular wave functions obtained at representative binding energies ($-4$, $-11$, and $-17$ MeV) and the internal hadron wave functions specified, we can evaluate the transition magnetic moments and subsequently the M1 radiative decay widths.
	
Except for the radiative decays, the magnetic moments provide another powerful probe of the internal structures. For a given hadronic molecule $H$, the total magnetic moment is defined as the expectation value of the magnetic operator in the state with the maximum $z$-component of the total angular momentum $J_{Hz}=J_H$ \cite{Li:2021ryu,Zhou:2022gra,Wang:2022tib,Wang:2023bek,Lai:2024jfe,Zhang:2025ame,Zhu:2025abk}:
\begin{align}\label{equ}
\mu_{H} = \left\langle J_{H}, J_{H} \left| \sum_{j} \hat{\mu}^{S}_{zj} + \hat{\mu}^{L}_{z} \right| J_{H}, J_{H} \right\rangle,
\end{align}
where the sum runs over all constituent hadrons (or quarks) in the molecule. The operators $\hat{\mu}^{S}_{zj}$ and $\hat{\mu}^{L}_{z}$ are the spin and orbital contributions defined earlier.

Within the constituent quark model, the masses of the constituent quarks serve as fundamental parameters for computing the hadronic electromagnetic  characteristics. In this work, we adopt the following masses: $m_u = 336$ MeV, $m_d = 336$ MeV, $m_s = 450$ MeV, and $m_c = 1680$ MeV, taken from Ref. \cite{Kumar:2005ei}. This set of parameters has been shown to effectively reproduce the experimental magnetic moments of the octet and decuplet baryons \cite{Schlumpf:1993rm, Kumar:2005ei, Ramalho:2009gk}. Moreover, it has been extensively used in recent systematic investigations of the electromagnetic  characteristics of the hadronic molecular states \cite{Wang:2023aob,Zhu:2025abk,Wang:2023ael,Sheng:2024hkf,Li:2021ryu,Zhou:2022gra,Wang:2022tib,Gao:2021hmv,Wang:2023bek,Wang:2022nqs,Lai:2024jfe,Wang:2024sbw,Zhang:2025ame}.
		
\subsection{Electromagnetic characteristics of the $K^{(*)}\bar{D}^{(*)}$ molecular candidates}
	
In the following, we present the numerical results for the electromagnetic  characteristics of the $K^{(*)}\bar{D}^{(*)}$ molecular candidates, based on the theoretical framework listed above. The discussion is structured around two physical observables. First, we analyse the M1 radiative decay widths of the $K^{(*)}\bar{D}^{(*)}$ molecular candidates. Second, we compute the magnetic moments of the $K^{(*)}\bar{D}^{(*)}$ molecular candidates. To maintain clarity in the coupled-channel analysis, we introduce the following simplified notations to label the coupled systems: $K\bar{D}[0(0^+)]$, $K\bar{D}^*[0(1^+)]$, $K\bar{D}^*[1(1^+)]$, and $K^*\bar{D}[0(1^+)]$ denote the $K\bar{D}/K^*\bar{D}^*$ coupled system with $I(J^P)=0(0^+)$, the $K\bar{D}^*/K^*\bar{D}/K^*\bar{D}^*$ coupled system with $I(J^P)=0(1^+)$, the $K\bar{D}^*/K^*\bar{D}/K^*\bar{D}^*$ coupled system with $I(J^P)=1(1^+)$, and the $K^*\bar{D}/K^*\bar{D}^*$ coupled system with $I(J^P)=0(1^+)$, respectively.

\begin{table}[!htbp]
\renewcommand\tabcolsep{0.10cm}
\renewcommand{\arraystretch}{1.50}
\caption{M1 radiative decay widths of the $K^{(*)}\bar{D}^{(*)}$ molecular candidates at three binding energies $E=-4,-11,-17$ MeV, obtained within the coupled-channel framework.}\label{KDbardecaywidth}
\begin{tabular*}{86mm}{@{\extracolsep{\fill}}cccc}
\toprule[1pt]
\toprule[1pt]
Processes&$I(J_H, J_{H^{\prime}})$&$I_3$&$\Gamma_{H\to H^{\prime}\gamma}$(keV)\\
\midrule[1.0pt]
\multirow{5}{*}{$K^*\bar{D}^*\to K\bar{D}^*+\gamma$}&0(0, 1)&0&0.08, 5.20, 12.68\\
\Xcline{2-4}{0.75pt}
~&0(1, 1)&0&0.77, 0.05, 0.02\\
\Xcline{2-4}{0.75pt}
~&\multirow{3}{*}{1(2, 1)}&1&214.59, 281.23, 316.45\\
~&~&0&0.83, 5.42, 9.22\\
~&~&$-1$&164.91, 147.22, 137.76\\
\midrule[1.0pt]
\multirow{5}{*}{$K^*\bar{D}^*\to K^*\bar{D}+\gamma$}&0(0, 1)&0&8.59, 10.94, 12.19\\
\Xcline{2-4}{0.75pt}
~&0(1, 1)&0&2.32, 1.25, 0.83\\
\Xcline{2-4}{0.75pt}
~&\multirow{3}{*}{1(2, 1)}&1&35.00, 35.38, 35.48\\
~&~&0&5.95, 6.56, 6.87\\
~&~&$-1$&1.08, 0.68, 0.52\\
\midrule[1.0pt]
$K\bar{D}^*\to K\bar{D}+\gamma$&0(1, 0)&0&3.21, 1.92, 1.20\\
\midrule[1.0pt]
$K^*\bar{D}\to K\bar{D}+\gamma$&0(1, 0)&0&9.54, 14.55, 18.30\\
\bottomrule[1pt]
\bottomrule[1pt]
\end{tabular*}
\end{table}

Table \ref{KDbardecaywidth} summarises the M1 radiative decay widths of the $K^{(*)}\bar{D}^{(*)}$ molecular candidates calculated for three representative binding energies: $-4$, $-11$, and $-17$ MeV. The obtained results exhibit a rich dependence on the spin-parity, the constituent hadron, the binding strength, and the isospin of the $K^{(*)}\bar{D}^{(*)}$ molecular candidates, offering clear experimental signatures.

A direct comparison between the processes $K^*\bar{D}^*[0(0^+)]\to K^*\bar{D}[0(1^+)]\gamma$ and $K^*\bar{D}^*[0(1^+)]\to K^*\bar{D}[0(1^+)]\gamma$ reveals a clear fingerprint: such two decays exhibit markedly different widths. This dichotomy provides a powerful tool for experimentally determining the spin-parity quantum numbers of the $K^*\bar{D}^*$ molecular candidates. Similarly, the $K\bar{D}^*[0(1^+)]\to K\bar{D}[0(0^+)]\gamma$ and $K^*\bar{D}[0(1^+)]\to K\bar{D}[0(0^+)]\gamma$ transitions show strikingly different magnitudes, allowing such two distinct $I(J^P)=0(1^+)$ configurations to be distinguished.

For the processes $K^*\bar{D}^*[1(2^+)]\to K\bar{D}^*[1(1^+)]\gamma$ and $K^*\bar{D}^*[1(2^+)]\to K^*\bar{D}[1(1^+)]\gamma$, the decay widths exhibit a strong dependence on the isospin projection $I_3$. The $I_3=1$ component overwhelmingly dominates: the $K^*\bar{D}^*[1(2^+)]\to K\bar{D}^*[1(1^+)]\gamma$ width increases from $214.59$ to $316.45$ keV as binding grows, while the $K^*\bar D^*[1(2^+)] \to K^*\bar D[1(1^+)]\gamma$ channel remains stable at approximately $35$ keV. In contrast, the $I_3=0$ and $-1$ components are orders of magnitude smaller. Specifically, for the $K^*\bar D^*[1(2^+)] \to K\bar D^*[1(1^+)]\gamma$ transition, the $I_3=0$ width rises from $0.83$ to $9.22$ keV, while the $I_3=-1$ width decreases from $164.91$ to $137.76$ keV. For the $K^*\bar{D}^*[1(2^+)]\to K^*\bar{D}[1(1^+)]\gamma$ channel, both remain below $10$ keV. This hierarchy originates from the different flavor wave functions of the $I=1$ $K^{(*)}\bar{D}^{(*)}$ states, which lead to different transition magnetic moments for distinct $I_3$.

\begin{figure}[htbp]
\centering
\includegraphics[width=8.6cm]{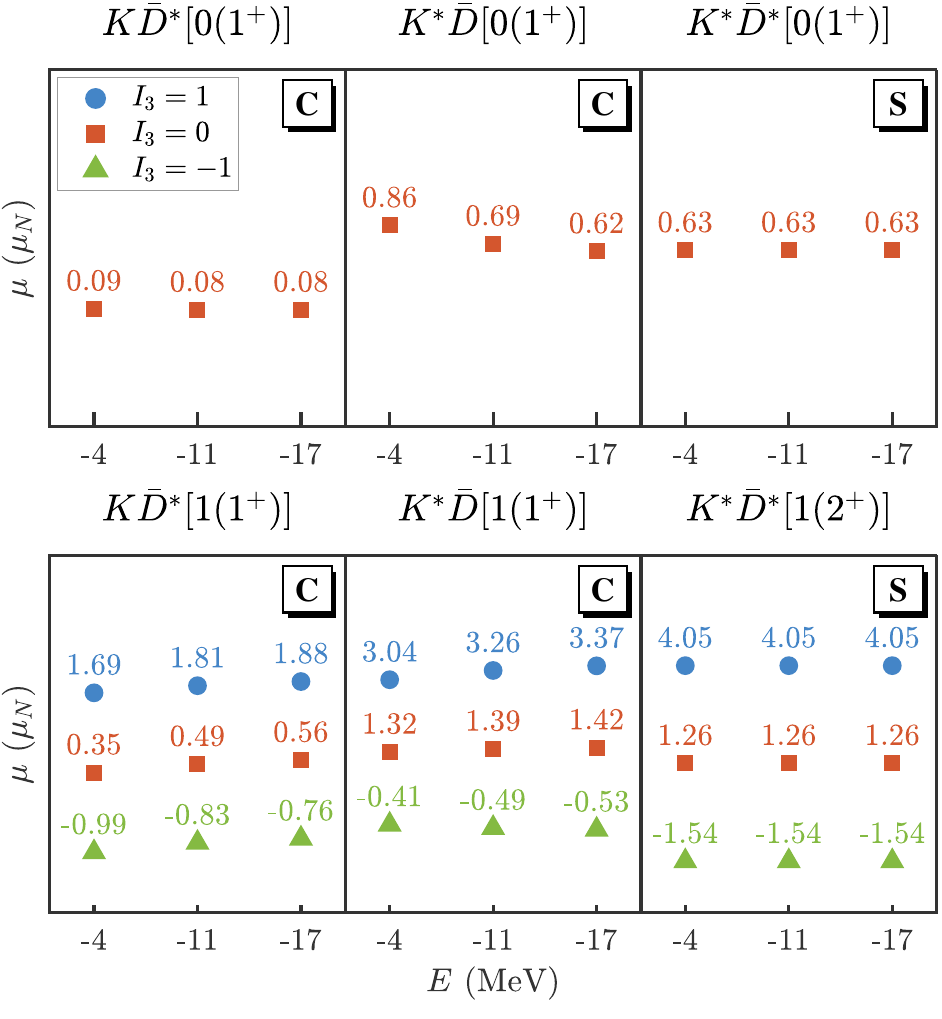}
\caption{Magnetic moments of the $K^{(*)}\bar{D}^{(*)}$ molecular candidates at the binding energies $-4$, $-11$, and $-17$ MeV. Indices ``S'' and ``C" denote the single-channel and coupled-channel analyses, respectively.}
\label{fig:magnetic_moments}
\end{figure}

Having comprehensively explored the radiative decay behaviors of the $K^{(*)}\bar{D}^{(*)}$ molecular candidates, we now shift our focus to their magnetic moments. Evaluating the magnetic moments of these molecular candidates provides a complementary and direct probe into their internal structures. Since the magnetic moment of a spin-$0$ hadron is identically zero, we restrict our discussion to the states with non-zero total angular momentum, as presented in Fig. \ref{fig:magnetic_moments}.

We first discuss the single-channel results, which are applicable only to the $K^*\bar{D}^*$ state with $I(J^P)=0(1^+)$ and the $K^*\bar{D}^*$ state with $I(J^P)=1(2^+)$. For the $K^*\bar{D}^*$ state with $I(J^P)=0(1^+)$, the magnetic moment is constant $0.63\ \mu_N$ across all binding energies ($-4$, $-11$, $-17$ MeV). Similarly, for the $K^*\bar{D}^*$ state with $I(J^P)=1(2^+)$, the magnetic moments are invariant: $4.05\ \mu_N$ ($I_3=1$), $1.26\ \mu_N$ ($I_3=0$), and $-1.54\ \mu_N$ ($I_3=-1$). This stability arises from the overwhelming dominance of the $S$-wave component, making the magnetic moment insensitive to modest changes in spatial extension with binding strength.

Coupled-channel results reveal a much richer behaviour, with the magnetic moments exhibiting varying degrees of binding-energy dependence that reflect the evolving composition of the molecular wave function as coupling between different $K^{(*)}\bar{D}^{(*)}$ channels changes with binding strength. The $K\bar{D}^*[0(1^+)]$ state shows negligible binding dependence: the magnetic moments are $0.09$, $0.08$, and $0.08\ \mu_N$ at $-4$, $-11$, and $-17$ MeV, respectively. This stability is due to the dominant $K\bar{D}^*$ component (over 90\%) which shields the magnetic moment from significant variation. In contrast, the $K^*\bar{D}[0(1^+)]$ state exhibits a clear decreasing trend: $0.86$, $0.69$, and $0.62\ \mu_N$ at the three binding energies. This trend arises from the growing admixture of the $K^*\bar{D}^*$ channel as binding strengthens, introducing a larger contribution from the spin-1 $K^*$ and $\bar{D}^*$ mesons. The $I=1$ $K^{(*)}\bar{D}^{(*)}$ coupled systems show even more pronounced binding-energy dependence, particularly for different isospin projections $I_3$. For the $K\bar{D}^*[1(1^+)]$ state, the magnetic moment increases from $1.69$ to $1.88\ \mu_N$ for $I_3=1$, increases from $0.35$ to $0.56\ \mu_N$ for $I_3=0$, and decreases from $-0.99$ to $-0.76\ \mu_N$ for $I_3=-1$. For the $K^*\bar{D}[1(1^+)]$ state, the magnetic moment increases from $3.04$ to $3.37\ \mu_N$ for $I_3=1$, increases from $1.32$ to $1.42\ \mu_N$ for $I_3=0$, and decreases from $-0.41$ to $-0.53\ \mu_N$ for $I_3=-1$. These trends are driven by the increasing proportion of the $K^*\bar{D}^*$ component as the binding energy deepens.

The inclusion of the coupled-channel effect introduces significant binding-energy dependence in the magnetic moments, absent in the single-channel approximation. This makes magnetic moments a sensitive probe of internal mixing between different components. In addition, the distinct values for $I_3=1$, $0$, $-1$ provide a clear signature for determining the isospin and flavour content. For example, a large positive magnetic moment ($\sim 3$-$4\,\mu_N$) would strongly favour the $I=1$ and $I_3=1$ $K^*\bar{D}[1(1^+)]$ or $K^*\bar{D}^*[1(2^+)]$ configurations.

\subsection{Electromagnetic  characteristics of the $K^{(*)}{D}^{(*)}$ molecular candidates}

\begin{table}[!htbp]
\renewcommand\tabcolsep{0.20cm}
\renewcommand{\arraystretch}{1.50}
\caption{M1 radiative decay widths of the $K^{(*)}D^{(*)}$ molecular candidates at three binding energies $E=-4,-11,-17$ MeV.}\label{KDdecaywidth}
\begin{tabular*}{86mm}{@{\extracolsep{\fill}}cc}
\toprule[1pt]
\toprule[1pt]
\multicolumn{2}{c}{Single-channel analysis}\\
\midrule[1.0pt]
Processes&$\Gamma_{H\to H^{\prime}\gamma}$(keV)\\
\midrule[1.0pt]
$K^*D^*[0(0^+)] \rightarrow K^*D[0(1^+)]\gamma$&4.22, 3.94, 3.80\\
$K^*D^*[0(1^+)] \rightarrow K^*D[0(1^+)]\gamma$&4.47, 4.33, 4.26\\
$K^*D^*[0(2^+)] \rightarrow K^*D[0(1^+)]\gamma$&4.62, 4.56, 4.53\\
\midrule[1.0pt]
\multicolumn{2}{c}{Coupled-channel analysis}\\
\midrule[1.0pt]
Processes&$\Gamma_{H\to H^{\prime}\gamma}$(keV)\\
\midrule[1.0pt]
$K^*D^*[0(0^+)] \rightarrow KD^*[0(1^+)]\gamma$&1.12, 0.22, 0.02\\
$K^*D^*[0(1^+)] \rightarrow KD^*[0(1^+)]\gamma$&10.94, 17.15, 21.17\\
$K^*D^*[0(2^+)] \rightarrow KD^*[0(1^+)]\gamma$&0.08, 0.53, 1.67\\
$K^*D^*[0(0^+)] \rightarrow K^*D[0(1^+)]\gamma$&3.34, 2.67, 2.37\\
$K^*D^*[0(1^+)] \rightarrow K^*D[0(1^+)]\gamma$&4.43, 5.32, 5.44\\
$K^*D^*[0(2^+)] \rightarrow K^*D[0(1^+)]\gamma$&5.29, 5.60, 5.75\\
$KD^*[0(1^+)] \rightarrow KD[0(0^+)]\gamma$&4.03, 3.60, 3.38\\
$K^*D[0(1^+)] \rightarrow KD[0(0^+)]\gamma$&2.61, 1.89, 1.51\\
\bottomrule[1pt]\bottomrule[1pt]
\end{tabular*}
\end{table}

In this subsection, we discuss the electromagnetic  characteristics of the $K^{(*)}D^{(*)}$ molecular candidates. The M1 radiative decay widths for the $K^{(*)}{D}^{(*)}$ molecular candidates are summarised in Table \ref{KDdecaywidth}.

Under the single-channel approximation, the radiative decay widths show remarkable stability as the binding energy varies. For the three transitions $K^*D^*[0(0^+)] \to K^*D[0(1^+)]\gamma$, $K^*D^*[0(1^+)] \to K^*D[0(1^+)]\gamma$, and $K^*D^*[0(2^+)] \to K^*D[0(1^+)]\gamma$, the widths change by less than $0.5$ keV over the full binding energy range. Specifically:
\begin{align*}
K^*D^*[0(0^+)] \to K^*D[0(1^+)]\gamma &: 4.22 \to 3.80\ \text{keV},\\
K^*D^*[0(1^+)] \to K^*D[0(1^+)]\gamma &: 4.47 \to 4.26\ \text{keV},\\
K^*D^*[0(2^+)] \to K^*D[0(1^+)]\gamma &: 4.62 \to 4.53\ \text{keV}.
\end{align*}
This near-constancy is a direct consequence of the dominant $S$-wave character of both the $K^*D^*$ and $K^*D$ wave functions, where the $S$-wave fraction remains above $90\%$ throughout. Including the coupled-channel dynamics introduces richer behaviour. Most transitions exhibit only modest changes (below $2$ keV) with binding energy, similar to the single-channel case. However, the $K^*D^*[0(1^+)] \to KD^*[0(1^+)]\gamma$ decay stands out with a pronounced enhancement: the width increases from $10.94$ keV at $E=-4$ MeV to $21.17$ keV at $E=-17$ MeV, nearly doubling. This growth signals a significant coupled-channel mixing effect: the admixture of other $K^{(*)}D^{(*)}$ channels amplifies the M1 transition amplitude.

The radiative decay widths of the $K^{(*)}D^{(*)}$ molecular candidates exhibit rich behavior that depends on the spin-parity quantum numbers, the constituent hadrons, the binding energy, and the coupled-channel mixing. These patterns provide unique fingerprints for identifying the internal structures and the quantum numbers of these discussed exotic hadrons. For instance, the widths of the two transitions $K^*D^*[0(0^+)]\to KD^*[0(1^+)]\gamma$ and $K^*D^*[0(1^+)]\to KD^*[0(1^+)]\gamma$ differ significantly in magnitude, offering a powerful tool for experimentally determining the spin-parity quantum numbers of the $K^*D^*$ molecular candidate.

\begin{figure}[htbp]
\centering
\includegraphics[width=8.6cm]{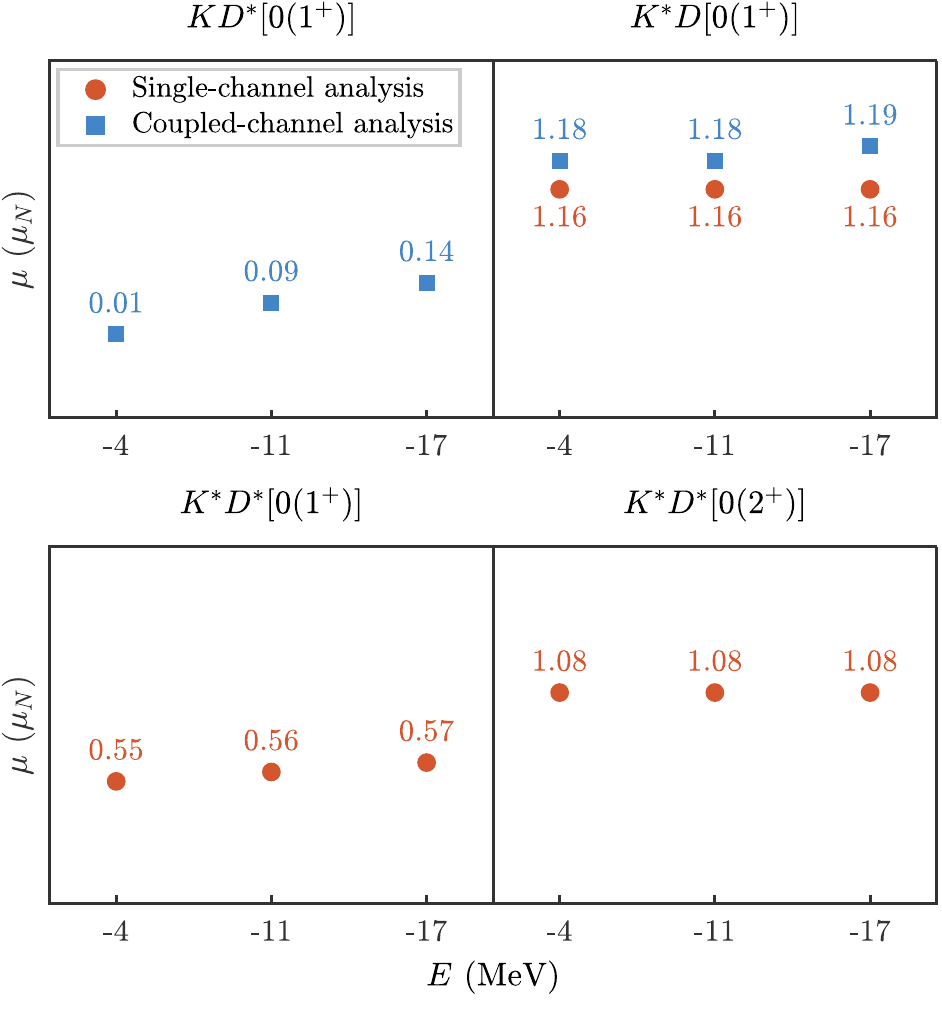}
\caption{Magnetic moments of the $K^{(*)}D^{(*)}$ molecular candidates at the binding energies of $-4$, $-11$, and $-17$ MeV.}
\label{fig:magnetic_moments_KD}
\end{figure}

Complementing our analysis of the radiative decays of the $K^{(*)}D^{(*)}$ molecular candidates, we now discuss their magnetic moments, and the corresponding  numerical results are summarized in Fig. \ref{fig:magnetic_moments_KD}.

Under the single-channel approximation, the magnetic moments of the $K^{(*)}D^{(*)}$ molecular candidates exhibit varying degrees of sensitivity to binding energy, directly reflecting the role of the $S$-$D$ wave mixing effect in the wave functions. For the $K^*D$ state with $I(J^P)=0(1^+)$, the magnetic moment is fixed at $1.16\ \mu_N$ across all binding energies. This perfect stability arises because the $K^*D$ system lacks a tensor force, resulting in negligible $D$-wave admixture even as the binding energy deepens. The magnetic moment is thus dominated by the intrinsic magnetic moments of the constituent $K^*$ and $D$ mesons, with no corrections from higher partial waves. The $K^*D^*$ state with $I(J^P)=0(2^+)$ also shows no binding energy dependence, with its magnetic moment remaining constant at $1.08\ \mu_N$. This behavior is consistent with a dominant $S$-wave configuration, where the $D$-wave fraction is too small to modify the magnetic moment significantly. In contrast, the $K^*D^*$ state with $I(J^P)=0(1^+)$ state exhibits a mild increase in its magnetic moment, from $0.55\ \mu_N$ at $E=-4$ MeV to $0.57\ \mu_N$ at $E=-17$ MeV. This trend signals a gradual growth of the $|^3D_1\rangle$ component in the wave function as the binding energy increases. The tensor force in the $K^*D^*$ system becomes more effective at stronger binding, enhancing the admixture of the $D$-channel and modifying the overall magnetic moment.

When the coupled-channel effect is taken into account, the magnetic moment of the $KD^*[0(1^+)]$ state shows a clear increasing trend with binding energy: $0.01\ \mu_N$ at $E=-4$ MeV, $0.09\ \mu_N$ at $E=-11$ MeV, and $0.14\ \mu_N$ at $E=-17$ MeV. This evolution directly reflects the growing admixture of the $K^*D^*$ channel in the coupled system as the binding strength increases. Since the $K^*D^*$ component has a larger intrinsic magnetic moment contribution than the $KD^*$, its rising proportion amplifies the overall magnetic moment. For the $K^*D[0(1^+)]$ state, the coupled-channel results ($1.18$, $1.18$, and $1.19\ \mu_N$ at the three binding energies) are nearly identical to the single-channel value of $1.16\ \mu_N$. The negligible difference indicates that the $K^*D^*$ channel admixture in the $K^*D/K^*D^*$ coupled system remains very small, even at the strongest binding. This observation aligns perfectly with our earlier mass spectrum analysis, which showed that the $K^*D$ component dominates the wave function with little change in composition as the binding energy deepens.

Notably, the magnetic moments exhibit pronounced differences between the hadronic molecular states composed of identical constituent hadrons but with distinct spin-parity quantum numbers, as illustrated by the $K^*D^*$ state with $I(J^P)=0(1^+)$ and the $K^*D^*$ state with $I(J^P)=0(2^+)$. Likewise, the hadronic molecular states sharing the same spin-parity but differing in their hadronic constituents, such as the $KD^*[0(1^+)]$ and $K^*D^*[0(1^+)]$ states, also yield clearly distinguishable magnetic moments. These systematic variations underscore that the magnetic moment is a sensitive discriminator, capable of resolving both the spin-parity assignment and the specific constituent composition of the $K^{(*)}D^{(*)}$ molecular candidates.

Before concluding this section, we assess how well the electromagnetic properties conform to a molecular interpretation of the predicted states. For a shallow bound state of two constituent hadrons, the impulse approximation is expected to be valid, implying that the overall electromagnetic response is predominantly determined by the individual constituents. This expectation is indeed supported by most of our single-channel calculations, which show only a mild dependence on the binding energy. Specifically, in the $K^{(*)}\bar{D}^{(*)}$ sector, the magnetic moments of the $K^*\bar{D}^*$ states with $I(J^P)=0(1^+)$ and $1(2^+)$ remain practically constant over the entire range of binding energies considered. Likewise, in the $K^{(*)}D^{(*)}$ sector, the magnetic moments of the $K^*D$ state with $I(J^P)=0(1^+)$ and the $K^*D^*$ state with $I(J^P)=0(2^+)$ exhibit no appreciable variation, whereas the spin-$0$ partners have identically zero moments. The stability of these static magnetic moments is traced to the dominant $S$-wave component of the wave function, in which the orbital contribution vanishes and only the intrinsic spins of the constituent mesons contribute. Changes in the binding energy modify the spatial size of the molecule, but they do not affect this leading-order contribution, a feature that is fully consistent with the loosely bound molecular scenario.

For the single-channel case, a residual binding-energy dependence is found only in the magnetic moment of the $K^*D^*$ state with $I(J^P)=0(1^+)$ and in the M1 radiative decay widths for $K^*D^*[0(0^+,1^+,2^+)]\to K^*D[0(1^+)]\gamma$. These effects originate from a gradual $D$-wave admixture, which is associated with the relative motion of the two constituents and thus gives only a subleading contribution. In the coupled-channel framework, the impulse approximation remains applicable in each channel, as the $S$-wave components continue to dominate. The electromagnetic observables then receive contributions from several coupled channels. A comparison with the single-channel results reveals that the additional binding-energy dependence in the coupled systems is not simply due to changes in the spatial extent; rather, it is mainly caused by the redistribution of channel fractions. For example, the magnetic moment of the $KD^*/K^*D/K^*D^*$ system with $I(J^P)=0(1^+)$ rises from $0.01\,\mu_N$ to $0.14\,\mu_N$ as the binding increases, reflecting an increasing admixture of the $K^*D$ and $K^*D^*$ components. In contrast, the magnetic moment of the $K^*D/K^*D^*$ system with the same quantum numbers remains nearly constant, around $1.18$--$1.19\,\mu_N$, which indicates the continued dominance of the $K^*D$ channel. These values, being close to those of the individual constituents, together with the correlated deviations induced by channel mixing, provide evidence that supports the molecular interpretation.

\begin{figure}[!htbp]
\centering
\includegraphics[width=7.5cm]{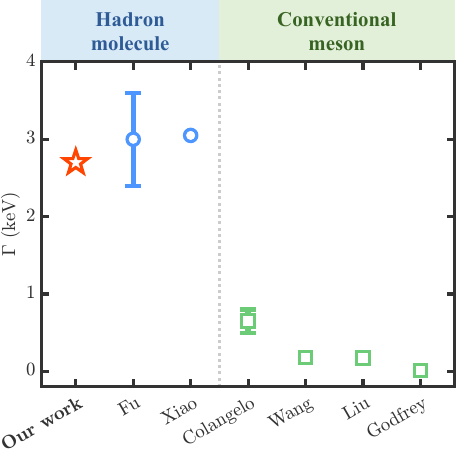}
\caption{Numerical results for the radiative decay width of the $D_{s1}(2460) \to D_{s0}^*(2317) \gamma$ in keV, obtained in the hadron molecule picture and the conventional meson picture.} \label{Comparison}
\end{figure}

Electromagnetic observables offer valuable insight into the nature and internal structure of the hadronic molecular states. As a concrete example, we present in Fig. \ref{Comparison} the radiative decay widths for $D_{s1}(2460)\to D_{s0}^*(2317)\gamma$ calculated under both the molecular scenario and the conventional meson assignment. Within the molecule picture, our prediction ($2.7$ keV) agrees well with previous estimates from Ref. \cite{Fu:2021wde} ($3.0\pm0.6$ keV) and Ref. \cite{Xiao:2016hoa} ($3.0\sim3.1$ keV). This consistency confirms that the constituent quark model, when applied to the electromagnetic properties of molecular states, yields reliable results. In contrast, the predictions from the conventional meson picture, $0.012$ keV \cite{Godfrey:2003kg}, $0.13\sim0.22$ keV~\cite{Liu:2006jx}, $0.18$ keV \cite{Wang:2006fg}, and $0.50\sim0.80$ keV \cite{Colangelo:2005hv}, are typically smaller by about one order of magnitude. Such a clear numerical disparity indicates that the radiative transition $D_{s1}(2460)\to D_{s0}^*(2317)\gamma$ can serve as an effective diagnostic for distinguishing the internal configurations of the $D_{s0}^*(2317)$ and $D_{s1}(2460)$ states. Therefore, the electromagnetic properties are of substantial value not only for probing the inner structure of exotic candidates but also for discriminating between different hadronic interpretations.

\section{Summary and outlook}\label{sec4}
	
Motivated by the observation of numerous charmed-strange hadrons near the $K^{(*)}\bar{D}^{(*)}$ and $K^{(*)}D^{(*)}$ thresholds, in this work we have performed a systematic study of the mass spectra and the electromagnetic  characteristics of the corresponding $K^{(*)}\bar{D}^{(*)}$ and $K^{(*)}D^{(*)}$ charmed-strange molecular tetraquark candidates, explicitly including the $S$-$D$ wave mixing and coupled-channel effects.

Starting from the OBE model, we derive the effective potentials for the $K^{(*)}\bar{D}^{(*)}$ systems, and then obtain those for the $K^{(*)}D^{(*)}$ systems via the $G$-parity rule. These interaction potentials are used to solve the coupled-channel Schr\"odinger equation. Our bound-state calculations predict several promising $K^{(*)}\bar{D}^{(*)}$ and $K^{(*)}D^{(*)}$ molecular tetraquarks awaiting future experimental confirmation. For the $K^{(*)}\bar{D}^{(*)}$ sector, these include $K\bar{D}/K^*\bar{D}^*$ coupled system with $I(J^P)=0(0^+)$, $K\bar{D}^*/K^*\bar{D}/K^*\bar{D}^*$ coupled system with $I(J^P)=0(1^+)$, $K\bar{D}^*/K^*\bar{D}/K^*\bar{D}^*$ coupled system with $I(J^P)=1(1^+)$, $K^*\bar{D}/K^*\bar{D}^*$ coupled system with $I(J^P)=0(1^+)$, $K^*\bar{D}/K^*\bar{D}^*$ coupled system with $I(J^P)=1(1^+)$, $K^*\bar{D}^*$ state with $I(J^P)=0(0^+)$, $K^*\bar{D}^*$ state with $I(J^P)=0(1^+)$,  $K^*\bar{D}^*$ state with $I(J^P)=0(2^+)$, and $K^*\bar{D}^*$ state with $I(J^P)=1(2^+)$. For the $K^{(*)}{D}^{(*)}$ sector, analogous predictions include $KD/K^*D^*$ coupled system with $I(J^P)=0(0^+)$, $K{D}^*/K^*{D}/K^*{D}^*$ coupled system with $I(J^P)=0(1^+)$, $K^*{D}/K^*{D}^*$ coupled system with $I(J^P)=0(1^+)$, $K^*{D}^*$ state with $I(J^P)=0(0^+)$, $K^*{D}^*$ state with $I(J^P)=0(1^+)$, $K^*{D}^*$ state with $I(J^P)=0(2^+)$, and $K^*{D}^*$ state with $I(J^P)=1(0^+)$. The coupled-channel dynamics is found to be essential: it not only enhances the binding in existing channels (e.g., the $K^{*}\bar{D}/K^{*}\bar{D}^{*}$ coupled system with $I(J^P)=0(1^{+})$ becomes bound at $\Lambda\approx1.04$ GeV, which is much lower than the single-channel $K^*\bar D$ state with $I(J^P)=0(1^+)$ at $2.14$ GeV), but also generates new loosely bound states that are completely absent in the single-channel approximation. Thus, ignoring the channel coupling would miss crucial binding mechanisms in the study of the mass spectra of the charm-strange molecular tetraquarks.

Using the constituent quark model, we compute the M1 radiative decay widths and magnetic moments of the predicted charm-strange molecular tetraquark states. These electromagnetic observables exhibit distinct patterns that are highly sensitive to their internal structures. For instance, our systematic analysis of the M1 decay widths reveals clear “fingerprints” for the spin-parity assignment: a direct comparison between $K^*\bar{D}^*[0(0^+)]\to K^*\bar{D}[0(1^+)]\gamma$ and $K^*\bar{D}^*[0(1^+)]\to K^*\bar{D}[0(1^+)]\gamma$ shows markedly different widths, offering a clean method to distinguish $0^+$ from $1^+$ in the $K^*\bar{D}^*$ molecular candidates. The magnetic moments also exhibit pronounced differences between states composed of identical hadrons but with different spin-parity, as illustrated by the $K^*D^*$ state with $I(J^P)=0(1^+)]$ and the $K^*D^*$ state with $I(J^P)=0(2^+)$. These findings collectively demonstrate that the electromagnetic observables are indispensable for unraveling the nature of the charmed-strange molecular tetraquark candidates.

This work presents a comprehensive study of both the mass spectra and the electromagnetic  characteristics of the $K^{(*)}\bar{D}^{(*)}$ and $K^{(*)}D^{(*)}$ molecular tetraquarks, with an emphasis on the critical role of the coupled-channel dynamics. The predicted states and their distinctive electromagnetic signatures provide a clear roadmap for future experimental searches at high-luminosity facilities such as LHCb. Our results encourage experimental collaborations to search for these predicted charmed-strange molecular tetraquark candidates. Of particular interest are the genuinely exotic $K^{(*)}\bar D^{(*)}$ states comprising four different flavors with the valence quark content $\bar c \bar s u d$, which have no conventional hadron counterparts and therefore represent unambiguous signatures of the hadronic molecules.

\section*{Acknowledgement}

F. L. Wang would like to thank Xiang Liu for helpful discussions. This work is supported by the Natural Science Foundation of Gansu Province (No. 26RCKA012), the National Natural Science Foundation of China under Grants No. 12335001 and No. 12405097, and the Talent Scientific Fund of Lanzhou University.

\appendix
	
\section{The effective potentials of the $K^{(*)}\bar{D}^{(*)}$ and $K^{(*)}D^{(*)}$ systems}\label{The effective potentials}
	
This appendix collects the expressions of the coordinate-space effective potentials for the $K^{(*)}\bar{D}^{(*)}$ systems, which read explicitly:
\begin{align}\label{equ}
\mathcal{V}^{K\bar{D}\rightarrow K\bar{D}}(\boldsymbol{r})&=-g_{\sigma}g_{\sigma}^{\prime}Y_{\sigma}+\frac{\beta\beta^{\prime}g_{V}g_{V}^{\prime}}{2}\mathcal{G}_{\mathbb{V}}(I)Y_{\mathbb{V}},\\
\mathcal{V}^{K\bar{D}^{*}\rightarrow K\bar{D}^{*}}(\boldsymbol{r})&=-g_{\sigma}g_{\sigma}^{\prime}\mathcal{O}_{1}[J]Y_{\sigma}+\frac{\beta\beta^{\prime}g_{V}g_{V}^{\prime}}{2}\mathcal{O}_{1}[J]\mathcal{G}_{\mathbb{V}}(I)Y_{\mathbb{V}},\\
\mathcal{V}^{K^{*}\bar{D}\rightarrow K^{*}\bar{D}}(\boldsymbol{r})&=-g_{\sigma}g_{\sigma}^{\prime}\mathcal{O}_{1}[J]Y_{\sigma}+\frac{\beta\beta^{\prime}g_{V}g_{V}^{\prime}}{2}\mathcal{O}_{1}[J]\mathcal{G}_{\mathbb{V}}(I)Y_{\mathbb{V}},\\
\nonumber
\mathcal{V}^{K^{*}\bar{D}^{*}\rightarrow K^{*}\bar{D}^{*}}(\boldsymbol{r})&=-g_{\sigma}g_{\sigma}^{\prime}\mathcal{O}_{2}[J]Y_{\sigma}\\
\nonumber
&-\frac{gg^{\prime}}{3f_{\pi}^{2}}(\mathcal{O}_{3}[J]\mathcal{Z}_{r}+\mathcal{O}_{4}[J]\mathcal{T}_{r})\mathcal{G}_{\mathbb{P}}(I)Y_{\mathbb{P}}\\
\nonumber
&+\frac{\beta\beta^{\prime}g_{V}g_{V}^{\prime}}{2}\mathcal{O}_{2}[J]\mathcal{G}_{\mathbb{V}}(I)Y_{\mathbb{V}}\\
&-\frac{2\lambda\lambda^{\prime}g_{V}g_{V}^{\prime}}{3}(2\mathcal{O}_{3}[J]\mathcal{Z}_{r}-\mathcal{O}_{4}[J]\mathcal{T}_{r})\mathcal{G}_{\mathbb{V}}(I)Y_{\mathbb{V}},\\
\nonumber
\mathcal{V}^{K^{*}\bar{D}\rightarrow K\bar{D}^{*}}(\boldsymbol{r})&=\frac{gg^{\prime}}{3f_{\pi}^{2}}(\mathcal{O}_{5}[J]\mathcal{Z}_{r}+\mathcal{O}_{6}[J]\mathcal{T}_{r})\mathcal{G}_{\mathbb{P}}(I)Y_{\mathbb{P}1}\\
&+\frac{2\lambda\lambda^{\prime}g_{V}g_{V}^{\prime}}{3}(2\mathcal{O}_{5}[J]\mathcal{Z}_{r}-\mathcal{O}_{6}[J]\mathcal{T}_{r})\mathcal{G}_{\mathbb{V}}(I)Y_{\mathbb{V}1},\\
\nonumber
\mathcal{V}^{K\bar{D}^{*}\rightarrow K^{*}\bar{D}}(\boldsymbol{r})&=\frac{gg^{\prime}}{3f_{\pi}^{2}}(\mathcal{O}_{5}[J]\mathcal{Z}_{r}+\mathcal{O}_{6}[J]\mathcal{T}_{r})\mathcal{G}_{\mathbb{P}}(I)Y_{\mathbb{P}2}\\
&+\frac{2\lambda\lambda^{\prime}g_{V}g_{V}^{\prime}}{3}(2\mathcal{O}_{5}[J]\mathcal{Z}_{r}-\mathcal{O}_{6}[J]\mathcal{T}_{r})\mathcal{G}_{\mathbb{V}}(I)Y_{\mathbb{V}2},\\
\nonumber
\mathcal{V}^{K^{*}\bar{D}^{*}\rightarrow K^{*}\bar{D}}(\boldsymbol{r})&=-\frac{gg^{\prime}}{3f_{\pi}^{2}}(\mathcal{O}_{7}[J]\mathcal{Z}_{r}+\mathcal{O}_{8}[J]\mathcal{T}_{r})\mathcal{G}_{\mathbb{P}}(I)Y_{\mathbb{P}3}\\
&-\frac{2\lambda\lambda^{\prime}g_{V}g_{V}^{\prime}}{3}(2\mathcal{O}_{7}[J]\mathcal{Z}_{r}-\mathcal{O}_{8}[J]\mathcal{T}_{r})\mathcal{G}_{\mathbb{V}}(I)Y_{\mathbb{V}3},\\
\nonumber
\mathcal{V}^{K^{*}\bar{D}^{*}\rightarrow K\bar{D}^{*}}(\boldsymbol{r})&=-\frac{gg^{\prime}}{3f_{\pi}^{2}}(\mathcal{O}_{7}[J]\mathcal{Z}_{r}+\mathcal{O}_{8}[J]\mathcal{T}_{r})\mathcal{G}_{\mathbb{P}}(I)Y_{\mathbb{P}4}\\
&-\frac{2\lambda\lambda^{\prime}g_{V}g_{V}^{\prime}}{3}(2\mathcal{O}_{7}[J]\mathcal{Z}_{r}-\mathcal{O}_{8}[J]\mathcal{T}_{r})\mathcal{G}_{\mathbb{V}}(I)Y_{\mathbb{V}4},\\
\nonumber
\mathcal{V}^{K^{*}\bar{D}\rightarrow K^{*}\bar{D}^{*}}(\boldsymbol{r})&=-\frac{gg^{\prime}}{3f_{\pi}^{2}}(\mathcal{O}_{9}[J]\mathcal{Z}_{r}+\mathcal{O}_{10}[J]\mathcal{T}_{r})\mathcal{G}_{\mathbb{P}}(I)Y_{\mathbb{P}5}\\
&-\frac{2\lambda\lambda^{\prime}g_{V}g_{V}^{\prime}}{3}(2\mathcal{O}_{9}[J]\mathcal{Z}_{r}-\mathcal{O}_{10}[J]\mathcal{T}_{r})\mathcal{G}_{\mathbb{V}}(I)Y_{\mathbb{V}5},\\
\nonumber
\mathcal{V}^{K\bar{D}^{*}\rightarrow K^{*}\bar{D}^{*}}(\boldsymbol{r})&=-\frac{gg^{\prime}}{3f_{\pi}^{2}}(\mathcal{O}_{9}[J]\mathcal{Z}_{r}+\mathcal{O}_{10}[J]\mathcal{T}_{r})\mathcal{G}_{\mathbb{P}}(I)Y_{\mathbb{P}6}\\
&-\frac{2\lambda\lambda^{\prime}g_{V}g_{V}^{\prime}}{3}(2\mathcal{O}_{9}[J]\mathcal{Z}_{r}-\mathcal{O}_{10}[J]\mathcal{T}_{r})\mathcal{G}_{\mathbb{V}}(I)Y_{\mathbb{V}6},\\
\nonumber
\mathcal{V}^{K\bar{D}\rightarrow K^{*}\bar{D}^{*}}(\boldsymbol{r})&=\frac{gg^{\prime}}{3f_{\pi}^{2}}(\mathcal{O}_{11}[J]\mathcal{Z}_{r}+\mathcal{O}_{12}[J]\mathcal{T}_{r})\mathcal{G}_{\mathbb{P}}(I)Y_{\mathbb{P}7}\\
&+\frac{2\lambda\lambda^{\prime}g_{V}g_{V}^{\prime}}{3}(2\mathcal{O}_{11}[J]\mathcal{Z}_{r}-\mathcal{O}_{12}[J]\mathcal{T}_{r})\mathcal{G}_{\mathbb{V}}(I)Y_{\mathbb{V}7},\\
\nonumber
\mathcal{V}^{K^{*}\bar{D}^{*}\rightarrow K\bar{D}}(\boldsymbol{r})&=\frac{gg^{\prime}}{3f_{\pi}^{2}}(\mathcal{O}_{13}[J]\mathcal{Z}_{r}+\mathcal{O}_{14}[J]\mathcal{T}_{r})\mathcal{G}_{\mathbb{P}}(I)Y_{\mathbb{P}8}\\
&+\frac{2\lambda\lambda^{\prime}g_{V}g_{V}^{\prime}}{3}(2\mathcal{O}_{13}[J]\mathcal{Z}_{r}-\mathcal{O}_{14}[J]\mathcal{T}_{r})\mathcal{G}_{\mathbb{V}}(I)Y_{\mathbb{V}8}.
\end{align}
The operators $\mathcal{Z}_r$ and $\mathcal{T}_r$ appearing above are defined through
\begin{eqnarray*}
\mathcal{Z}_{r}=\frac{1}{r^{2}}\frac{\partial}{\partial r}r^{2}\frac{\partial}{\partial r}\quad\text{and}\quad
\mathcal{T}_{r}=\frac{1}{r}\frac{\partial}{\partial r}\frac{1}{r}\frac{\partial}{\partial r}.
\end{eqnarray*}
The function $Y_i$ takes the following piecewise form:
\begin{eqnarray*}
\renewcommand{\arraystretch}{2.50}
Y_{i}=\left\{
\begin{array}{l}
\lvert q_{i} \rvert \leqslant m,\displaystyle{\frac{e^{-m_{i}r}-e^{-\Lambda_{i} r}}{4\pi r}-\frac{\Lambda_{i}^{2}-m_{i}^{2}}{8\pi \Lambda_{i}}e^{-\Lambda_{i}r}},\\
\lvert q_{i} \rvert \textgreater m,\displaystyle{\frac{\cos(m_{i}^{\prime}r)-e^{-\Lambda_{i} r}}{4\pi r}-\frac{\Lambda_{i}^{2}+m_{i}^{\prime2}}{8\pi\Lambda_{i}}e^{-\Lambda_{i}r}}.\\
\end{array}
\right.
\end{eqnarray*}
For a given transition $AB \to CD$, the quantity $q_i$ is given by \cite{Wang:2023ael}
\begin{eqnarray*}
q_{i}^{2}=\displaystyle{\frac{m_{A}^{2}+m_{D}^{2}-m_{B}^{2}-m_{C}^{2}}{(2m_{C}+2m_{D})^{2}}}.
\end{eqnarray*}
Based on $q_i$, the redefined mass and cutoff parameters $m_i$, $m'_i$, and $\Lambda_i$ are given by
\begin{eqnarray*}
m_{i}=\sqrt{m^{2}-q_{i}^{2}},\quad
m_{i}^{\prime}=\sqrt{q_{i}^{2}-m^{2}},\quad\text{and}\quad
\Lambda_{i}=\sqrt{\Lambda^{2}-q_{i}^{2}}.
\end{eqnarray*}
For the specific transitions $K^{(*)}\bar D^{(*)}\to K^{(*)}\bar D^{(*)}$, the numerical values of $q_i$ are $q_1 = 0.220$ GeV, $q_2 = 0.200$ GeV, $q_3 = 0.099$ GeV, $q_4 = 0.111$ GeV, $q_5 = 0.094$ GeV, $q_6 = 0.096$ GeV, $q_7 = 0.001$ GeV, and $q_8 = 0.002$ GeV. The meson masses employed in this work are $m_\sigma = 600.00$ MeV, $m_\pi = 137.27$ MeV, $m_\eta = 547.86$ MeV, $m_\rho = 769.00$ MeV, $m_\omega = 782.66$ MeV, $m_{{D}} = 1867.25$ MeV, $m_{{D}^*} = 2008.56$ MeV, $m_K = 495.64$ MeV, and $m_{K^*} = 895.55$ MeV \cite{ParticleDataGroup:2024cfk}.
	
The isospin factors entering the effective potentials for the $K^{(*)}\bar D^{(*)}\to K^{(*)}\bar D^{(*)}$ processes are
\begin{eqnarray*}
\mathcal{G}_{\mathbb{P}}(0)Y_{\mathbb{P}}&=-\displaystyle{\frac{3}{2}}Y_{\pi}+\frac{1}{6}Y_{\eta},~~~~~
\mathcal{G}_{\mathbb{P}}(1)Y_{\mathbb{P}}=\frac{1}{2}Y_{\pi}+\frac{1}{6}Y_{\eta},\nonumber\\
\mathcal{G}_{\mathbb{V}}(0)Y_{\mathbb{V}}&=-\displaystyle{\frac{3}{2}}Y_{\rho}+\frac{1}{2}Y_{\omega},~~~~~
\mathcal{G}_{\mathbb{V}}(1)Y_{\mathbb{V}}=\frac{1}{2}Y_{\rho}+\frac{1}{2}Y_{\omega}.
\end{eqnarray*}

The spin-dependent operators $\mathcal{O}_i[J]$ appearing in the effective potentials are defined as:
\begin{align*}
\mathcal{O}_{1}[J]&= \bm{\epsilon}_{3}^{\dag}\cdot\bm{\epsilon}_{1}, &
\mathcal{O}_{2}[J]&= (\bm{\epsilon}_{3}^{\dag}\cdot\bm{\epsilon}_{1})(\bm{\epsilon}_{4}^{\dag}\cdot\bm{\epsilon}_{2}), \notag\\[0.8ex]
\mathcal{O}_{3}[J]&= (\bm{\epsilon}_{3}^{\dag} \times \bm{\epsilon}_{1})\cdot(\bm{\epsilon}_{4}^{\dag}\times\bm{\epsilon}_{2}), &
\mathcal{O}_{4}[J]&= S(\bm{\epsilon}_{3}^{\dag} \times \bm{\epsilon}_{1},\bm{\epsilon}_{4}^{\dag}\times\bm{\epsilon}_{2},\bm{\hat{r}}), \notag\\[0.8ex]
\mathcal{O}_{5}[J]&= \bm{\epsilon}_{4}^{\dag}\cdot\bm{\epsilon}_{1}, &
\mathcal{O}_{6}[J]&= S(\bm{\epsilon}_{4}^{\dag},\bm{\epsilon}_{1},\bm{\hat{r}}), \notag\\[0.8ex]
\mathcal{O}_{7}[J]&= (\bm{\epsilon}_{3}^{\dag} \times \bm{\epsilon}_{1})\cdot\bm{\epsilon}_{2}, &
\mathcal{O}_{8}[J]&= S(\bm{\epsilon}_{2},\bm{\epsilon}_{3}^{\dag} \times \bm{\epsilon}_{1},\bm{\hat{r}}), \notag\\[0.8ex]
\mathcal{O}_{9}[J]&= (\bm{\epsilon}_{3}^{\dag} \times \bm{\epsilon}_{1})\cdot\bm{\epsilon}_{4}^{\dag}, &
\mathcal{O}_{10}[J]&= S(\bm{\epsilon}_{4}^{\dag},\bm{\epsilon}_{3}^{\dag} \times \bm{\epsilon}_{1},\bm{\hat{r}}), \notag\\[0.8ex]
\mathcal{O}_{11}[J]&= \bm{\epsilon}_{4}^{\dag}\cdot\bm{\epsilon}_{3}^{\dag}, &
\mathcal{O}_{12}[J]&= S(\bm{\epsilon}_{4}^{\dag},\bm{\epsilon}_{3}^{\dag},\bm{\hat{r}}), \notag\\[0.8ex]
\mathcal{O}_{13}[J]&= \bm{\epsilon}_{2}\cdot\bm{\epsilon}_{1}, &
\mathcal{O}_{14}[J]&= S(\bm{\epsilon}_{2},\bm{\epsilon}_{1},\bm{\hat{r}}),
\end{align*}
where the tensor operator $S(\bm{x},\bm{y},\bm{\hat{r}})$ is
$S(\bm{x},\bm{y},\bm{\hat{r}})=3(\bm{\hat{r}}\cdot\bm{x})(\bm{\hat{r}}\cdot\bm{y})-\bm{x}\cdot\bm{y}$.

Using the spin-orbital wave functions of the $K^{(*)}\bar{D}^{(*)}$ systems in Eqs. (\ref{spin0})–(\ref{spin2}), we evaluate the numerical matrix elements $\langle f|\mathcal{O}_{i}[J]|i\rangle$ in the following:
\begin{align*}
\mathcal{O}_{1}[1] &= \mathrm{diag}(1, 1), &
\mathcal{O}_{2}[0] &= \mathrm{diag}(1, 1), \notag\\[0.8ex]
\mathcal{O}_{2}[1] &= \mathrm{diag}(1, 1, 1), &
\mathcal{O}_{2}[2] &= \mathrm{diag}(1, 1, 1, 1), \notag\\[0.8ex]
\mathcal{O}_{3}[0] &= \mathrm{diag}(2, -1), &
\mathcal{O}_{3}[1] &= \mathrm{diag}(1, 1, -1), \notag\\[0.8ex]
\mathcal{O}_{3}[2] &= \mathrm{diag}(-1, 2, 1, -1), &
\mathcal{O}_{5}[1] &= \mathrm{diag}(1, 1), \notag\\[0.8ex]
\mathcal{O}_{4}[0] &= \begin{pmatrix} 0 & \sqrt{2} \\ \sqrt{2} & 2 \end{pmatrix}, &
\mathcal{O}_{6}[1] &= \begin{pmatrix} 0 & -\sqrt{2} \\ -\sqrt{2} & 1 \end{pmatrix}, \notag\\[0.8ex]
\mathcal{O}_{4}[1] &= \begin{pmatrix} 0 & -\sqrt{2} & 0 \\ -\sqrt{2} & 1 & 0 \\ 0 & 0 & 1 \end{pmatrix}, & \mathcal{O}_{7}[1] &= \begin{pmatrix}
    \sqrt{2} & 0 & 0\\
    0 & \sqrt{2} & 0
\end{pmatrix}, \notag\\[0.8ex]
\mathcal{O}_{4}[2] &= \begin{pmatrix} 0 & \frac{\sqrt{2}}{\sqrt{5}} & 0 & -\frac{\sqrt{14}}{\sqrt{5}} \\ \frac{\sqrt{2}}{\sqrt{5}} & 0 & 0 & -\frac{2}{\sqrt{7}} \\ 0 & 0 & -1 & 0 \\ -\frac{\sqrt{14}}{\sqrt{5}} & -\frac{2}{\sqrt{7}} & 0 & -\frac{3}{7} \end{pmatrix}, &
\mathcal{O}_{8}[1] &= \begin{pmatrix}
    0 & -1 & \sqrt{3}\\
    -1 & \frac{1}{\sqrt{2}} & \frac{\sqrt{3}}{\sqrt{2}}
\end{pmatrix}, \notag\\[0.8ex]
\mathcal{O}_{9}[1] &= \begin{pmatrix}
    \sqrt{2} & 0\\
    0 & \sqrt{2}\\
    0 & 0
\end{pmatrix},&
\mathcal{O}_{10}[1] &= \begin{pmatrix}
    0 & -1\\
    -1 & \frac{1}{\sqrt{2}}\\
    \sqrt{3} & \frac{\sqrt{3}}{\sqrt{2}}
\end{pmatrix},\notag\\[0.8ex]
\mathcal{O}_{11}[0] &= \begin{pmatrix}
    -\sqrt{3} \\ 0
\end{pmatrix}, &
\mathcal{O}_{12}[0] &= \begin{pmatrix}
    0 \\ \sqrt{6}
\end{pmatrix},\notag\\[0.8ex]
\mathcal{O}_{13}[0] &= \begin{pmatrix}
    -\sqrt{3} & 0
\end{pmatrix}, &
\mathcal{O}_{14}[0] &= \begin{pmatrix}
    0 & \sqrt{6}
\end{pmatrix}.
\end{align*}

Following the $G$-parity rule \cite{Klempt:2002ap}, the effective interactions for the $K^{(*)}D^{(*)} \to K^{(*)}D^{(*)}$ processes through the light meson exchanges can be subsequently obtained.


\begin{thebibliography}{99}
		
%\cite{Belle:2003nnu}
\bibitem{Belle:2003nnu}
S.~K.~Choi \textit{et al.} [Belle],
Observation of a narrow charmonium-like state in exclusive $B^\pm \to K^\pm \pi^+ \pi^- J/\psi$ decays,	
\href{https://journals.aps.org/prl/abstract/10.1103/PhysRevLett.91.262001}{Phys. Rev. Lett. \textbf{91}, 262001 (2003)}.
%doi:10.1103/PhysRevLett.91.262001
%[arXiv:hep-ex/0309032 [hep-ex]].
%2913 citations counted in INSPIRE as of 10 Mar 2026

%\cite{Liu:2013waa}
\bibitem{Liu:2013waa}
X.~Liu,
An overview of $XYZ$ new particles,
\href{http://dx.doi.org/10.1007/s11434-014-0407-2}{Chin.\ Sci.\ Bull.\  {\bf 59}, 3815 (2014)}.
%doi:10.1007/s11434-014-0407-2
%\href{https://arxiv.org/abs/1312.7408}{[arXiv:1312.7408 [hep-ph]]}.
%%CITATION = doi:10.1007/s11434-014-0407-2;%%
%67 citations counted in INSPIRE as of 19 Jul 2017

%\cite{Hosaka:2016pey}
\bibitem{Hosaka:2016pey}
A.~Hosaka, T.~Iijima, K.~Miyabayashi, Y.~Sakai and S.~Yasui,
Exotic hadrons with heavy flavors: $X$, $Y$, $Z$, and related states,
\href{http://dx.doi.org/10.1093/ptep/ptw045}{Prog. Theor. Exp. Phys. {\bf 2016}, 062C01 (2016)}.
% doi:10.1093/ptep/ptw045
%\href{https://arxiv.org/abs/1603.09229}{[arXiv:1603.09229 [hep-ph]]}.
%%CITATION = doi:10.1093/ptep/ptw045;%%
%29 citations counted in INSPIRE as of 25 Jul 2017

\bibitem{Richard:2016eis}
J.~M.~Richard,
Exotic hadrons: review and perspectives,
\href{https://link.springer.com/article/10.1007/s00601-016-1159-0}{Few Body Syst. \textbf{57}, 1185-1212 (2016)}.
%doi:10.1007/s00601-016-1159-0
%[arXiv:1606.08593 [hep-ph]].
%76 citations counted in INSPIRE as of 08 Jun 2021

%\cite{Chen:2016qju}
\bibitem{Chen:2016qju}
H.~X.~Chen, W.~Chen, X.~Liu and S.~L.~Zhu,
The hidden-charm pentaquark and tetraquark states,
\href{http://linkinghub.elsevier.com/retrieve/pii/S037015731630103X}{Phys.\ Rep.\  {\bf 639}, 1 (2016)}.
% doi:10.1016/j.physrep.2016.05.004
%\href{http://arxiv.org/abs/arXiv:1601.02092}{[arXiv:1601.02092 [hep-ph]]}.
%%CITATION = doi:10.1016/j.physrep.2016.05.004;%%
%175 citations counted in INSPIRE as of 19 Jul 2017

%\cite{Lebed:2016hpi}
\bibitem{Lebed:2016hpi}
R.~F.~Lebed, R.~E.~Mitchell and E.~S.~Swanson,
Heavy-Quark QCD Exotica,
\href{https://www.sciencedirect.com/science/article/pii/S0146641016300734?via\%3Dihub}{Prog. Part. Nucl. Phys. \textbf{93}, 143-194 (2017)}.
%doi:10.1016/j.ppnp.2016.11.003
%[arXiv:1610.04528 [hep-ph]].
%435 citations counted in INSPIRE as of 07 Jul 2022

%\cite{Olsen:2017bmm}
\bibitem{Olsen:2017bmm}
S.~L.~Olsen, T.~Skwarnicki and D.~Zieminska,
Nonstandard heavy mesons and baryons: Experimental evidence,
\href{https://journals.aps.org/rmp/abstract/10.1103/RevModPhys.90.015003}{Rev.\ Mod.\ Phys.\  {\bf 90}, 015003 (2018)}.
%doi:10.1103/RevModPhys.90.015003
%% [arXiv:1708.04012 [hep-ph]].
%%CITATION = doi:10.1103/RevModPhys.90.015003;%%
%196 citations counted in INSPIRE as of 17 Dec 2019

%\cite{Guo:2017jvc}
\bibitem{Guo:2017jvc}
F.~K.~Guo, C.~Hanhart, U.~G.~Mei$\ss$ner, Q.~Wang, Q.~Zhao and B.~S.~Zou,
Hadronic molecules,
\href{https://journals.aps.org/rmp/abstract/10.1103/RevModPhys.90.015004}{Rev.\ Mod.\ Phys.\  {\bf 90}, 015004 (2018)}.
%%doi:10.1103/RevModPhys.90.015004
%%[arXiv:1705.00141 [hep-ph]].
%%CITATION = doi:10.1103/RevModPhys.90.015004;%%
%322 citations counted in INSPIRE as of 17 Dec 2019

%\cite{Liu:2019zoy}
\bibitem{Liu:2019zoy}
Y.~R.~Liu, H.~X.~Chen, W.~Chen, X.~Liu and S.~L.~Zhu,
Pentaquark and tetraquark states,
\href{https://www.sciencedirect.com/science/article/pii/S0146641019300304?via\%3Dihub}{Prog.\ Part.\ Nucl.\ Phys.\  {\bf 107}, 237 (2019)}.
%%doi:10.1016/j.ppnp.2019.04.003
%%[arXiv:1903.11976 [hep-ph]].
%%CITATION = doi:10.1016/j.ppnp.2019.04.003;%%
%30 citations counted in INSPIRE as of 29 Jun 2019

%\cite{Brambilla:2019esw}
\bibitem{Brambilla:2019esw}
N.~Brambilla, S.~Eidelman, C.~Hanhart, A.~Nefediev, C.~P.~Shen, C.~E.~Thomas, A.~Vairo and C.~Z.~Yuan,
The $XYZ$ states: Experimental and theoretical status and perspectives,
\href{https://www.sciencedirect.com/science/article/pii/S0370157320301915?via\%3Dihub}{Phys. Rep. \textbf{873}, 1 (2020)}.
%doi:10.1016/j.physrep.2020.05.001
%[arXiv:1907.07583 [hep-ex]].
%99 citations counted in INSPIRE as of 24 Aug 2020

%\cite{Chen:2022asf}
\bibitem{Chen:2022asf}
H.~X.~Chen, W.~Chen, X.~Liu, Y.~R.~Liu and S.~L.~Zhu,
An updated review of the new hadron states,
\href{https://iopscience.iop.org/article/10.1088/1361-6633/aca3b6}{Rept. Prog. Phys. \textbf{86}, 026201 (2023)}.
%doi:10.1088/1361-6633/aca3b6
%[arXiv:2204.02649 [hep-ph]].
%164 citations counted in INSPIRE as of 02 Jul 2023

%\cite{Meng:2022ozq}
\bibitem{Meng:2022ozq}
L.~Meng, B.~Wang, G.~J.~Wang and S.~L.~Zhu,
Chiral perturbation theory for heavy hadrons and chiral effective field theory for heavy hadronic molecules,
\href{https://www.sciencedirect.com/science/article/pii/S0370157323001679?via\%3Dihub}{Phys. Rept. \textbf{1019}, 1-149 (2023)}.
%doi:10.1016/j.physrep.2023.04.003
%[arXiv:2204.08716 [hep-ph]].
%63 citations counted in INSPIRE as of 02 Jul 2023

%\cite{Liu:2024uxn}
\bibitem{Liu:2024uxn}
M.~Z.~Liu, Y.~W.~Pan, Z.~W.~Liu, T.~W.~Wu, J.~X.~Lu and L.~S.~Geng,
Three ways to decipher the nature of exotic hadrons: Multiplets, three-body hadronic molecules, and correlation functions,
\href{https://www.sciencedirect.com/science/article/pii/S0370157324004277?via\%3Dihub}{Phys. Rept. \textbf{1108}, 1-108 (2025)}.
%doi:10.1016/j.physrep.2024.12.001
%[arXiv:2404.06399 [hep-ph]].
%114 citations counted in INSPIRE as of 07 Oct 2025

%\cite{Wang:2025sic}
\bibitem{Wang:2025sic}
Z.~G.~Wang,
Review of the QCD sum rules for exotic states,
\href{https://journal.hep.com.cn/fop/EN/10.15302/frontphys.2026.016300}{Front. Phys. (Beijing) \textbf{21}, 016300 (2026)}.
%doi:10.15302/frontphys.2026.016300
%[arXiv:2502.11351 [hep-ph]].
%32 citations counted in INSPIRE as of 07 Oct 2025

%\cite{Wang:2025dur}
\bibitem{Wang:2025dur}
X.~Wang, X.~Liu and Y.~Gao,
Colloquium: Hadron production in open-charm meson pairs at $e^+e^-$ colliders,
\href{https://journals.aps.org/rmp/abstract/10.1103/2mrp-chly}{Rev. Mod. Phys. \textbf{98}, 021001 (2026)}.
%doi:10.1103/2mrp-chly
%[arXiv:2502.15117 [hep-ex]].
%15 citations counted in INSPIRE as of 22 May 2026

%\cite{Bai:2026atm}
\bibitem{Bai:2026atm}
Z.~Y.~Bai, D.~Y.~Chen, Qi-Huang, X.~Liu, S.~Q.~Luo and J.~Z.~Wang,
Unquenched charmonium and beyond,
\href{https://arxiv.org/abs/2602.19887}{arXiv:2602.19887}.
%0 citations counted in INSPIRE as of 23 Mar 2026
		
%\cite{BaBar:2003oey}
\bibitem{BaBar:2003oey}
B.~Aubert \textit{et al.} [BaBar],
Observation of a narrow meson decaying to $D_s^+ \pi^0$ at a mass of 2.32-GeV/c$^2$,
\href{https://journals.aps.org/prl/abstract/10.1103/PhysRevLett.90.242001}{Phys. Rev. Lett. \textbf{90}, 242001 (2003)}.
%doi:10.1103/PhysRevLett.90.242001
%[arXiv:hep-ex/0304021 [hep-ex]].
%1067 citations counted in INSPIRE as of 01 Apr 2026
		
%\cite{CLEO:2003ggt}
\bibitem{CLEO:2003ggt}
D.~Besson \textit{et al.} [CLEO],
Observation of a narrow resonance of mass 2.46-GeV/c$^2$ decaying to $D_s^{*+} \pi^0$ and confirmation of the $D^*_{sJ}(2317)$ state,
\href{https://journals.aps.org/prd/abstract/10.1103/PhysRevD.68.032002}{Phys. Rev. D \textbf{68}, 032002 (2003)}.
%[erratum: Phys. Rev. D \textbf{75}, 119908 (2007)]
%doi:10.1103/PhysRevD.68.032002
%[arXiv:hep-ex/0305100 [hep-ex]].
%764 citations counted in INSPIRE as of 10 Mar 2026

%\cite{Godfrey:1985xj}
\bibitem{Godfrey:1985xj}
S.~Godfrey and N.~Isgur,
Mesons in a Relativized Quark Model with Chromodynamics,
\href{https://journals.aps.org/prd/abstract/10.1103/PhysRevD.32.189}{Phys. Rev. D \textbf{32}, 189-231 (1985)}.
%doi:10.1103/PhysRevD.32.189
%3524 citations counted in INSPIRE as of 09 May 2026

%\cite{Godfrey:1986wj}
\bibitem{Godfrey:1986wj}
S.~Godfrey and R.~Kokoski,
The Properties of p Wave Mesons with One Heavy Quark,
\href{https://journals.aps.org/prd/abstract/10.1103/PhysRevD.43.1679}{Phys. Rev. D \textbf{43}, 1679-1687 (1991)}.
%doi:10.1103/PhysRevD.43.1679
%539 citations counted in INSPIRE as of 09 May 2026

%\cite{Ebert:2009ua}
\bibitem{Ebert:2009ua}
D.~Ebert, R.~N.~Faustov and V.~O.~Galkin,
Heavy-light meson spectroscopy and Regge trajectories in the relativistic quark model,
\href{https://link.springer.com/article/10.1140/epjc/s10052-010-1233-6}{Eur. Phys. J. C \textbf{66}, 197-206 (2010)}.
%doi:10.1140/epjc/s10052-010-1233-6
%[arXiv:0910.5612 [hep-ph]].
%297 citations counted in INSPIRE as of 09 May 2026

%\cite{Song:2015nia}
\bibitem{Song:2015nia}
Q.~T.~Song, D.~Y.~Chen, X.~Liu and T.~Matsuki,
Charmed-strange mesons revisited: mass spectra and strong decays,
\href{https://journals.aps.org/prd/abstract/10.1103/PhysRevD.91.054031}{Phys. Rev. D \textbf{91}, 054031 (2015)}.
%doi:10.1103/PhysRevD.91.054031
%[arXiv:1501.03575 [hep-ph]].
%109 citations counted in INSPIRE as of 09 May 2026

%\cite{Godfrey:2015dva}
\bibitem{Godfrey:2015dva}
S.~Godfrey and K.~Moats,
Properties of Excited Charm and Charm-Strange Mesons,
\href{https://journals.aps.org/prd/abstract/10.1103/PhysRevD.93.034035}{Phys. Rev. D \textbf{93}, 034035 (2016)}.
%doi:10.1103/PhysRevD.93.034035
%[arXiv:1510.08305 [hep-ph]].
%216 citations counted in INSPIRE as of 09 May 2026
%%%%%%%%%%%%%%%%%%%%%%%%%%%
%\cite{Barnes:2003dj}
\bibitem{Barnes:2003dj}
T.~Barnes, F.~E.~Close and H.~J.~Lipkin,
Implications of a $DK$ molecule at 2.32-GeV,
\href{https://journals.aps.org/prd/abstract/10.1103/PhysRevD.68.054006}{Phys. Rev. D \textbf{68}, 054006 (2003)}.
%doi:10.1103/PhysRevD.68.054006
%[arXiv:hep-ph/0305025 [hep-ph]].
%479 citations counted in INSPIRE as of 10 Mar 2026\

%\cite{Kolomeitsev:2003ac}
\bibitem{Kolomeitsev:2003ac}
E.~E.~Kolomeitsev and M.~F.~M.~Lutz,
On Heavy light meson resonances and chiral symmetry,
\href{https://www.sciencedirect.com/science/article/abs/pii/S0370269303019105?via%3Dihub}{Phys. Lett. B \textbf{582}, 39-48 (2004)}.
%doi:10.1016/j.physletb.2003.10.118
%[arXiv:hep-ph/0307133 [hep-ph]].
%370 citations counted in INSPIRE as of 17 May 2026

%\cite{Hofmann:2003je}
\bibitem{Hofmann:2003je}
J.~Hofmann and M.~F.~M.~Lutz,
Open charm meson resonances with negative strangeness,
\href{https://www.sciencedirect.com/science/article/abs/pii/S0375947403019766?via%3Dihub}{Nucl. Phys. A \textbf{733}, 142-152 (2004)}.
%doi:10.1016/j.nuclphysa.2003.12.013
%[arXiv:hep-ph/0308263 [hep-ph]].
%191 citations counted in INSPIRE as of 17 May 2026

%\cite{Chen:2004dy}
\bibitem{Chen:2004dy}
Y.~Q.~Chen and X.~Q.~Li,
A Comprehensive four-quark interpretation of $D_s(2317)$, $D_s(2457)$ and $D_s(2632)$,
\href{https://journals.aps.org/prl/abstract/10.1103/PhysRevLett.93.232001}{Phys. Rev. Lett. \textbf{93}, 232001 (2004)}.
%doi:10.1103/PhysRevLett.93.232001
%[arXiv:hep-ph/0407062 [hep-ph]].
%177 citations counted in INSPIRE as of 10 Mar 2026

%1
%\cite{Sassen:2005ej}
\bibitem{Sassen:2005ej}
F.~P.~Sassen and S.~Krewald,
An investigation of the $D_s^+ \pi^0 - D K - D_s^+ \eta$ coupled channel dynamics,
\href{https://www.worldscientific.com/doi/abs/10.1142/S0217751X05022226}{Int. J. Mod. Phys. A \textbf{20}, 705-707 (2005)}.
%doi:10.1142/S0217751X05022226
%6 citations counted in INSPIRE as of 16 May 2026

%\cite{Zhang:2006ix}
\bibitem{Zhang:2006ix}
Y.~J.~Zhang, H.~C.~Chiang, P.~N.~Shen and B.~S.~Zou,
Possible S-wave bound-states of two pseudoscalar mesons,
\href{https://journals.aps.org/prd/abstract/10.1103/PhysRevD.74.014013}{Phys. Rev. D \textbf{74}, 014013 (2006)}.
%doi:10.1103/PhysRevD.74.014013
%[arXiv:hep-ph/0604271 [hep-ph]].
%95 citations counted in INSPIRE as of 17 May 2026
		
%\cite{Guo:2006fu}
\bibitem{Guo:2006fu}
F.~K.~Guo, P.~N.~Shen, H.~C.~Chiang, R.~G.~Ping and B.~S.~Zou,
Dynamically generated $0^+$ heavy mesons in a heavy chiral unitary approach,
\href{https://www.sciencedirect.com/science/article/pii/S0370269306011221?via%3Dihub}{Phys. Lett. B \textbf{641}, 278-285 (2006)}.
%doi:10.1016/j.physletb.2006.08.064
%[arXiv:hep-ph/0603072 [hep-ph]].
%338 citations counted in INSPIRE as of 10 Mar 2026

%\cite{Guo:2006rp}
\bibitem{Guo:2006rp}
F.~K.~Guo, P.~N.~Shen and H.~C.~Chiang,
Dynamically generated $1^+$ heavy mesons,
\href{https://www.sciencedirect.com/science/article/pii/S0370269307001724?via%3Dihub}{Phys. Lett. B \textbf{647}, 133-139 (2007)}.
%doi:10.1016/j.physletb.2007.01.050
%[arXiv:hep-ph/0610008 [hep-ph]].
%199 citations counted in INSPIRE as of 10 Mar 2026

%\cite{Gamermann:2006nm}
\bibitem{Gamermann:2006nm}
D.~Gamermann, E.~Oset, D.~Strottman and M.~J.~Vicente Vacas,
Dynamically generated open and hidden charm meson systems,
\href{https://journals.aps.org/prd/abstract/10.1103/PhysRevD.76.074016}{Phys. Rev. D \textbf{76}, 074016 (2007)}.
%doi:10.1103/PhysRevD.76.074016
%[arXiv:hep-ph/0612179 [hep-ph]].
%387 citations counted in INSPIRE as of 10 Mar 2026

%\cite{Shen:2007zza}
\bibitem{Shen:2007zza}
P.~N.~Shen, B.~S.~Zou, F.~K.~Guo and H.~C.~Chiang,
$0^+$ and $1^+$ heavy mesons in heavy chiral unitary approach,
\href{https://www.sciencedirect.com/science/article/pii/S0375947407003417?via%3Dihub}{Nucl. Phys. A \textbf{790}, 477-480 (2007)}.
%doi:10.1016/j.nuclphysa.2007.03.081
%0 citations counted in INSPIRE as of 17 May 2026

%\cite{Flynn:2007ki}
\bibitem{Flynn:2007ki}
J.~M.~Flynn and J.~Nieves,
Elastic s-wave $B \pi$, $D \pi$, $D K$ and $K \pi$ scattering from lattice calculations of scalar form-factors in semileptonic decays,
\href{https://journals.aps.org/prd/abstract/10.1103/PhysRevD.75.074024}{Phys. Rev. D \textbf{75}, 074024 (2007)}.
%doi:10.1103/PhysRevD.75.074024
%[arXiv:hep-ph/0703047 [hep-ph]].
%68 citations counted in INSPIRE as of 17 May 2026

%\cite{Gamermann:2007fi}
\bibitem{Gamermann:2007fi}
D.~Gamermann and E.~Oset,
Axial resonances in the open and hidden charm sectors,
\href{https://link.springer.com/article/10.1140/epja/i2007-10435-1}{Eur. Phys. J. A \textbf{33}, 119-131 (2007)}.
%doi:10.1140/epja/i2007-10435-1
%[arXiv:0704.2314 [hep-ph]].
%200 citations counted in INSPIRE as of 17 May 2026

%\cite{Faessler:2007us}
\bibitem{Faessler:2007us}
A.~Faessler, T.~Gutsche, V.~E.~Lyubovitskij and Y.~L.~Ma,
$D^* K$ molecular structure of the $D_{s1}(2460)$ meson,
\href{https://journals.aps.org/prd/abstract/10.1103/PhysRevD.76.114008}{Phys. Rev. D \textbf{76}, 114008 (2007)}.
%doi:10.1103/PhysRevD.76.114008
%[arXiv:0709.3946 [hep-ph]].
%133 citations counted in INSPIRE as of 17 May 2026

%\cite{Xie:2010zza}
\bibitem{Xie:2010zza}
Z.~X.~Xie, G.~Q.~Feng and X.~H.~Guo,
Analyzing $D_{s0}^*(2317)^+$ in the $DK$ molecule picture in the Beth-Salpeter approach,
\href{https://journals.aps.org/prd/abstract/10.1103/PhysRevD.81.036014}{Phys. Rev. D \textbf{81}, 036014 (2010)}.
%doi:10.1103/PhysRevD.81.036014
%51 citations counted in INSPIRE as of 10 Mar 2026

%\cite{Wang:2012bu}
\bibitem{Wang:2012bu}
P.~Wang and X.~G.~Wang,
Study on $0^+$ states with open charm in unitarized heavy meson chiral approach,
\href{https://journals.aps.org/prd/abstract/10.1103/PhysRevD.86.014030}{Phys. Rev. D \textbf{86}, 014030 (2012)}.
%doi:10.1103/PhysRevD.86.014030
%[arXiv:1204.5553 [hep-ph]].
%42 citations counted in INSPIRE as of 09 May 2026

%\cite{Feng:2012zze}
\bibitem{Feng:2012zze}
G.~Q.~Feng, X.~H.~Guo and Z.~H.~Zhang,
Studying the $D^* K$ molecular structure of $D_{s1}^+(2460)$ in the Bethe-Salpeter approach,
\href{https://link.springer.com/article/10.1140/epjc/s10052-012-2033-y}{Eur. Phys. J. C \textbf{72}, 2033 (2012)}.
%doi:10.1140/epjc/s10052-012-2033-y
%15 citations counted in INSPIRE as of 10 Mar 2026

%\cite{Feng:2012zzf}
\bibitem{Feng:2012zzf}
G.~Q.~Feng and X.~H.~Guo,
$DK$ molecule in the Bethe-Salpeter equation approach in the heavy quark limit,
\href{https://journals.aps.org/prd/abstract/10.1103/PhysRevD.86.036004}{Phys. Rev. D \textbf{86}, 036004 (2012)}.
%doi:10.1103/PhysRevD.86.036004
%11 citations counted in INSPIRE as of 18 May 2026

%\cite{Liu:2012zya}
\bibitem{Liu:2012zya}
L.~Liu, K.~Orginos, F.~K.~Guo, C.~Hanhart and U.~G.~Meissner,
Interactions of charmed mesons with light pseudoscalar mesons from lattice QCD and implications on the nature of the $D_{s0}^*(2317)$,
\href{https://journals.aps.org/prd/abstract/10.1103/PhysRevD.87.014508}{Phys. Rev. D \textbf{87}, 014508 (2013)}.
%doi:10.1103/PhysRevD.87.014508
%[arXiv:1208.4535 [hep-lat]].
%255 citations counted in INSPIRE as of 10 May 2026

%\cite{Agadjanov:2014ana}
\bibitem{Agadjanov:2014ana}
D.~Agadjanov, F.~K.~Guo, G.~R{\'\i}os and A.~Rusetsky,
Bound states on the lattice with partially twisted boundary conditions,
\href{https://link.springer.com/article/10.1007/JHEP01(2015)118}{JHEP \textbf{01}, 118 (2015)}.
%doi:10.1007/JHEP01(2015)118
%[arXiv:1411.1859 [hep-lat]].
%17 citations counted in INSPIRE as of 18 May 2026

% Coupled channel
%\cite{MartinezTorres:2014kpc}
\bibitem{MartinezTorres:2014kpc}
A.~Mart{\'\i}nez Torres, E.~Oset, S.~Prelovsek and A.~Ramos,
Reanalysis of lattice QCD spectra leading to the $D_{s0}^*(2317)$ and $D_{s1}^*(2460)$,
\href{https://link.springer.com/article/10.1007/JHEP05(2015)153}{JHEP \textbf{05}, 153 (2015)}.
%doi:10.1007/JHEP05(2015)153
%[arXiv:1412.1706 [hep-lat]].
%131 citations counted in INSPIRE as of 18 May 2026

%\cite{Navarra:2015iea}
\bibitem{Navarra:2015iea}
F.~S.~Navarra, M.~Nielsen, E.~Oset and T.~Sekihara,
Testing the molecular nature of $D_{s0}^*(2317)$ and $D_0^*(2400)$ in semileptonic $B_s$ and $B$ decays,
\href{https://journals.aps.org/prd/abstract/10.1103/PhysRevD.92.014031}{Phys. Rev. D \textbf{92}, 014031 (2015)}.
%doi:10.1103/PhysRevD.92.014031
%[arXiv:1501.03422 [hep-ph]].
%48 citations counted in INSPIRE as of 09 May 2026

%\cite{Guo:2015dha}
\bibitem{Guo:2015dha}
Z.~H.~Guo, U.~G.~Mei{\ss}ner and D.~L.~Yao,
New insights into the $D^{*}_{s0}(2317)$ and other charm scalar mesons,
\href{https://journals.aps.org/prd/abstract/10.1103/PhysRevD.92.094008}{Phys. Rev. D \textbf{92}, 094008 (2015)}.
%doi:10.1103/PhysRevD.92.094008
%[arXiv:1507.03123 [hep-ph]].
%87 citations counted in INSPIRE as of 18 May 2026

%\cite{Ortega:2016mms}
\bibitem{Ortega:2016mms}
P.~G.~Ortega, J.~Segovia, D.~R.~Entem and F.~Fernandez,
Molecular components in $P$-wave charmed-strange mesons,
\href{https://journals.aps.org/prd/abstract/10.1103/PhysRevD.94.074037}{Phys. Rev. D \textbf{94}, 074037 (2016)}.
%doi:10.1103/PhysRevD.94.074037
%[arXiv:1603.07000 [hep-ph]].
%92 citations counted in INSPIRE as of 18 May 2026

%\cite{Albaladejo:2016hae}
\bibitem{Albaladejo:2016hae}
M.~Albaladejo, D.~Jido, J.~Nieves and E.~Oset,
$D^*_{s0}(2317)$ and $DK$ scattering in $B$ decays from BaBar and LHCb data,
\href{https://link.springer.com/article/10.1140/epjc/s10052-016-4144-3}{Eur. Phys. J. C \textbf{76}, 300 (2016)}.
%doi:10.1140/epjc/s10052-016-4144-3
%[arXiv:1604.01193 [hep-ph]].
%56 citations counted in INSPIRE as of 18 May 2026

%\cite{Albaladejo:2016jhb}
\bibitem{Albaladejo:2016jhb}
M.~Albaladejo, M.~Nielsen and E.~Oset,
Exploring the decay $\bar{B}^{0}_{s} \to D^{-}(DK)^+$ to study $DK$ scattering and the $D^{*}_{s0}$(2317) state,
\href{https://pubs.aip.org/aip/acp/article/1735/1/050016/1018350/Exploring-the-decay-B-s0-Ds-DK-to-study-DK}{AIP Conf. Proc. \textbf{1735}, 050016 (2016)}.
%doi:10.1063/1.4949435
%0 citations counted in INSPIRE as of 18 May 2026

%\cite{Du:2017ttu}
\bibitem{Du:2017ttu}
M.~L.~Du, F.~K.~Guo, U.~G.~Mei{\ss}ner and D.~L.~Yao,
Study of open-charm $0^+$ states in unitarized chiral effective theory with one-loop potentials,
\href{https://link.springer.com/article/10.1140/epjc/s10052-017-5287-6}{Eur. Phys. J. C \textbf{77}, 728 (2017)}.
%doi:10.1140/epjc/s10052-017-5287-6
%[arXiv:1703.10836 [hep-ph]].
%55 citations counted in INSPIRE as of 19 May 2026

%\cite{Bali:2017pdv}
\bibitem{Bali:2017pdv}
G.~S.~Bali, S.~Collins, A.~Cox and A.~Sch{\"a}fer,
Masses and decay constants of the $D_{s0}^*(2317)$ and $D_{s1}(2460)$ from $N_f=2$ lattice QCD close to the physical point,
\href{https://journals.aps.org/prd/abstract/10.1103/PhysRevD.96.074501}{Phys. Rev. D \textbf{96}, 074501 (2017)}.
%doi:10.1103/PhysRevD.96.074501
%[arXiv:1706.01247 [hep-lat]].
%138 citations counted in INSPIRE as of 19 May 2026

%\cite{Albaladejo:2018mhb}
\bibitem{Albaladejo:2018mhb}
M.~Albaladejo, P.~Fernandez-Soler, J.~Nieves and P.~G.~Ortega,
Contribution of constituent quark model $c\bar{s}$ states to the dynamics of the $D_{s0}^*(2317)$ and $D_{s1}(2460)$ resonances,
\href{https://link.springer.com/article/10.1140/epjc/s10052-018-6176-3}{Eur. Phys. J. C \textbf{78}, 722 (2018)}.
%doi:10.1140/epjc/s10052-018-6176-3
%[arXiv:1805.07104 [hep-ph]].
%72 citations counted in INSPIRE as of 19 May 2026

% Coupled channel
%\cite{Guo:2018tjx}
\bibitem{Guo:2018tjx}
Z.~H.~Guo, L.~Liu, U.~G.~Mei{\ss}ner, J.~A.~Oller and A.~Rusetsky,
Towards a precise determination of the scattering amplitudes of the charmed and light-flavor pseudoscalar mesons,
\href{https://link.springer.com/article/10.1140/epjc/s10052-018-6518-1}{Eur. Phys. J. C \textbf{79}, 13 (2019)}.
%doi:10.1140/epjc/s10052-018-6518-1
%[arXiv:1811.05585 [hep-ph]].
%80 citations counted in INSPIRE as of 19 May 2026

% Coupled channel
%\cite{Matuschek:2020gqe}
\bibitem{Matuschek:2020gqe}
I.~Matuschek, V.~Baru, F.~K.~Guo and C.~Hanhart,
On the nature of near-threshold bound and virtual states,
\href{https://link.springer.com/article/10.1007/JHEP02(2021)100}{Eur. Phys. J. A \textbf{57}, 101 (2021)}.
%doi:10.1140/epja/s10050-021-00413-y
%[arXiv:2007.05329 [hep-ph]].
%109 citations counted in INSPIRE as of 19 May 2026

%\cite{Cheung:2020mql}
\bibitem{Cheung:2020mql}
G.~K.~C.~Cheung \textit{et al.} [Hadron Spectrum],
$DK$ $I = 0$,$ D\bar{K} $$I = 0, 1$ scattering and the $ {D}_{s0}^{\ast } $(2317) from lattice QCD,
\href{https://link.springer.com/article/10.1007/JHEP02(2021)100}{JHEP \textbf{02}, 100 (2021)}.
%doi:10.1007/JHEP02(2021)100
%[arXiv:2008.06432 [hep-lat]].
%95 citations counted in INSPIRE as of 19 May 2026

% Coupled channel
%\cite{Kong:2021ohg}
\bibitem{Kong:2021ohg}
S.~Y.~Kong, J.~T.~Zhu, D.~Song and J.~He,
Heavy-strange meson molecules and possible candidates $D_{s0}^*(2317)$, $D_{s1}(2460)$, and $X_0(2900)$,
\href{https://journals.aps.org/prd/abstract/10.1103/PhysRevD.104.094012}{Phys. Rev. D \textbf{104}, 094012 (2021)}.
%doi:10.1103/PhysRevD.104.094012
%[arXiv:2106.07272 [hep-ph]].
%58 citations counted in INSPIRE as of 19 May 2026

%\cite{Huang:2021fdt}
\bibitem{Huang:2021fdt}
B.~L.~Huang, Z.~Y.~Lin and S.~L.~Zhu,
Light pseudoscalar meson and heavy meson scattering lengths to $\mathcal{O}(p^4)$ in heavy meson chiral perturbation theory,
\href{https://journals.aps.org/prd/abstract/10.1103/PhysRevD.105.036016}{Phys. Rev. D \textbf{105}, 036016 (2022)}.
%doi:10.1103/PhysRevD.105.036016
%[arXiv:2112.13702 [hep-ph]].
%22 citations counted in INSPIRE as of 19 May 2026

%\cite{Albaladejo:2022sux}
\bibitem{Albaladejo:2022sux}
M.~Albaladejo and J.~Nieves,
Compositeness of $S$-wave weakly-bound states from next-to-leading order Weinberg's relations,
\href{https://link.springer.com/article/10.1140/epjc/s10052-022-10695-1}{Eur. Phys. J. C \textbf{82}, 724 (2022)}.
%doi:10.1140/epjc/s10052-022-10695-1
%[arXiv:2203.04864 [hep-ph]].
%52 citations counted in INSPIRE as of 19 May 2026

%\cite{Guo:2022izw}
\bibitem{Guo:2022izw}
F.~K.~Guo,
Resolving the mysteries of the positive-parity charm mesons,
\href{https://www.sciengine.com/CSB/doi/10.1360/TB-2022-0510}{Chin. Sci. Bull. \textbf{67}, 4344-4355 (2022)}.
%doi:10.1360/TB-2022-0510
%2 citations counted in INSPIRE as of 19 May 2026

%\cite{Chen:2022svh}
\bibitem{Chen:2022svh}
R.~Chen and Q.~Huang,
From the isovector molecular explanation of the newly $T_{c\bar{s}}^{a0(++)}(2900)$ to possible charmed-strange molecular pentaquarks,
\href{https://arxiv.org/abs/2208.10196}{arXiv:2208.10196}.
%26 citations counted in INSPIRE as of 19 May 2026

%\cite{Liu:2023uly}
\bibitem{Liu:2023uly}
Z.~W.~Liu, J.~X.~Lu and L.~S.~Geng,
Study of the $DK$ interaction with femtoscopic correlation functions,
\href{https://journals.aps.org/prd/abstract/10.1103/PhysRevD.107.074019}{Phys. Rev. D \textbf{107}, 074019 (2023)}.
%doi:10.1103/PhysRevD.107.074019
%[arXiv:2302.01046 [hep-ph]].
%62 citations counted in INSPIRE as of 19 May 2026

% Coupled channel
%\cite{Ikeno:2023ojl}
\bibitem{Ikeno:2023ojl}
N.~Ikeno, G.~Toledo and E.~Oset,
Model independent analysis of femtoscopic correlation functions: An application to the $D_{s0}^*(2317)$,
\href{https://www.sciencedirect.com/science/article/pii/S0370269323006159?via%3Dihub}{Phys. Lett. B \textbf{847}, 138281 (2023)}.
%doi:10.1016/j.physletb.2023.138281
%[arXiv:2305.16431 [hep-ph]].
%59 citations counted in INSPIRE as of 19 May 2026

%\cite{Gil-Dominguez:2023puj}
\bibitem{Gil-Dominguez:2023puj}
F.~Gil-Dom{\'\i}nguez and R.~Molina,
Quark mass dependence of the $D_{s0}^*(2317)$ and $D_{s1}(2460)$ resonances,
\href{https://journals.aps.org/prd/abstract/10.1103/PhysRevD.109.096002}{Phys. Rev. D \textbf{109}, 096002 (2024)}.
%doi:10.1103/PhysRevD.109.096002
%[arXiv:2306.01848 [hep-ph]].
%19 citations counted in INSPIRE as of 19 May 2026

% Coupled channel
%\cite{Kim:2023htt}
\bibitem{Kim:2023htt}
H.~J.~Kim and H.~C.~Kim,
$D^*_{s0}(2317)$ and $B^*_{s0}$ as Molecular States,
\href{https://academic.oup.com/ptep/article/2024/7/073D01/7695915?login=false}{PTEP \textbf{2024}, 073D01 (2024)}.
%doi:10.1093/ptep/ptae095
%[arXiv:2310.13370 [hep-ph]].
%7 citations counted in INSPIRE as of 19 May 2026

%\cite{Montesinos:2024uhq}
\bibitem{Montesinos:2024uhq}
V.~Montesinos, M.~Albaladejo, J.~Nieves and L.~Tolos,
Charge-conjugation asymmetry and molecular content: The $D_{s0}^*(2317)^{\pm}$ in matter,
\href{https://www.sciencedirect.com/science/article/pii/S0370269324002144?via%3Dihub}{Phys. Lett. B \textbf{853}, 138656 (2024)}
%doi:10.1016/j.physletb.2024.138656
%[arXiv:2403.00451 [hep-ph]].
%6 citations counted in INSPIRE as of 19 May 2026

%\cite{Yeo:2024chk}
\bibitem{Yeo:2024chk}
J.~D.~E.~Yeo \textit{et al.} [Hadron Spectrum],
$DK/D\pi$ scattering and an exotic virtual bound state at the $SU(3)$ flavour symmetric point from lattice QCD,
\href{https://link.springer.com/article/10.1007/JHEP07(2024)012}{JHEP \textbf{07}, 012 (2024)}.
%doi:10.1007/JHEP07(2024)012
%[arXiv:2403.10498 [hep-lat]].
%27 citations counted in INSPIRE as of 19 May 2026

%\cite{Li:2024rlw}
\bibitem{Li:2024rlw}
H.~P.~Li, W.~H.~Liang, C.~W.~Xiao, J.~J.~Xie and E.~Oset,
Determination of the binding and $DK$ probability of the $D^{*}_{s0}(2317)$ from the $(\bar{D}\bar{K})^-$ mass distributions in $\Lambda _{b}\rightarrow \Lambda _{c} (\bar{D}\bar{K})^-$ decays,
\href{https://link.springer.com/article/10.1140/epjc/s10052-025-14343-2}{Eur. Phys. J. C \textbf{85}, 616 (2025)}.
%doi:10.1140/epjc/s10052-025-14343-2
%[arXiv:2411.17098 [hep-ph]].
%3 citations counted in INSPIRE as of 19 May 2026

%\cite{Xiao:2024kfq}
\bibitem{Xiao:2024kfq}
C.~W.~Xiao and J.~J.~Wu,
Searching for bound states in the open strangeness systems,
\href{https://link.springer.com/article/10.1140/epja/s10050-025-01648-9}{Eur. Phys. J. A \textbf{61}, 179 (2025)}.
%doi:10.1140/epja/s10050-025-01648-9
%[arXiv:2406.08313 [hep-ph]].
%2 citations counted in INSPIRE as of 20 May 2026

% Coupled channel
%\cite{Shen:2025qpj}
\bibitem{Shen:2025qpj}
Y.~b.~Shen, Z.~W.~Liu, J.~X.~Lu, M.~Z.~Liu and L.~S.~Geng,
Probing the structure of exotic hadrons through correlation functions,
\href{https://arxiv.org/abs/2506.23476}{arXiv:2506.23476}.
%9 citations counted in INSPIRE as of 19 May 2026

%\cite{Zhou:2025yjb}
\bibitem{Zhou:2025yjb}
Z.~Zhou, G.~L.~Yu, Z.~G.~Wang, J.~Lu and B.~Wu,
Analysis of the charm-strange hadrons and their bottom analogs with QCD sum rules,
\href{https://link.springer.com/article/10.1140/epja/s10050-026-01805-8}{Eur. Phys. J. A \textbf{62}, 39 (2026)}.
%doi:10.1140/epja/s10050-026-01805-8
%[arXiv:2508.00402 [hep-ph]].
%4 citations counted in INSPIRE as of 19 May 2026
		
%\cite{LHCb:2020bls}
\bibitem{LHCb:2020bls}
R.~Aaij \textit{et al.} [LHCb],
A model-independent study of resonant structure in $B^+\to D^+D^-K^+$ decays,
\href{https://journals.aps.org/prl/abstract/10.1103/PhysRevLett.125.242001}{Phys. Rev. Lett. \textbf{125}, 242001 (2020)}.
%doi:10.1103/PhysRevLett.125.242001
%[arXiv:2009.00025 [hep-ex]].
%262 citations counted in INSPIRE as of 10 Mar 2026
		
%\cite{LHCb:2020pxc}
\bibitem{LHCb:2020pxc}
R.~Aaij \textit{et al.} [LHCb],
Amplitude analysis of the $B^+\to D^+D^-K^+$ decay,
\href{https://journals.aps.org/prd/abstract/10.1103/PhysRevD.102.112003}{Phys. Rev. D \textbf{102}, 112003 (2020)}.
%doi:10.1103/PhysRevD.102.112003
%[arXiv:2009.00026 [hep-ex]].
%314 citations counted in INSPIRE as of 10 Mar 2026

%\cite{Molina:2010tx}
\bibitem{Molina:2010tx}
R.~Molina, T.~Branz and E.~Oset,
A new interpretation for the $D^*_{s2}(2573)$ and the prediction of novel exotic charmed mesons,
\href{https://journals.aps.org/prd/abstract/10.1103/PhysRevD.82.014010}{Phys. Rev. D \textbf{82}, 014010 (2010)}.
%doi:10.1103/PhysRevD.82.014010
%[arXiv:1005.0335 [hep-ph]].
%158 citations counted in INSPIRE as of 23 May 2026
		
%\cite{Molina:2020hde}
\bibitem{Molina:2020hde}
R.~Molina and E.~Oset,
Molecular picture for the $X_0(2866)$ as a $D^* \bar{K}^*$ $J^P=0^+$ state and related $1^+,2^+$ states,
\href{https://www.sciencedirect.com/science/article/pii/S0370269320306730?via%3Dihub}{Phys. Lett. B \textbf{811}, 135870 (2020)}.
%[erratum: Phys. Lett. B \textbf{837}, 137645 (2023)]
%doi:10.1016/j.physletb.2020.135870
%[arXiv:2008.11171 [hep-ph]].
%89 citations counted in INSPIRE as of 10 Mar 2026
		
%\cite{Liu:2020nil}
\bibitem{Liu:2020nil}
M.~Z.~Liu, J.~J.~Xie and L.~S.~Geng,
$X_0(2866)$ as a $D^*\bar{K}^*$ molecular state,
\href{https://journals.aps.org/prd/abstract/10.1103/PhysRevD.102.091502}{Phys. Rev. D \textbf{102}, 091502 (2020)}.
%doi:10.1103/PhysRevD.102.091502
%[arXiv:2008.07389 [hep-ph]].
%107 citations counted in INSPIRE as of 10 Mar 2026
		
%\cite{Chen:2020aos}
\bibitem{Chen:2020aos}
H.~X.~Chen, W.~Chen, R.~R.~Dong and N.~Su,
$X_0$(2900) and $X_1$(2900): Hadronic Molecules or Compact Tetraquarks,
\href{https://iopscience.iop.org/article/10.1088/0256-307X/37/10/101201}{Chin. Phys. Lett. \textbf{37}, 101201 (2020)}.
%doi:10.1088/0256-307X/37/10/101201
%[arXiv:2008.07516 [hep-ph]].
%97 citations counted in INSPIRE as of 10 Mar 2026

%\cite{Hu:2020mxp}
\bibitem{Hu:2020mxp}
M.~W.~Hu, X.~Y.~Lao, P.~Ling and Q.~Wang,
$X_0$(2900) and its heavy quark spin partners in molecular picture,
\href{https://iopscience.iop.org/article/10.1088/1674-1137/abcfaa}{Chin. Phys. C \textbf{45}, 021003 (2021)}.
%doi:10.1088/1674-1137/abcfaa
%[arXiv:2008.06894 [hep-ph]].
%74 citations counted in INSPIRE as of 20 May 2026

%\cite{Agaev:2020nrc}
\bibitem{Agaev:2020nrc}
S.~S.~Agaev, K.~Azizi and H.~Sundu,
New scalar resonance $X_0(2900)$ as a $\bar{D}^*K^*$ molecule: mass and width,
\href{https://iopscience.iop.org/article/10.1088/1361-6471/ac0b31}{J. Phys. G \textbf{48}, 085012 (2021)}.
%doi:10.1088/1361-6471/ac0b31
%[arXiv:2008.13027 [hep-ph]].
%66 citations counted in INSPIRE as of 20 May 2026
		
%\cite{He:2020btl}
\bibitem{He:2020btl}
J.~He and D.~Y.~Chen,
Molecular picture for $X_0(2900)$ and $X_1(2900)$,
\href{https://iopscience.iop.org/article/10.1088/1674-1137/abeda8}{Chin. Phys. C \textbf{45}, 063102 (2021)}.
%doi:10.1088/1674-1137/abeda8
%[arXiv:2008.07782 [hep-ph]].
%63 citations counted in INSPIRE as of 10 Mar 2026
		
%\cite{Wang:2021lwy}
\bibitem{Wang:2021lwy}
B.~Wang and S.~L.~Zhu,
How to understand the $X$(2900)?,
\href{https://link.springer.com/article/10.1140/epjc/s10052-022-10396-9}{Eur. Phys. J. C \textbf{82}, 419 (2022)}.
%doi:10.1140/epjc/s10052-022-10396-9
%[arXiv:2107.09275 [hep-ph]].
%43 citations counted in INSPIRE as of 10 Mar 2026

%\cite{Dai:2022htx}
\bibitem{Dai:2022htx}
L.~R.~Dai, R.~Molina and E.~Oset,
Looking for the exotic $X_0(2866)$ and its $J^P=1^+$ partner in the $\bar{B}^0 \to D^{(*)+}K^-K^{(*)0}$ reactions,
\href{https://journals.aps.org/prd/abstract/10.1103/PhysRevD.105.096022}{Phys. Rev. D \textbf{105}, 096022 (2022)}.
%doi:10.1103/PhysRevD.105.096022
%[arXiv:2202.11973 [hep-ph]].
%19 citations counted in INSPIRE as of 20 May 2026

%\cite{Ke:2022ocs}
\bibitem{Ke:2022ocs}
H.~W.~Ke, Y.~F.~Shi, X.~H.~Liu and X.~Q.~Li,
Possible molecular states of $\bar{D}^*K^*$ ($D^*K^*$) and new exotic states $X_0(2900)$, $X_1(2900)$ ($T_{cs0}^a(2900)^0$ and $T_{cs0}^a(2900)^{++}$),
\href{https://journals.aps.org/prd/abstract/10.1103/PhysRevD.106.114032}{Phys. Rev. D \textbf{106}, 114032 (2022)}.
%doi:10.1103/PhysRevD.106.114032
%[arXiv:2210.06215 [hep-ph]].
%27 citations counted in INSPIRE as of 10 May 2026

%\cite{Chen:2023syh}
\bibitem{Chen:2023syh}
Y.~K.~Chen, W.~L.~Wu, L.~Meng and S.~L.~Zhu,
Unified description of the $Qs\bar{q}\bar{q}$ molecular bound states, molecular resonances, and compact tetraquark states in the quark potential model,
\href{https://journals.aps.org/prd/abstract/10.1103/PhysRevD.109.014010}{Phys. Rev. D \textbf{109}, 014010 (2024)}.
%doi:10.1103/PhysRevD.109.014010
%[arXiv:2310.14597 [hep-ph]].
%22 citations counted in INSPIRE as of 20 May 2026

%\cite{Ding:2024dif}
\bibitem{Ding:2024dif}
Z.~M.~Ding, Q.~Huang and J.~He,
$X_0(2900)$ and $\chi _{c0}(3930)$ in process $B^+\rightarrow D^+ D^- K^+$,
\href{https://link.springer.com/article/10.1140/epjc/s10052-024-13214-6}{Eur. Phys. J. C \textbf{84}, 822 (2024)}.
%doi:10.1140/epjc/s10052-024-13214-6
%[arXiv:2407.13503 [hep-ph]].
%9 citations counted in INSPIRE as of 20 May 2026

%\cite{Ding:2025uhh}
\bibitem{Ding:2025uhh}
Z.~M.~Ding, Q.~Huang and J.~He,
Roles of $\bar{D}^{*}K^{*}$ and $D^{*}\bar{D}$ molecular states in decay $B^+ \rightarrow D^{*+} D^{-} K^{+}$,
\href{https://link.springer.com/article/10.1140/epjc/s10052-025-14882-8}{Eur. Phys. J. C \textbf{85}, 1133 (2025)}.
%doi:10.1140/epjc/s10052-025-14882-8
%[arXiv:2508.12686 [hep-ph]].
%5 citations counted in INSPIRE as of 20 May 2026

%\cite{LHCb:2022lzp}
\bibitem{LHCb:2022lzp}
R.~Aaij \textit{et al.} [LHCb],
Amplitude analysis of $B^0 \to \bar{D}^0D_s^+\pi^-$ and $B^+ \to D^-D_s^+\pi^+$ decays,
\href{https://journals.aps.org/prd/abstract/10.1103/PhysRevD.108.012017}{Phys. Rev. D \textbf{108}, 012017 (2023)}.
%doi:10.1103/PhysRevD.108.012017
%[arXiv:2212.02717 [hep-ex]].
%110 citations counted in INSPIRE as of 09 May 2026
		
%\cite{LHCb:2022sfr}
\bibitem{LHCb:2022sfr}
R.~Aaij \textit{et al.} [LHCb],
First Observation of a Doubly Charged Tetraquark and Its Neutral Partner,
\href{https://journals.aps.org/prl/abstract/10.1103/PhysRevLett.131.041902}{Phys. Rev. Lett. \textbf{131}, 041902 (2023)}.
%doi:10.1103/PhysRevLett.131.041902
%[arXiv:2212.02716 [hep-ex]].
%131 citations counted in INSPIRE as of 10 Mar 2026

%\cite{Agaev:2022eyk}
\bibitem{Agaev:2022eyk}
S.~S.~Agaev, K.~Azizi and H.~Sundu,
Modeling the resonance $T_{cs0}^a(2900)^{++}$ as a hadronic molecule $D^{*+}K^{*+}$,
\href{https://journals.aps.org/prd/abstract/10.1103/PhysRevD.107.094019}{Phys. Rev. D \textbf{107}, 094019 (2023)}.
%doi:10.1103/PhysRevD.107.094019
%[arXiv:2212.12001 [hep-ph]].
%33 citations counted in INSPIRE as of 10 May 2026

%\cite{Duan:2023lcj}
\bibitem{Duan:2023lcj}
M.~Y.~Duan, M.~L.~Du, Z.~H.~Guo, E.~Wang and D.~Y.~Chen,
Coupled-channel $D^\ast K^\ast -D_s^\ast \rho$ interactions and the origin of $T_{c\bar{s}0}(2900)$,
\href{https://journals.aps.org/prd/abstract/10.1103/PhysRevD.108.074006}{Phys. Rev. D \textbf{108}, 074006 (2023)}.
%doi:10.1103/PhysRevD.108.074006
%[arXiv:2307.04092 [hep-ph]].
%23 citations counted in INSPIRE as of 10 May 2026

%\cite{Wang:2023hpp}
\bibitem{Wang:2023hpp}
B.~Wang, K.~Chen, L.~Meng and S.~L.~Zhu,
Spectrum of the molecular tetraquarks: Unraveling the $T_{cs0}(2900)$ and $T_{c\bar{s}0}^a(2900)$,
\href{https://journals.aps.org/prd/abstract/10.1103/PhysRevD.109.034027}{Phys. Rev. D \textbf{109}, 034027 (2024)}.
%doi:10.1103/PhysRevD.109.034027
%[arXiv:2309.02191 [hep-ph]].
%32 citations counted in INSPIRE as of 10 May 2026

%\cite{Molina:2023ghu}
\bibitem{Molina:2023ghu}
R.~Molina and E.~Oset,
The $T_{c\bar{s}} (2900)$ as a threshold effect from the interaction of the $D^*K^*$, $D_s^*\rho$ channels,
\href{https://www.epj-conferences.org/articles/epjconf/abs/2024/01/epjconf_meson2023_03010/epjconf_meson2023_03010.html}{EPJ Web Conf. \textbf{291}, 03010 (2024)}.
%doi:10.1051/epjconf/202429103010
%[arXiv:2310.09794 [hep-ph]].
%3 citations counted in INSPIRE as of 21 May 2026

%\cite{Duan:2023qsg}
\bibitem{Duan:2023qsg}
M.~Y.~Duan, E.~Wang and D.~Y.~Chen,
Searching for the open flavor tetraquark $T_{c\bar{s}0}(2900)^{++}$ in the process $B^+\rightarrow K^+ D^+ D^-$,
\href{https://link.springer.com/article/10.1140/epjc/s10052-024-13044-6}{Eur. Phys. J. C \textbf{84}, 681 (2024)}.
%doi:10.1140/epjc/s10052-024-13044-6
%[arXiv:2305.09436 [hep-ph]].
%27 citations counted in INSPIRE as of 10 May 2026

%\cite{Yeo:2026dgo}
\bibitem{Yeo:2026dgo}
J.~D.~E.~Yeo \textit{et al.} [Hadron Spectrum],
Exotic $T^*_{csJ}$ and $T^*_{c\bar{s}J}$ states and coupled-channel scattering at the $SU(3)$ flavour symmetric point from lattice QCD,
\href{https://arxiv.org/abs/2604.19553}{arXiv:2604.19553}.
%1 citations counted in INSPIRE as of 21 May 2026

%\cite{LHCb:2019kea}
\bibitem{LHCb:2019kea}
R.~Aaij \textit{et al.} [LHCb],
Observation of a narrow pentaquark state, $P_c(4312)^+$, and of two-peak structure of the $P_c(4450)^+$,
\href{https://journals.aps.org/prl/abstract/10.1103/PhysRevLett.122.222001}{Phys. Rev. Lett. \textbf{122}, 222001 (2019)}.
%doi:10.1103/PhysRevLett.122.222001
%[arXiv:1904.03947 [hep-ex]].
%1043 citations counted in INSPIRE as of 21 May 2026

%\cite{LHCb:2020jpq}
\bibitem{LHCb:2020jpq}
R.~Aaij \textit{et al.} [LHCb],
Evidence of a $J/\psi\Lambda$ structure and observation of excited $\Xi^-$ states in the $\Xi^-_b \to J/\psi\Lambda K^-$ decay,
\href{https://www.sciencedirect.com/science/article/pii/S2095927321001717?via%3Dihub}{Sci. Bull. \textbf{66}, 1278-1287 (2021)}.
%doi:10.1016/j.scib.2021.02.030
%[arXiv:2012.10380 [hep-ex]].
%355 citations counted in INSPIRE as of 21 May 2026

%\cite{Chen:2019asm}
\bibitem{Chen:2019asm}
R.~Chen, Z.~F.~Sun, X.~Liu and S.~L.~Zhu,
Strong LHCb evidence supporting the existence of the hidden-charm molecular pentaquarks,
\href{https://journals.aps.org/prd/abstract/10.1103/PhysRevD.100.011502}{Phys. Rev. D \textbf{100}, 011502 (2019)}.
%doi:10.1103/PhysRevD.100.011502
%[arXiv:1903.11013 [hep-ph]].
%248 citations counted in INSPIRE as of 10 May 2026

%\cite{Chen:2020kco}
\bibitem{Chen:2020kco}
R.~Chen,
Can the newly reported $P_{cs}(4459)$ be a strange hidden-charm $\Xi_c\bar D^*$ molecular pentaquark?,
\href{https://journals.aps.org/prd/abstract/10.1103/PhysRevD.103.054007}{Phys. Rev. D \textbf{103}, 054007 (2021)}.
%doi:10.1103/PhysRevD.103.054007
%[arXiv:2011.07214 [hep-ph]].
%82 citations counted in INSPIRE as of 10 May 2026

%\cite{Wang:2022mxy}
\bibitem{Wang:2022mxy}
F.~L.~Wang and X.~Liu,
Emergence of molecular-type characteristic spectrum of hidden-charm pentaquark with strangeness embodied in the $P_{\psi s}^{\Lambda}(4338)$ and $P_{cs}(4459)$,
\href{https://www.sciencedirect.com/science/article/pii/S0370269322007171?via%3Dihub}{Phys. Lett. B \textbf{835}, 137583 (2022)}.
%doi:10.1016/j.physletb.2022.137583
%[arXiv:2207.10493 [hep-ph]].
%56 citations counted in INSPIRE as of 24 May 2026

%\cite{Schlumpf:1992vq}
\bibitem{Schlumpf:1992vq}
F.~Schlumpf,
Relativistic constituent quark model of electroweak properties of baryons,
\href{https://journals.aps.org/prd/abstract/10.1103/PhysRevD.47.4114}{Phys. Rev. D \textbf{47}, 4114 (1993)}.
%[erratum: Phys. Rev. D \textbf{49}, 6246 (1994)]
%doi:10.1103/PhysRevD.47.4114
%[arXiv:hep-ph/9212250 [hep-ph]].
%137 citations counted in INSPIRE as of 18 Mar 2026

%\cite{Schlumpf:1993rm}
\bibitem{Schlumpf:1993rm}
F.~Schlumpf,
Magnetic moments of the baryon decuplet in a relativistic quark model,
\href{https://journals.aps.org/prd/abstract/10.1103/PhysRevD.48.4478}{Phys. Rev. D \textbf{48}, 4478-4480 (1993)}.
%doi:10.1103/PhysRevD.48.4478
%[arXiv:hep-ph/9305293 [hep-ph]].
%107 citations counted in INSPIRE as of 18 Mar 2026
		
%\cite{Cheng:1997kr}
\bibitem{Cheng:1997kr}
T.~P.~Cheng and L.~F.~Li,
Why naive quark model can yield a good account of the baryon magnetic moments,
\href{https://journals.aps.org/prl/abstract/10.1103/PhysRevLett.80.2789}{Phys. Rev. Lett. \textbf{80}, 2789-2792 (1998)}.
%doi:10.1103/PhysRevLett.80.2789
%[arXiv:hep-ph/9709295 [hep-ph]].
%93 citations counted in INSPIRE as of 18 Mar 2026
		
%\cite{Ha:1998gf}
\bibitem{Ha:1998gf}
P.~Ha and L.~Durand,
Baryon magnetic moments in a QCD based quark model with loop corrections,
\href{https://journals.aps.org/prd/abstract/10.1103/PhysRevD.58.093008}{Phys. Rev. D \textbf{58}, 093008 (1998)}.
%doi:10.1103/PhysRevD.58.093008
%[arXiv:hep-ph/9804382 [hep-ph]].
%30 citations counted in INSPIRE as of 18 Mar 2026
		
%\cite{Liu:2003ab}
\bibitem{Liu:2003ab}
Y.~R.~Liu, P.~Z.~Huang, W.~Z.~Deng, X.~L.~Chen and S.~L.~Zhu,
Pentaquark magnetic moments in different models,
\href{https://journals.aps.org/prc/abstract/10.1103/PhysRevC.69.035205}{Phys. Rev. C \textbf{69}, 035205 (2004)}.
%doi:10.1103/PhysRevC.69.035205
%[arXiv:hep-ph/0312074 [hep-ph]].
%65 citations counted in INSPIRE as of 18 Mar 2026
		
%\cite{Huang:2004tn}
\bibitem{Huang:2004tn}
P.~Z.~Huang, Y.~R.~Liu, W.~Z.~Deng, X.~L.~Chen and S.~L.~Zhu,
Heavy pentaquarks,
\href{https://journals.aps.org/prd/abstract/10.1103/PhysRevD.70.034003}{Phys. Rev. D \textbf{70}, 034003 (2004)}.
%doi:10.1103/PhysRevD.70.034003
%[arXiv:hep-ph/0401191 [hep-ph]].
%37 citations counted in INSPIRE as of 18 Mar 2026
		
%\cite{Zhu:2004xa}
\bibitem{Zhu:2004xa}
S.~L.~Zhu,
Pentaquarks,
\href{https://www.worldscientific.com/doi/abs/10.1142/S0217751X04019676}{Int. J. Mod. Phys. A \textbf{19}, 3439-3469 (2004)}.
%doi:10.1142/S0217751X04019676
%[arXiv:hep-ph/0406204 [hep-ph]].
%88 citations counted in INSPIRE as of 18 Mar 2026

%\cite{Wang:2016dzu}
\bibitem{Wang:2016dzu}
G.~J.~Wang, R.~Chen, L.~Ma, X.~Liu and S.~L.~Zhu,
Magnetic moments of the hidden-charm pentaquark states,
\href{https://journals.aps.org/prd/abstract/10.1103/PhysRevD.94.094018}{Phys. Rev. D \textbf{94}, 094018 (2016)}.
%doi:10.1103/PhysRevD.94.094018
%[arXiv:1605.01337 [hep-ph]].
%75 citations counted in INSPIRE as of 18 Mar 2026

%\cite{Ozdem:2018qeh}
\bibitem{Ozdem:2018qeh}
U.~{\"O}zdem and K.~Azizi,
Electromagnetic multipole moments of the $P_c^+(4380)$ pentaquark in light-cone QCD,
\href{https://link.springer.com/article/10.1140/epjc/s10052-018-5873-2}{Eur. Phys. J. C \textbf{78}, 379 (2018)}.
%doi:10.1140/epjc/s10052-018-5873-2
%[arXiv:1803.06831 [hep-ph]].
%33 citations counted in INSPIRE as of 12 May 2026

%\cite{Xu:2020flp}
\bibitem{Xu:2020flp}
Y.~J.~Xu, Y.~L.~Liu and M.~Q.~Huang,
The magnetic moment of $P_{c}(4312)$ as a $\bar{D}\Sigma _{c}$ molecular state,
\href{https://link.springer.com/article/10.1140/epjc/s10052-021-09211-8}{Eur. Phys. J. C \textbf{81}, 421 (2021)}.
%doi:10.1140/epjc/s10052-021-09211-8
%[arXiv:2008.07937 [hep-ph]].
%36 citations counted in INSPIRE as of 12 May 2026

%\cite{Ozdem:2021btf}
\bibitem{Ozdem:2021btf}
U.~{\"O}zdem,
Electromagnetic properties of the $P_c$ (4312) pentaquark state,
\href{https://iopscience.iop.org/article/10.1088/1674-1137/abd01c}{Chin. Phys. C \textbf{45}, 023119 (2021)}.
%doi:10.1088/1674-1137/abd01c
%22 citations counted in INSPIRE as of 12 May 2026

%\cite{Ozdem:2021ugy}
\bibitem{Ozdem:2021ugy}
U.~{\"O}zdem,
Magnetic dipole moments of the hidden-charm pentaquark states: $P_c(4440)$, $P_c(4457)$ and $P_{cs}(4459)$,
\href{https://link.springer.com/article/10.1140/epjc/s10052-021-09070-3}{Eur. Phys. J. C \textbf{81}, 277 (2021)}.
%doi:10.1140/epjc/s10052-021-09070-3
%[arXiv:2102.01996 [hep-ph]].
%58 citations counted in INSPIRE as of 12 May 2026

%\cite{Ozdem:2021yvo}
\bibitem{Ozdem:2021yvo}
U.~{\"O}zdem and K.~Azizi,
Magnetic dipole moment of the $Z_{cs}(3985)$ state: diquark{\textendash}antidiquark and molecular pictures,
\href{https://link.springer.com/article/10.1140/epjp/s13360-021-01977-w}{Eur. Phys. J. Plus \textbf{136}, 968 (2021)}.
%doi:10.1140/epjp/s13360-021-01977-w
%[arXiv:2102.09231 [hep-ph]].
%37 citations counted in INSPIRE as of 12 May 2026

%\cite{Ozdem:2021hka}
\bibitem{Ozdem:2021hka}
U.~{\"O}zdem and A.~K.~Y{\i}ld{\i}r{\i}m,
Magnetic dipole moments of the $Z_c(4020)^+$, $Z_c(4200)^+$, $Z_{cs}(4000)^+$ and $Z_{cs}(4220)^+$ states in light-cone QCD,
\href{https://journals.aps.org/prd/abstract/10.1103/PhysRevD.104.054017}{Phys. Rev. D \textbf{104}, 054017 (2021)}.
%doi:10.1103/PhysRevD.104.054017
%[arXiv:2104.13074 [hep-ph]].
%22 citations counted in INSPIRE as of 12 May 2026

%\cite{Li:2021ryu}
\bibitem{Li:2021ryu}
M.~W.~Li, Z.~W.~Liu, Z.~F.~Sun and R.~Chen,
Magnetic moments and transition magnetic moments of $P_c$ and $P_{cs}$ states,
\href{https://journals.aps.org/prd/abstract/10.1103/PhysRevD.104.054016}{Phys. Rev. D \textbf{104}, 054016 (2021)}.
%doi:10.1103/PhysRevD.104.054016
%[arXiv:2106.15053 [hep-ph]].
%48 citations counted in INSPIRE as of 18 Mar 2026

%\cite{Deng:2021gnb}%69
\bibitem{Deng:2021gnb}
C.~Deng and S.~L.~Zhu,
$T_{cc}^+$ and its partners,
\href{https://journals.aps.org/prd/abstract/10.1103/PhysRevD.105.054015}{Phys. Rev. D \textbf{105}, 054015 (2022)}.
%doi:10.1103/PhysRevD.105.054015
%[arXiv:2112.12472 [hep-ph]].
%93 citations counted in INSPIRE as of 18 Mar 2026

%\cite{Wang:2022tib}
\bibitem{Wang:2022tib}
F.~L.~Wang, H.~Y.~Zhou, Z.~W.~Liu and X.~Liu,
What can we learn from the electromagnetic properties of hidden-charm molecular pentaquarks with single strangeness?,
\href{https://journals.aps.org/prd/abstract/10.1103/PhysRevD.106.054020}{Phys. Rev. D \textbf{106}, 054020 (2022)}.
%doi:10.1103/PhysRevD.106.054020
%[arXiv:2208.10756 [hep-ph]].
%41 citations counted in INSPIRE as of 18 Mar 2026

%\cite{Ozdem:2021vry}
\bibitem{Ozdem:2021vry}
U.~{\"O}zdem,
Magnetic moment of the $\Xi _b(6227)$ as a molecular pentaquark state,
\href{https://link.springer.com/article/10.1140/epjp/s13360-022-02339-w}{Eur. Phys. J. Plus \textbf{137}, 103 (2022)}.
%doi:10.1140/epjp/s13360-022-02339-w
%[arXiv:2109.09313 [hep-ph]].
%10 citations counted in INSPIRE as of 12 May 2026

%\cite{Gao:2021hmv}
\bibitem{Gao:2021hmv}
F.~Gao and H.~S.~Li,
Magnetic moments of hidden-charm strange pentaquark states,
\href{https://iopscience.iop.org/article/10.1088/1674-1137/ac8651}{Chin. Phys. C \textbf{46}, 123111 (2022)}.
%doi:10.1088/1674-1137/ac8651
%[arXiv:2112.01823 [hep-ph]].
%48 citations counted in INSPIRE as of 08 May 2026
		
%\cite{Zhou:2022gra}
\bibitem{Zhou:2022gra}
H.~Y.~Zhou, F.~L.~Wang, Z.~W.~Liu and X.~Liu,
Probing the electromagnetic properties of the $\Sigma_c^{(*)}D^{(*)}$-type doubly charmed molecular pentaquarks,
\href{https://journals.aps.org/prd/abstract/10.1103/PhysRevD.106.034034}{Phys. Rev. D \textbf{106}, 034034 (2022)}.
%doi:10.1103/PhysRevD.106.034034
%[arXiv:2207.08660 [hep-ph]].
%21 citations counted in INSPIRE as of 18 Mar 2026

%\cite{Ozdem:2022ylm}
\bibitem{Ozdem:2022ylm}
U.~{\"O}zdem,
Electromagnetic properties of the $ D \bar{D}^* K$ molecular hexaquark state,
\href{https://link.springer.com/article/10.1140/epjp/s13360-022-03124-5}{Eur. Phys. J. Plus \textbf{137}, 908 (2022)}.
%doi:10.1140/epjp/s13360-022-03124-5
%[arXiv:2207.14549 [hep-ph]].
%3 citations counted in INSPIRE as of 12 May 2026

%\cite{Ozdem:2022yhi}
\bibitem{Ozdem:2022yhi}
U.~{\"O}zdem,
Magnetic dipole moments of states,
\href{https://iopscience.iop.org/article/10.1088/1674-1137/ac8653}{Chin. Phys. C \textbf{46}, 113106 (2022)}.
%doi:10.1088/1674-1137/ac8653
%[arXiv:2203.07759 [hep-ph]].
%10 citations counted in INSPIRE as of 12 May 2026

%\cite{Wu:2022gie}
\bibitem{Wu:2022gie}
T.~W.~Wu and Y.~L.~Ma,
Doubly heavy tetraquark multiplets as heavy antiquark-diquark symmetry partners of heavy baryons,
\href{https://journals.aps.org/prd/abstract/10.1103/PhysRevD.107.L071501}{Phys. Rev. D \textbf{107}, L071501 (2023)}.
%doi:10.1103/PhysRevD.107.L071501
%[arXiv:2211.15094 [hep-ph]].
%32 citations counted in INSPIRE as of 18 Mar 2026

%\cite{An:2022qpt}
\bibitem{An:2022qpt}
H.~T.~An, S.~Q.~Luo, Z.~W.~Liu and X.~Liu,
Spectroscopic behavior of fully heavy tetraquarks,
\href{https://link.springer.com/article/10.1140/epjc/s10052-023-11847-7}{Eur. Phys. J. C \textbf{83}, 740 (2023)}.
%doi:10.1140/epjc/s10052-023-11847-7
%[arXiv:2208.03899 [hep-ph]].
%45 citations counted in INSPIRE as of 18 Mar 2026

%\cite{Ozdem:2023eyz}
\bibitem{Ozdem:2023eyz}
U.~Ozdem,
Electromagnetic properties of the $\Sigma _{c}(2800)^+$ and $\Lambda _c(2940)^+$ states via light-cone QCD,
\href{https://link.springer.com/article/10.1140/epjc/s10052-023-12251-x}{Eur. Phys. J. C \textbf{83}, 1077 (2023)}.
%doi:10.1140/epjc/s10052-023-12251-x
%[arXiv:2309.00959 [hep-ph]].
%12 citations counted in INSPIRE as of 12 May 2026

%\cite{Yue:2023qgx}
\bibitem{Yue:2023qgx}
Z.~L.~Yue, C.~J.~Xiao and D.~Y.~Chen,
Pionic and radiative transitions from $T_{c\bar{s}0}^+(2900)$ to $D_{s1}^+(2460)$ as a probe of the structure of $D_{s1}^+(2460)$,
\href{https://link.springer.com/article/10.1140/epjc/s10052-023-11948-3}{Eur. Phys. J. C \textbf{83}, 769 (2023)}.
%doi:10.1140/epjc/s10052-023-11948-3
%[arXiv:2308.15355 [hep-ph]].
%5 citations counted in INSPIRE as of 23 May 2026

%\cite{Li:2023wgq}
\bibitem{Li:2023wgq}
X.~J.~Li, Y.~S.~Li, F.~L.~Wang and X.~Liu,
Spectroscopic survey of higher-lying states of $B_c$ meson family,
\href{https://link.springer.com/article/10.1140/epjc/s10052-023-12237-9}{Eur. Phys. J. C \textbf{83}, 1080 (2023)}.
%doi:10.1140/epjc/s10052-023-12237-9
%[arXiv:2308.07206 [hep-ph]].
%30 citations counted in INSPIRE as of 23 May 2026

%\cite{Wang:2023aob}
\bibitem{Wang:2023aob}
F.~L.~Wang and X.~Liu,
New type of doubly charmed molecular pentaquarks containing most strange quarks: Mass spectra, radiative decays, and magnetic moments,
\href{https://journals.aps.org/prd/abstract/10.1103/PhysRevD.108.074022}{Phys. Rev. D \textbf{108}, 074022 (2023)}.
%doi:10.1103/PhysRevD.108.074022
%[arXiv:2308.15255 [hep-ph]].
%15 citations counted in INSPIRE as of 18 Mar 2026

%\cite{Wang:2022nqs}
\bibitem{Wang:2022nqs}
F.~L.~Wang, S.~Q.~Luo, H.~Y.~Zhou, Z.~W.~Liu and X.~Liu,
Exploring the electromagnetic properties of the $\Xi_c^{(',*)}\bar{D}_s^*$ and $\Omega_c^{(*)}\bar{D}_s^*$ molecular states,
\href{https://journals.aps.org/prd/abstract/10.1103/PhysRevD.108.034006}{Phys. Rev. D \textbf{108}, 034006 (2023)}.
%doi:10.1103/PhysRevD.108.034006
%[arXiv:2210.02809 [hep-ph]].
%30 citations counted in INSPIRE as of 18 Mar 2026 127

%\cite{Wang:2023bek}
\bibitem{Wang:2023bek}
F.~L.~Wang, S.~Q.~Luo and X.~Liu,
Radiative decays and magnetic moments of the predicted Bc-like molecules,
\href{https://journals.aps.org/prd/abstract/10.1103/PhysRevD.107.114017}{Phys. Rev. D \textbf{107}, 114017 (2023)}.
%doi:10.1103/PhysRevD.107.114017
%[arXiv:2303.04542 [hep-ph]].
%21 citations counted in INSPIRE as of 18 Mar 2026

%\cite{Guo:2023fih}
\bibitem{Guo:2023fih}
F.~Guo and H.~S.~Li,
Analysis of the hidden-charm pentaquark states based on magnetic moment and transition magnetic moment,
\href{https://link.springer.com/article/10.1140/epjc/s10052-024-12699-5}{Eur. Phys. J. C \textbf{84}, 392 (2024)}.
%doi:10.1140/epjc/s10052-024-12699-5
%[arXiv:2304.10981 [hep-ph]].
%29 citations counted in INSPIRE as of 12 May 2026

%\cite{Lei:2023ttd}
\bibitem{Lei:2023ttd}
Y.~D.~Lei and H.~S.~Li,
Electromagnetic properties of the $T_{cc}^+$ molecular states,
\href{https://journals.aps.org/prd/abstract/10.1103/PhysRevD.109.076014}{Phys. Rev. D \textbf{109}, 076014 (2024)}.
%doi:10.1103/PhysRevD.109.076014
%[arXiv:2312.01332 [hep-ph]].
%10 citations counted in INSPIRE as of 12 May 2026

%\cite{Li:2024wxr}
\bibitem{Li:2024wxr}
H.~S.~Li, F.~Guo, Y.~D.~Lei and F.~Gao,
Magnetic moments and axial charges of the octet hidden-charm molecular pentaquark family
\href{https://journals.aps.org/prd/abstract/10.1103/PhysRevD.109.094027}{Phys. Rev. D \textbf{109}, 094027 (2024)}.
%doi:10.1103/PhysRevD.109.094027
%[arXiv:2401.14767 [hep-ph]].
%32 citations counted in INSPIRE as of 12 May 2026

%\cite{Lei:2024geu}
\bibitem{Lei:2024geu}
Y.~D.~Lei and H.~S.~Li,
Radiative decay and axial-vector decay behaviors of octet pentaquark states,
\href{https://journals.aps.org/prd/abstract/10.1103/PhysRevD.110.056026}{Phys. Rev. D \textbf{110}, 056026 (2024)}.
%doi:10.1103/PhysRevD.110.056026
%[arXiv:2406.18857 [hep-ph]].
%3 citations counted in INSPIRE as of 12 May 2026

%\cite{Wang:2023ael}
\bibitem{Wang:2023ael}
F.~L.~Wang and X.~Liu,
Surveying the mass spectra and the electromagnetic properties of the $\Xi_c^{(',*)}D^{(*)}$ molecular pentaquarks,
\href{https://journals.aps.org/prd/abstract/10.1103/PhysRevD.109.014043}{Phys. Rev. D \textbf{109}, 014043 (2024)}.
%doi:10.1103/PhysRevD.109.014043
%[arXiv:2311.13968 [hep-ph]].
%21 citations counted in INSPIRE as of 18 Mar 2026

%\cite{Zhang:2024usz}
\bibitem{Zhang:2024usz}
Z.~L.~Zhang, Z.~W.~Liu, S.~Q.~Luo, P.~Chen and Z.~H.~Guo,
Masses and radiative decay widths of $D_{s0}^*(2317)$ and $D_{s1}^{\prime}(2460)$ and their bottom analogs,
\href{https://journals.aps.org/prd/abstract/10.1103/PhysRevD.110.094037}{Phys. Rev. D \textbf{110}, 094037 (2024)}.
%doi:10.1103/PhysRevD.110.094037
%[arXiv:2409.05337 [hep-ph]].
%12 citations counted in INSPIRE as of 23 May 2026

%\cite{Ozdem:2024ydl}
\bibitem{Ozdem:2024ydl}
U.~{\"O}zdem,
Analysis of the $\Xi _c^* {\bar{K}}$ molecular pentaquark state by its electromagnetic properties,
\href{https://link.springer.com/article/10.1140/epjc/s10052-024-13134-5}{Eur. Phys. J. C \textbf{84}, 765 (2024)}.
%doi:10.1140/epjc/s10052-024-13134-5
%[arXiv:2407.08635 [hep-ph]].
%6 citations counted in INSPIRE as of 12 May 2026

%\cite{Ozdem:2024jty}
\bibitem{Ozdem:2024jty}
U.~{\"O}zdem,
Analysis of the isospin eigenstate $\bar{D} \Sigma _c$, $\bar{D}^{*} \Sigma _c$, and $\bar{D} \Sigma _c^{*}$ pentaquarks by their electromagnetic properties,
\href{https://link.springer.com/article/10.1140/epjc/s10052-024-13124-7}{Eur. Phys. J. C \textbf{84}, 769 (2024)}.
%doi:10.1140/epjc/s10052-024-13124-7
%[arXiv:2401.12678 [hep-ph]].
%18 citations counted in INSPIRE as of 12 May 2026

%\cite{Li:2024jlq}
\bibitem{Li:2024jlq}
H.~S.~Li,
Molecular pentaquark magnetic moments in heavy pentaquark chiral perturbation theory,
\href{https://journals.aps.org/prd/abstract/10.1103/PhysRevD.109.114039}{Phys. Rev. D \textbf{109}, 114039 (2024)}.
%doi:10.1103/PhysRevD.109.114039
%[arXiv:2401.14759 [hep-ph]].
%28 citations counted in INSPIRE as of 22 May 2026

%\cite{Mutuk:2024ltc}
\bibitem{Mutuk:2024ltc}
H.~Mutuk and X.~W.~Kang,
Unveiling the structure of hidden-bottom strange pentaquarks via magnetic moments,
\href{https://www.sciencedirect.com/science/article/pii/S0370269324003307?via%3Dihub}{Phys. Lett. B \textbf{855}, 138772 (2024)}.
%doi:10.1016/j.physletb.2024.138772
%[arXiv:2405.07066 [hep-ph]].
%15 citations counted in INSPIRE as of 22 May 2026

%\cite{Mutuk:2024jxf}
\bibitem{Mutuk:2024jxf}
H.~Mutuk,
Magnetic moments of hidden-bottom pentaquark states,
\href{https://link.springer.com/article/10.1140/epjc/s10052-024-13263-x}{Eur. Phys. J. C \textbf{84}, 874 (2024)}.
%doi:10.1140/epjc/s10052-024-13263-x
%[arXiv:2403.16616 [hep-ph]].
%15 citations counted in INSPIRE as of 22 May 2026

%\cite{Mutuk:2024elj}
\bibitem{Mutuk:2024elj}
H.~Mutuk,
Magnetic Moment of $\Xi_b(6227)$ as Molecular Pentaquark State,
\href{https://arxiv.org/abs/2403.13896}{arXiv:2403.13896}.
%2 citations counted in INSPIRE as of 22 May 2026

%\cite{Lai:2024jfe}
\bibitem{Lai:2024jfe}
B.~J.~Lai, F.~L.~Wang and X.~Liu,
Investigating the M1 radiative decay behaviors and the magnetic moments of the predicted triple-charm molecular-type pentaquarks,
\href{https://journals.aps.org/prd/abstract/10.1103/PhysRevD.109.054036}{Phys. Rev. D \textbf{109}, 054036 (2024)}.
%doi:10.1103/PhysRevD.109.054036
%[arXiv:2402.07195 [hep-ph]].
%10 citations counted in INSPIRE as of 18 Mar 2026

%\cite{Sheng:2024hkf}
\bibitem{Sheng:2024hkf}
L.~C.~Sheng, J.~Y.~Huo, R.~Chen, F.~L.~Wang and X.~Liu,
Exploring the mass spectrum and electromagnetic property of the $\Xi_{cc}K^{(*)}$ and $\Xi_{cc}\bar{K}^{(*)}$ molecules,
\href{https://journals.aps.org/prd/abstract/10.1103/PhysRevD.110.054044}{Phys. Rev. D \textbf{110}, 054044 (2024)}.
%doi:10.1103/PhysRevD.110.054044
%[arXiv:2406.16115 [hep-ph]].
%9 citations counted in INSPIRE as of 13 Mar 2026

%\cite{Wang:2024kke}
\bibitem{Wang:2024kke}
F.~L.~Wang, S.~Q.~Luo, R.~Q.~Qian and X.~Liu,
Spectroscopic properties of double-strangeness molecular tetraquarks,
\href{https://journals.aps.org/prd/abstract/10.1103/PhysRevD.110.114041}{Phys. Rev. D \textbf{110}, 114041 (2024)}.
%doi:10.1103/PhysRevD.110.114041
%[arXiv:2410.15339 [hep-ph]].
%7 citations counted in INSPIRE as of 13 Mar 2026
		
%\cite{Wang:2024sbw}
\bibitem{Wang:2024sbw}
F.~L.~Wang, S.~Q.~Luo and X.~Liu,
Unveiling the composition of the single-charm molecular pentaquarks: insights from radiative decay and magnetic moment,
\href{https://link.springer.com/article/10.1140/epjc/s10052-025-13891-x}{Eur. Phys. J. C \textbf{85}, 216 (2025)}.
%doi:10.1140/epjc/s10052-025-13891-x
%[arXiv:2403.13532 [hep-ph]].
%5 citations counted in INSPIRE as of 18 Mar 2026

%\cite{Ozdem:2024yel}
\bibitem{Ozdem:2024yel}
U.~{\"O}zdem,
Investigation on the electromagnetic properties of the $ D^{(*)} \Sigma _c^{(*)}$ molecules,
\href{https://link.springer.com/article/10.1140/epja/s10050-024-01477-2}{Eur. Phys. J. A \textbf{61}, 10 (2025)}.
%doi:10.1140/epja/s10050-024-01477-2
%[arXiv:2405.07273 [hep-ph]].
%12 citations counted in INSPIRE as of 12 May 2026

%\cite{Ozdem:2025olj}
\bibitem{Ozdem:2025olj}
U.~{\"O}zdem,
Electromagnetic properties of the $D_{s1}^+(2460)$, $D_{s1}^+(2536)$, and their bottom partners in a molecular configuration,
\href{https://journals.aps.org/prd/abstract/10.1103/y77p-ch1k}{Phys. Rev. D \textbf{112}, 114013 (2025)}.
%doi:10.1103/y77p-ch1k
%[arXiv:2510.17477 [hep-ph]].
%3 citations counted in INSPIRE as of 12 May 2026

%\cite{Zhang:2025ame}
\bibitem{Zhang:2025ame}
C.~K.~Zhang, F.~L.~Wang, S.~Q.~Luo and X.~Liu,
Electromagnetic properties of possible triple-charm molecular hexaquarks,
\href{https://link.springer.com/article/10.1140/epjc/s10052-025-14317-4}{Eur. Phys. J. C \textbf{85}, 582 (2025)}.
%doi:10.1140/epjc/s10052-025-14317-4
%[arXiv:2505.10318 [hep-ph]].
%4 citations counted in INSPIRE as of 22 May 2026

%\cite{Li:2025ddx}
\bibitem{Li:2025ddx}
H.~S.~Li,
Axial charges and magnetic moments of the decuplet pentaquark family,
\href{https://journals.aps.org/prd/abstract/10.1103/32n9-j3pp}{Phys. Rev. D \textbf{113}, 056017 (2026)}.
%doi:10.1103/32n9-j3pp
%[arXiv:2511.12858 [hep-ph]].
%3 citations counted in INSPIRE as of 23 May 2026

%\cite{Fu:2025lfo}
\bibitem{Fu:2025lfo}
H.~L.~Fu, F.~K.~Guo, C.~Hanhart and A.~Nefediev,
What can we learn from the radiative decays of the Ds1(2460) meson?,
\href{https://journals.aps.org/prd/abstract/10.1103/786l-tjny}{Phys. Rev. D \textbf{113}, 074026 (2026)}.
%doi:10.1103/786l-tjny
%[arXiv:2512.05476 [hep-ph]].
%3 citations counted in INSPIRE as of 23 May 2026

%\cite{Zhu:2025abk}
\bibitem{Zhu:2025abk}
S.~H.~Zhu, F.~L.~Wang and X.~Liu,
Electromagnetic characteristics as probes into the inner structures of the predicted $\Xi _c^{(',*)}D^{(*)}_s$ molecular states,
\href{https://link.springer.com/article/10.1140/epjc/s10052-026-15552-z}{Eur. Phys. J. C \textbf{86}, 385 (2026)}.
%doi:10.1140/epjc/s10052-026-15552-z
%[arXiv:2510.18492 [hep-ph]].
%5 citations counted in INSPIRE as of 22 May 2026

%\cite{Ozdem:2026wmf}
\bibitem{Ozdem:2026wmf}
U.~{\"O}zdem,
Structural dissection of hadronic molecules: The $D^{(*)}\bar{K}^{(*)}$ family under QCD light-cone sum rules,
\href{https://arxiv.org/abs/2602.10638}{arXiv:2602.10638}.
%1 citations counted in INSPIRE as of 24 May 2026

%\cite{Vujmilovic:2025czt}
\bibitem{Vujmilovic:2025czt}
I.~Vujmilovic, S.~Collins, L.~Leskovec and S.~Prelovsek,
Electromagnetic Form Factors and Structure of the $T_{bb}$ Tetraquark from Lattice QCD,
\href{https://journals.aps.org/prl/abstract/10.1103/jnsy-5nhj}{Phys. Rev. Lett. \textbf{136}, 161901 (2026)}.
%doi:10.1103/jnsy-5nhj
%[arXiv:2510.17549 [hep-lat]].
%9 citations counted in INSPIRE as of 24 May 2026

%\cite{Klempt:2002ap}
\bibitem{Klempt:2002ap}
E.~Klempt, F.~Bradamante, A.~Martin and J.~M.~Richard,
Antinucleon nucleon interaction at low energy: Scattering and protonium,
\href{https://www.sciencedirect.com/science/article/abs/pii/S0370157302001448?via%3Dihub}{Phys. Rept. \textbf{368}, 119-316 (2002)}.
%doi:10.1016/S0370-1573(02)00144-8
%215 citations counted in INSPIRE as of 17 Mar 2026

%\cite{Ding:2008gr}
\bibitem{Ding:2008gr}
G.~J.~Ding,
Are $Y(4260)$ and $Z^+_2(4250)$ $D_1D$ or $D_0 D^*$ Hadronic Molecules?,
\href{https://journals.aps.org/prd/abstract/10.1103/PhysRevD.79.014001}{Phys. Rev. D \textbf{79}, 014001 (2009)}.
%doi:10.1103/PhysRevD.79.014001
%[arXiv:0809.4818 [hep-ph]].
%234 citations counted in INSPIRE as of 21 May 2026
		
%\cite{Wang:2024ukc}
\bibitem{Wang:2024ukc}
J.~Z.~Wang, Z.~Y.~Lin, B.~Wang, L.~Meng and S.~L.~Zhu,
Double pole structures of $X_1(2900)$ as the P-wave $\bar{D}^{*}K^{*}$ resonances,
\href{https://journals.aps.org/prd/abstract/10.1103/PhysRevD.110.114003}{Phys. Rev. D \textbf{110}, 114003 (2024)}.
%doi:10.1103/PhysRevD.110.114003
%[arXiv:2408.08965 [hep-ph]].
%10 citations counted in INSPIRE as of 12 Mar 2026

%\cite{Wang:2025wpc}
\bibitem{Wang:2025wpc}
F.~L.~Wang, S.~Q.~Luo and X.~Liu,
Predicting charmed-strange molecular tetraquarks with $K^{(*)}$ and $T$-doublet (anti-)charmed meson,
\href{https://arxiv.org/abs/2510.17244}{arXiv:2510.17244}.
%0 citations counted in INSPIRE as of 13 Mar 2026
		
%\cite{Riska:2000gd}
\bibitem{Riska:2000gd}
D.~O.~Riska and G.~E.~Brown,
Nucleon resonance transition couplings to vector mesons,
\href{https://www.sciencedirect.com/science/article/abs/pii/S0375947400003626?via%3Dihub}{Nucl. Phys. A \textbf{679}, 577-596 (2001)}.
%doi:10.1016/S0375-9474(00)00362-6
%[arXiv:nucl-th/0005049 [nucl-th]].
%151 citations counted in INSPIRE as of 13 Mar 2026

%\cite{Isola:2003fh}
\bibitem{Isola:2003fh}
C.~Isola, M.~Ladisa, G.~Nardulli and P.~Santorelli,
Charming penguins in $B \to K^* \pi$, $K(\rho, \omega, \phi)$ decays,
\href{https://journals.aps.org/prd/abstract/10.1103/PhysRevD.68.114001}{Phys. Rev. D \textbf{68}, 114001 (2003)}.
%doi:10.1103/PhysRevD.68.114001
%[arXiv:hep-ph/0307367 [hep-ph]].
%226 citations counted in INSPIRE as of 13 Mar 2026

%\cite{Cleven:2016qbn}
\bibitem{Cleven:2016qbn}
M.~Cleven and Q.~Zhao,
Cross section line shape of $e^+e^-\to\chi_{c0}\omega$ around the $Y(4260)$ mass region,
\href{https://linkinghub.elsevier.com/retrieve/pii/S0370269317301545}{Phys.\ Lett.\ B {\bf 768}, 52 (2017)}.
%doi:10.1016/j.physletb.2017.02.041
%[arXiv:1611.04408 [hep-ph]].
%%CITATION = doi:10.1016/j.physletb.2017.02.041;%%
%9 citations counted in INSPIRE as of 18 Apr 2019

%\cite{ParticleDataGroup:2022pth}
\bibitem{ParticleDataGroup:2022pth}
R.~L.~Workman \textit{et al.} [Particle Data Group],
Review of Particle Physics,
\href{https://academic.oup.com/ptep/article/2022/8/083C01/6651666?login=false}{PTEP \textbf{2022}, 083C01 (2022)}.
%doi:10.1093/ptep/ptac097
%5494 citations counted in INSPIRE as of 13 Mar 2026

%\cite{Bando:1987br}
\bibitem{Bando:1987br}
M.~Bando, T.~Kugo and K.~Yamawaki,
Nonlinear Realization and Hidden Local Symmetries,
\href{https://www.sciencedirect.com/science/article/abs/pii/0370157388900191?via%3Dihub}{Phys. Rept. \textbf{164}, 217-314 (1988)}.
%doi:10.1016/0370-1573(88)90019-1
%1492 citations counted in INSPIRE as of 11 May 2026
		
%\cite{Berestetskii:1982qgu}
\bibitem{Berestetskii:1982qgu}
V.~B.~Berestetskii, E.~M.~Lifshitz and L.~P.~Pitaevskii,
Quantum Electrodynamics,
Pergamon Press, 1982,
ISBN 978-0-7506-3371-0.
%53 citations counted in INSPIRE as of 21 May 2026
		
%\cite{Tornqvist:1993ng}
\bibitem{Tornqvist:1993ng}
N.~A.~Tornqvist,
From the deuteron to deusons, an analysis of deuteron - like meson meson bound states,
\href{https://link.springer.com/article/10.1007/BF01413192}{Z. Phys. C \textbf{61}, 525-537 (1994)}.
%doi:10.1007/BF01413192
%[arXiv:hep-ph/9310247 [hep-ph]].
%591 citations counted in INSPIRE as of 14 Mar 2026
		
%\cite{Tornqvist:1993vu}
\bibitem{Tornqvist:1993vu}
N.~A.~Tornqvist,
On deusons or deuteron - like meson meson bound states,
\href{https://link.springer.com/article/10.1007/BF02734018}{Nuovo Cim. A \textbf{107}, 2471-2476 (1994)}.
%doi:10.1007/BF02734018
%[arXiv:hep-ph/9310225 [hep-ph]].
%100 citations counted in INSPIRE as of 14 Mar 2026

%\cite{Kumar:2005ei}
\bibitem{Kumar:2005ei}
S.~Kumar, R.~Dhir and R.~C.~Verma,
Magnetic moments of charm baryons using effective mass and screened charge of quarks,
\href{https://iopscience.iop.org/article/10.1088/0954-3899/31/2/006}{J. Phys. G \textbf{31}, 141-147 (2005)}.
%doi:10.1088/0954-3899/31/2/006
%42 citations counted in INSPIRE as of 11 May 2026

%\cite{Ramalho:2009gk}
\bibitem{Ramalho:2009gk}
G.~Ramalho, K.~Tsushima and F.~Gross,
A Relativistic quark model for the Omega- electromagnetic form factors,
\href{https://journals.aps.org/prd/abstract/10.1103/PhysRevD.80.033004}{Phys. Rev. D \textbf{80}, 033004 (2009)}.
%doi:10.1103/PhysRevD.80.033004
%[arXiv:0907.1060 [hep-ph]].
%79 citations counted in INSPIRE as of 11 May 2026

%\cite{ParticleDataGroup:2024cfk}
\bibitem{ParticleDataGroup:2024cfk}
S.~Navas \textit{et al.} [Particle Data Group],
Review of particle physics,
\href{https://journals.aps.org/prd/abstract/10.1103/PhysRevD.110.030001}{Phys. Rev. D \textbf{110}, 030001 (2024)}.
%doi:10.1103/PhysRevD.110.030001
%4374 citations counted in INSPIRE as of 18 Mar 2026

\bibitem{Quantum Theory of Angular Momentum}
V.~K.~Khersonskii, A.~N.~Moskalev and D.~A.~Varshalovich,
Quantum Theory of Angular Momentum,
\href{https://www.worldscientific.com/worldscibooks/10.1142/0270#t=aboutBook}{World Scientific Publishing Company, Singapore, 1998}.

%\cite{Fu:2021wde}
\bibitem{Fu:2021wde}
H.~L.~Fu, H.~W.~Grie{\ss}hammer, F.~K.~Guo, C.~Hanhart and U.~G.~Mei{\ss}ner,
Update on strong and radiative decays of the $D_{s0}^*(2317)$ and $D_{s1}(2460)$ and their bottom cousins,
\href{https://link.springer.com/article/10.1140/epja/s10050-022-00724-8}{Eur. Phys. J. A \textbf{58}, 70 (2022)}.
%doi:10.1140/epja/s10050-022-00724-8
%[arXiv:2111.09481 [hep-ph]].
%37 citations counted in INSPIRE as of 08 May 2026

%\cite{Xiao:2016hoa}
\bibitem{Xiao:2016hoa}
C.~J.~Xiao, D.~Y.~Chen and Y.~L.~Ma,
Radiative and pionic transitions from the $D_{s1}(2460)$ to the $D_{s0}^\ast(2317)$,
\href{https://journals.aps.org/prd/abstract/10.1103/PhysRevD.93.094011}{Phys. Rev. D \textbf{93}, 094011 (2016)}.
%doi:10.1103/PhysRevD.93.094011
%[arXiv:1601.06399 [hep-ph]].
%42 citations counted in INSPIRE as of 08 May 2026

%\cite{Godfrey:2003kg}
\bibitem{Godfrey:2003kg}
S.~Godfrey,
Testing the nature of the $D_{sJ}^*(2317)^+$ and $D_{sJ}(2463)^+$ states using radiative transitions,
\href{https://www.sciencedirect.com/science/article/pii/S0370269303009493?via%3Dihub}{Phys. Lett. B \textbf{568}, 254-260 (2003)}.
%doi:10.1016/j.physletb.2003.06.049
%[arXiv:hep-ph/0305122 [hep-ph]].
%211 citations counted in INSPIRE as of 08 May 2026

%\cite{Liu:2006jx}
\bibitem{Liu:2006jx}
X.~Liu, Y.~M.~Yu, S.~M.~Zhao and X.~Q.~Li,
Study on decays of $D^*_{sJ}(2317)$ and $D_{sJ}(2460)$ in terms of the CQM model,
\href{https://link.springer.com/article/10.1140/epjc/s2006-02564-0}{Eur. Phys. J. C \textbf{47}, 445-452 (2006)}.
%doi:10.1140/epjc/s2006-02564-0
%[arXiv:hep-ph/0601017 [hep-ph]].
%35 citations counted in INSPIRE as of 08 May 2026

%\cite{Wang:2006fg}
\bibitem{Wang:2006fg}
F.~L.~Wang, X.~L.~Chen, D.~H.~Lu, S.~L.~Zhu and W.~Z.~Deng,
Decays of $D^*_{sj}(2317)$ and $D_{sj}(2460)$ Mesons in the Quark Model,
\href{https://zgwlc.xml-journal.net/fileZGWLC/journal/article/zgwlc/2006/11/PDF/2006-0108.pdf#2#1}{HEPNP \textbf{30}, 1041-1047 (2006)}.
%[arXiv:hep-ph/0604090 [hep-ph]].
%13 citations counted in INSPIRE as of 08 May 2026

%\cite{Colangelo:2005hv}
\bibitem{Colangelo:2005hv}
P.~Colangelo, F.~De Fazio and A.~Ozpineci,
Radiative transitions of $D^*_{sJ}(2317)$ and $D_{sJ}(2460)$,
\href{https://journals.aps.org/prd/abstract/10.1103/PhysRevD.72.074004}{Phys. Rev. D \textbf{72}, 074004 (2005)}.
%doi:10.1103/PhysRevD.72.074004
%[arXiv:hep-ph/0505195 [hep-ph]].
%138 citations counted in INSPIRE as of 08 May 2026


\end{thebibliography}
\end{document}